\documentclass[manuscript,screen,table,nonacm]{acmart}

\AtBeginDocument{%
  \providecommand\BibTeX{{%
    \normalfont B\kern-0.5em{\scshape i\kern-0.25em b}\kern-0.8em\TeX}}}

\setcopyright{acmcopyright}
\copyrightyear{2018}
\acmYear{2023}
\acmDOI{XXXXXXX.XXXXXXX}

\usepackage{amsmath,amsfonts}
\usepackage{enumerate}
\usepackage{bbding}

\usepackage{bm}
\usepackage{xcolor}
\usepackage{diagbox}
\usepackage{makecell}
\usepackage{booktabs} 
\usepackage{multirow}

\begin{document}

\title{A Survey on Privacy-Preserving Caching at Network Edge: Classification, Solutions, and Challenges}

\author{Xianzhi Zhang}
\email{zhangxzh9@mail2.sysu.edu.cn}
\affiliation{
  \institution{School of Computer Science and Engineering, Sun Yat-sen University}
  \city{Guangzhou}
  \country{China}
}
\affiliation{
\institution{School of Computing, Macquarie University}
\city{Sydney}
\country{Australia}
}

\author{Yipeng Zhou}
\email{yipeng.zhou@mq.edu.au}
\affiliation{
  \institution{School of Computing, Macquarie University}
  \city{Sydney}
  \country{Australia}}

\author{Di Wu}
\email{wudi27@mail.sysu.edu.cn}
\authornote{Corrosponding Author.}
\affiliation{
    \institution{School of Computer Science and Engineering, Sun Yat-sen University}
    \city{Guangzhou}
    \country{China}
}

\author{Quan Z. Sheng}
\email{michael.sheng@mq.edu.au}
\author{Shazia Riaz }
\email{Shazia.Riaz@mq.edu.au}
\affiliation{
   \institution{School of Computing, Macquarie University}
   \city{Sydney}
   \country{Australia}
}

\author{Miao Hu}
\email{humiao5@mail.sysu.edu.cn}
\author{Linchang Xiao}
\email{xiaolch3@mail2.sysu.edu.cn}
\affiliation{%
    \institution{School of Computer Science and Engineering, Sun Yat-sen University}
    \city{Guangzhou}
    \country{China}
  }

\renewcommand{\shortauthors}{Zhang and Zhou, et al.}

\begin{abstract}
Caching content at the edge network is a popular and effective technique widely deployed to alleviate the burden of network backhaul, shorten service delay and improve service quality. 
However, there has been some controversy over privacy violations in caching content at the edge network. 
On the one hand, the multi-access open edge network provides an ideal entrance or interface for external attackers to obtain private data from edge caches by extracting sensitive information. 
On the other hand, privacy can be infringed on by curious edge caching providers through caching trace analysis targeting the achievement of better caching performance or higher profits.
Therefore, an in-depth understanding of privacy issues in edge caching networks is vital and indispensable for creating a privacy-preserving caching service at the edge network. In this article, we are among the first to fill this gap by examining privacy-preserving techniques for caching content at the edge network. Firstly, we provide an introduction to the background of privacy-preserving edge caching (PPEC). Next, we summarize the key privacy issues and present a taxonomy for caching at the edge network from the perspective of private information. Additionally, we conduct a retrospective review of the state-of-the-art countermeasures against privacy leakage from content caching at the edge network.
Finally, we conclude the survey and envision challenges for future research.
\end{abstract}




\maketitle

\section{Introduction}
Content caching at the edge network is driven by two factors. First, the population of networked devices has become astronomical due to advances in intelligent terminals and the broad deployment of the Internet of Things (IoT)~\cite{Cui2022, Guo2022, Cui2023, Zhao2023}. Second, the Internet content market is blooming due to the proliferation of various multimedia content~\cite{Zhang2022, Ni2021}. 
\textcolor{black}{According to the report by Splunk, a Cisco company, there will be approximately 5.44 billion Internet users worldwide in 2024, including 5.07 billion social media users~\cite{splunk2023internet}.}
As a result, network-based content delivery services are extremely bandwidth-consuming. 
\textcolor{black}{At the same time, emerging network technologies, such as Gigabit Ethernet, and 5G and beyond, are expected to provide extremely high data transmission rates and low access delays for terminal devices at the edge network to support time-sensitive services such as autonomous driving, industrial automation, high-quality video streaming, and virtual/enhanced emerging applications.}

Such a vast data flow brings two main challenges to the established networks: (1) It brings a heavy communication burden to the Internet core network links. During the peak hours of network usage, a large amount of content transmission will inevitably aggravate the link burden of the core network, causing network congestion and increasing network operating costs; (2) It will also prolong the service delay of content transmission from remote servers to end devices, which will adversely influence users' service Quality-of-Experience (QoE) or even ruin the reliability of delay-sensitive applications.


\textcolor{black}{Edge Caching is a technique that involves storing content in close proximity to end users, typically at or near the point of user access or ahead of the core network~\cite{Ni2021}.} Its primary objective is to shorten service latency and enhance content delivery performance by bringing content closer to the users who request it.
When users request content that is available in \textcolor{black}{edge caches (ECs)}, their requests can be directly served at the edge network with a high Quality-of-Service (QoS). 
However, if the requested content is not available in the EC, it can be redirected to a remote server, such as a data center.
{\color{black} Here, we make a brief introduction to edge caching from five aspects:}

\textcolor{black}{\textit{\textbf{Benefit of edge caching.}}} Caching content at the edge network is effective in reducing the burden of network backhaul~\cite{Yang2019, Jiang2017, Qiao2022}, shortening service latency~\cite{Zhang2022d, Qiao2022, Cui2020c}, and diminishing resource cost~\cite{Jiang2017, Hassanpour2023}. First,  it is common to cache popular content at the edge network through which the edge network can offload the access of requests and hence reduce the backhaul data flow.
Even though the caching capability is limited at the edge network, edge caches can offload up to 35\% of the traffic burden over backhaul links ~\cite{Ni2021}.  
Second,  the service latency can be shortened by caching content on edge devices near end users. In particular, a shortened latency is critical for content delivery of latency-sensitive applications~\cite{Ni2021}. 
Third, edge networks can make content access inexpensive since caching content at edge devices can avoid the bottleneck. For example, in wireless edge networks, spectral efficiency and energy efficiency can be improved by about 900\% and 500\%, respectively, by using edge networks for caching content~\cite{Liu2016}.

{\color{black}
\textbf{Where to cache:} Building on the work of Ni~\emph{et al.}~\cite{Ni2021}, we further identify three main entities in edge networks for edge caching as follows: 
    (1) \textit{End devices} (e.g., smartphones, laptops, intelligent vehicles, and industrial IoT devices) carried by users will generate requests for downloading content via networks~\cite{Cui2022, Guo2022}. It is possible that end devices can share content through Device-to-Device (D2D) communications with licensed-band or unlicensed-band protocols.
    (2) \textit{Access infrastructures}, utilizing wired and/or wireless communication technologies, can support end devices in accessing the Internet. These infrastructures include 5G small base stations (SBSs)~\cite{Xu2019}, WiFi routers, local switches, and roadside units (RSUs) in the Internet of Vehicles (IoV)~\cite{Cui2020, Zhang2022b}. Popular content can be cached within these access infrastructures to promptly serve user requests.
    (3) \textit{Edge servers} (ESs) positioned ahead of the core network, such as edge nodes (ENs) in the Content Delivery Network (CDN)~\cite{Cui2020c}, edge routers in the Information-Centric Network (ICN)~\cite{Sivaraman2021, Xue2019, Xue2018}, and macro base stations operated by Internet service providers (ISPs)~\cite{Araldo2018}, can be utilized as ECs, a concept known as \textit{in-network-edge caching}. These ESs, typically maintained by third-party suppliers, are the core points for multi-access edge networks, enhancing various content delivery applications. 

\textbf{What to Cache.} In edge caching systems, determining what content to cache is crucial for optimizing cache space utilization and reducing latency. The content to be cached generally falls into three categories:
(1) \textit{User-related popular content:} This includes content that is frequently requested by end users, such as web pages, videos, images, and other multimedia files~\cite{Cui2020c,Wu2016}. Caching such content at the edge improves user experience by reducing service delay when users access commonly accessed content.
(2) \textit{Public and static content:} This category includes high-reuse, non-user-specific content associated with applications~\cite{Ni2021}, such as JavaScript files, CSS stylesheets, icons, PDF documents, and API responses. Caching these static resources decreases application load times and reduces the burden on central servers.
(3) \textit{Edge-computable and storable content:} This includes data that can be computed and stored directly at the edge, such as model parameters for federated learning~\cite{Liu2022,Qiao2022}, user patterns and content popularity for edge caching decisions~\cite{Cui2020,Cui2020c}, and IoT sensor data awaiting processing~\cite{Wang2020,Yu2021b}. Caching such content helps minimize backhaul traffic, enables efficient edge processing, and reduces overall latency by avoiding redundant computations.

\textbf{How to Cache.} Edge caching strategies can be broadly classified into reactive and proactive approaches.
(1) \textit{Reactive caching} employs eviction-based methods that decide whether to cache a specific content item only after it has been requested. This approach often relies on empirical formulas and classical caching algorithms, such as LRU (Least Recently Used) and LFU (Least Frequently Used), as well as their variants~\cite{Famaey2013, Shafiq2014}. While these algorithms are simple and efficient, they frequently encounter challenges in selecting optimal parameter values, which can limit their performance in dynamic and diverse edge environments.
(2) \textit{Proactive caching} involves predictive methods to determine what content should be cached before any user requests are made. This approach leverages content popularity predictions and user behaviour profiles to make caching decisions at edge networks in advance. Advanced machine learning models, such as LSTM (Long Short-Term Memory) networks~\cite{Feng2019}, are often employed to forecast content demand based on historical request patterns in proactive caching. These learning-driven methods generally offer superior caching performance compared to classical algorithms by automatically adjusting model parameters. However, they may require extensive computational capacity and high-quality training data, both of which may be often limited at the edge.}

\textcolor{black}{\textit{\textbf{Privacy concerns of edge caching.}}} 
Despite the enormous benefits brought by caching content at the edge network, there has been some controversy over privacy violations brought by such caching. 
The concerns can be illustrated from two aspects. 
(1) The first privacy threat comes from external attackers, such as malicious user devices~\cite{ Sivaraman2021, Liang2019, Acs2019, Qian2020, Cui2020c, Tong2022}. The multi-access open property of the edge network provides an ideal entrance or interface for external attackers to obtain the cached content from the edge cache to extract sensitive information of end users~\cite{Xu2020}. Adversaries can obtain user-sensitive information by launching cache side-channel attacks~\cite{Sivaraman2021, Liang2019, Acs2019} and cache tampering attacks~\cite{Qian2020, Cui2020c, Tong2022}. 
However, it is non-trivial to embed advanced privacy protection mechanisms into edge networks due to the limited computing capacity, energy power, and storage space of edge devices.
(2) Second, user privacy can be infringed by curious edge caching providers by analyzing traces and management records. Due to limited caching space relative to the rapidly growing user population and the scale of content~\cite{Zhou2019}, edge network providers have a strong motivation to spy on user privacy in order to improve their resource utilization.  In other words, if content popularity can be accurately predicted, the right content can be cached by edge devices just before the surge of requests towards content~\cite{Zhang2022}. Hence, edge network providers are curious about users' personal interests and confidential information to infer their request behaviours, which can be extracted from users' historical request traces (e.g., request patterns~\cite{Zhang2022, Cui2020, Cui2020c}, identifiable information~\cite{Zhang2022b, Zhu2021, Araldo2018, Cui2020, Cui2020c}).
Edge network providers can implement monitoring attacks~\cite{Xue2018, Zhang2022b} and inference attacks~\cite{Qiao2022, Liu2022} in their systems to compromise users' privacy based on collected request information from users.   
Therefore, an in-depth understanding of privacy risks in privacy-preserving edge caching (PPEC) is crucial for the design of feasible solutions to achieve privacy-preserving content cache at the edge network.

\textcolor{black}{\textit{\textbf{Our contributions.}} Recently, significant progress has been made in enhancing privacy protection for content caching at edge networks.  
However, these works fail to provide a comprehensive discussion of the privacy issues in edge caching systems. For instance, Ren \emph{et al.}~\cite{Ren2019} primarily discuss the state-of-the-art researches on caching and privacy, respectively, in emerging edge computing paradigms. Besides, most surveys only discuss privacy issues for particular scenarios, 
such as IoT~\cite{Lidia2023, Kinza2021}, edge intelligence~\cite{Zhang2021}, the metaverse~\cite{Yazan2024}, and federated learning (FL)~\cite{Paolo2021}, while overlooking the distinct aspects of edge caching. The surveys in~\cite{Ni2021, Xiao2018a} addressed privacy-preserving solutions and countermeasures for edge caching without covering all relevant issues in a thorough manner.
In particular, there has been a lack of comprehensive discussion of protection methods targeting different types of private information in PPEC.
Given these limitations and the absence of comprehensive literature reviews, this article aims to thoroughly examine and categorize current works on privacy issues in edge caching scenarios.}
The main contributions of this article are summarized as follows:
(1) We make in-depth discussions on sensitive information in edge caching and propose a taxonomy from a private information perspective to classify existing works. 
To the best of our knowledge, this is the first such comprehensive exposition.
(2) We conduct a thorough review of recent high-quality research, diving into the background of privacy attacks and mitigation methods in the realm of edge caching. Our review encompasses the latest solutions proposed for enhancing privacy in edge caching, \textcolor{black}{which have been published in 
leading 
conferences and journals in the fields of computing networks, architecture, and privacy, such as CCS, INFOCOM, ToN, JSAC, TPDS, TIFS, and TDSC, as well as other top venues.} 
Based on different kinds of privacy information and attacks towards each kind of privacy information, we respectively review countermeasures to defend against attacks for protecting each kind of infringed privacy.
(3) Based on open problems outlined in existing works, we envision privacy-related open challenges in PPEC to provide insights for inspiring future research. 

\textcolor{black}{\textit{\textbf{Paper outline.}}} The remainder of this article is organized as follows.
Section~\ref{sec: Classification for data issues} provides an introduction to the taxonomy of privacy-preserving solutions that are based on the protection of sensitive information data in edge caching.
Section~\ref{sec: issues} provides a background discussion on privacy issues in the edge caching paradigm from two plain perspectives, i.e., privacy attacks and mitigation methods. 
From Section~\ref{sec: user privacy} to Section~\ref{sec: knowledge privacy}, we describe the possible privacy mitigation solutions for edge caching in correspondence with three main classes of privacy, i.e., user privacy, content privacy and knowledge privacy, respectively. 
Section~\ref{sec: Future Directions} provides open challenges and future research directions.
Finally, we make a summary in Section~\ref{sec: Conclusion}. To facilitate readability, we have compiled a summary of commonly used abbreviations for the solutions in Table~\ref{tab:abbreviations} in Appendix.

\section{Overview of Private Information in Edge Caching}\label{sec: Classification for data issues}

In this section, we overview sensitive information that should be protected to avoid privacy leakage in PPEC. 
In the realm of edge caching, sensitive information can be exposed by either users~\cite{Acs2019} unconsciously or edge servers~\cite{Cui2020,Araldo2018}. Specifically, users' sensitive information includes personal information, browsing history, location, and private content data, through their request traces to the ES or other service providers. 
Similarly, edge servers can leak their private information and extract knowledge from a collection of users who have interacted with edge servers\cite{Cui2020,Cui2020c}. 
Therefore, to build a privacy-preserving content caching system, the first step is to understand what private information can be exposed by users and edge servers. 
In Fig.~\ref{fig: Private data in edge cache}, we outline all kinds of sensitive information that should be protected in PPEC. We will elaborate on each kind of private information in this section.

\subsection{User Privacy}

In PPEC, all information related to users but not directly related to cached content is regarded as user privacy such as users' historical records, age, gender, and location. For our discussion, we classify all user privacy information into three types: \textit{request trace}, \textit{personal information} and \textit{location}. 

\subsubsection{\textbf{Request trace}}
{\color{black}
A request trace refers to a sequence of content requests and responses between an end device and ESs or service providers.
These traces often contain private information such as request patterns~\cite{Acs2019, Liang2019}, preferences~\cite{Cui2020, Qian2020}, and interests~\cite{Sivaraman2021, Cui2020c}. 
Advertisers or malicious attackers can exploit such information to make profits or harm.
Additionally, user request traces are valuable assets to service providers and caching systems. 
Service or content providers can analyze these request traces to infer users' behaviour patterns, such as the type of websites or applications they frequently use and the content they prefer to consume. ESs can maintain and analyze request traces to improve caching performance by predicting future requests, allowing for prefetching and caching popular content in advance. 

There are primary two risks associated with request traces: \textit{interception} and \textit{misuse}. First, request records can be intercepted and sniffed by other users and external attackers. For example, malicious users can use timing attacks~\cite{Sivaraman2021, Acs2019} to impersonate legitimate users, sending requests to the server. Attackers may then infer user request traces by exploiting the timing difference between cached and non-cached responses~\cite{Sivaraman2021, Acs2019}, facilitating illegal advertising and cache pollution attacks~\cite{Wu2016}. 
Second, edge servers and service providers, curious about user interest patterns, may misuse request traces for their purposes. For example, request traces can be exploited to develop trace-driven content caching algorithms, posing privacy threats from untrusted or profit-driven third-party edge servers~\cite{Araldo2018, Cui2020, Schlegel2022, Tong2022}. 

However, designing methods to preserve user privacy in edge caching systems is non-trivial. Most existing privacy-enhancing approaches fail to effectively address the privacy leakage risks users face in caching systems, as request records cannot be arbitrarily altered or obfuscated by users and must remain visible to service providers and edge caching servers to provide reliable services.
}

\begin{figure}[!tb]
\centering
\includegraphics[width=0.85\linewidth]{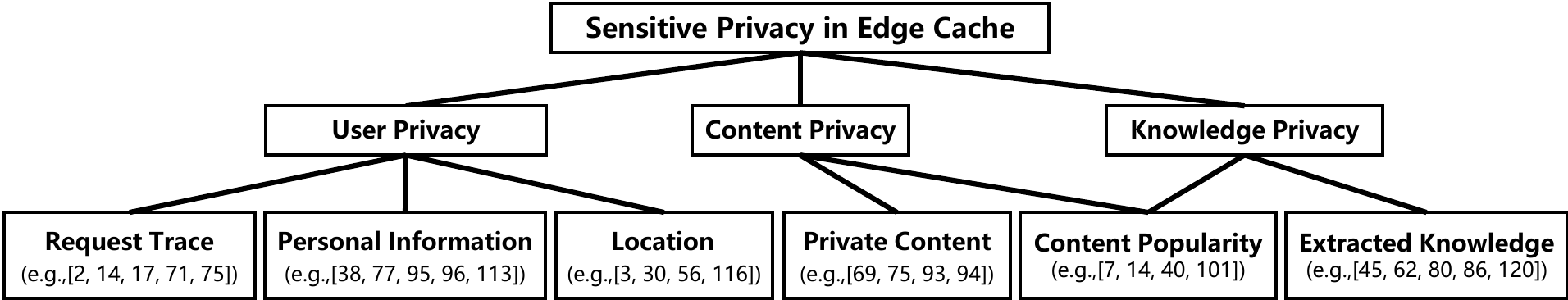}
\Description{The framework for PPEC encompasses six distinct types of data concerns: request traces, personal information, location data, machine learning knowledge, private content, and content popularity. These concerns can be primarily classified into three categories of private information: \textit{user privacy}, \textit{content privacy}, and \textit{knowledge privacy}.}
\vspace{-2mm}
\caption{The framework for PPEC encompasses six distinct types of data concerns, which can be primarily classified into three categories of private information: \textit{user privacy}, \textit{content privacy}, and \textit{knowledge privacy}.}
\vspace{-4mm}
\label{fig: Private data in edge cache}
\end{figure}




\subsubsection{\textbf{Personal information}} 

Personal information is a type of private information that can be 
mined to identify a specific end device or user in the network. 
Edge caching servers and service providers can obtain various types of personal information from users, depending on the specific context and implementation of the edge caching system. Typical examples of personal information that can be compromised in edge caching include:
(1) Identifier information such as pseudonyms and IP addresses. 
In particular, through IP addresses, we can identify a user's Internet service provider (ISP), approximate location, and other information, with which the edge cache (EC) can carry out sensitive operations, such as integrity verification~\cite{Tong2022} and cache admission control~\cite{Xue2019, Xue2018}. 
(2) Device information such as the operating system, connection type, browser type, and version, which is also essential for edge servers to provide high-performance edge caching and tailored content to users~\cite{Zhang2022b, Cui2022}. 
(3) Account-related information such as email address, gender, age, payment, and social relation, which can be captured by ECs or service providers when a user logs in or creates an account to access the service, potentially revealing more personal privacy~\cite{Zhang2022b, Cui2020c}.

Excessively exposing personal information by edge caching can result in annoying tracking and profiling. 
\textcolor{black}{When personal information is collected, edge caching servers and service providers can create detailed user profiles, encompassing browsing habits and interests. By identifying specific users or user groups, service providers can accurately predict future requests, allowing for content prefetching to reduce latency and improve Quality-of-Service (QoS). Additionally, detailed profiles facilitate targeted advertising and personalized recommendations, potentially increasing revenue. However, these practices raise ethical concerns~\cite{Zhang2022b, Cui2022}, including the potential for manipulation or discrimination against certain user groups.}
In addition, malicious nodes and attackers can take advantage of excessive disclosure of personal information to gain unauthorized access to user accounts~\cite{Xue2019, Xue2018} and pull off cache tampering attacks~\cite{Cui2022, Cui2020c, Tong2022}, resulting in financial losses and other harms.

\subsubsection{\textbf{Location}}
Location information is a critical type of privacy data carrying location,  spatial coordinates, and the current time of moving objects. In edge caching systems, there are two fundamental types of location information: users'  location information and Point of Interests (POIs).
When users access edge caching systems, they may unconsciously expose private location information in the following processes: (1) A user's geographic location can be exposed to the EC when accessing content or services directly from the EC~\cite{Cui2020b}; (2) Content providers (CP) and Edge Caching providers can proactively collect users' geographic location information to provide better content distribution services, such as predicting user moving patterns~\cite{Zhang2022a}; (3) In location-based services (LBS), users may provide their private geographic information and POIs to search for their interests in the EC~\cite{Amini2011, Cui2020b, GUYi-mingBAIGuang-weiSHENHang, Nisha2022}. This information can be abused, resulting in undesired tracking and profiling or even more severe consequences, such as location-based attacks.

Location information is sensitive and can be utilized to learn an individual's daily routine and movements. Service providers can use this information to deliver more relevant advertisements and cached content to users, potentially boosting profits. Yet, if malicious attackers obtain location information, it can put users at risk of physical harm. Malicious attackers can use location information to track a user's movement trajectories and potentially cause harm, particularly in the case of stalking or other criminal activities. 


\subsection{Content Privacy}

Content privacy refers to privacy information contained by the content stored and transmitted through edge caching systems, mainly including \textit{private content} and \textit{content popularity}. 

\subsubsection{\textbf{Private content}}

\textcolor{black}{
Private content refers to sensitive and confidential data that is stored and potentially cached by edge systems. We name such sensitive content data in edge caching systems as \textit{``private content"}. Given its sensitive nature, private content requires strict privacy protections to prevent unauthorized access and misuse. This type of content includes but is not limited to, video clips, photos, social media, and textual data from users, copyrighted materials, confidential business documents, and government secrets.}
For example, in mobile social networks, each user can be regarded as a content provider who can produce fresh content desiring that their content can be efficiently and accurately delivered to consumers~\cite{Xu2019, Zhou2019}. In this case,  edge computing is a feasible architecture for caching and delivering the content. Consumers in proximity ~\cite{Zhang2022a, Xu2020} or with close social relations~\cite{Wang2019} to a particular user content provider in social networks are more likely to request this content.  Thereby, using an edge server to cache and deliver content in mobile social networks can diminish bandwidth costs, which however raises privacy leakage risks. 

Briefly speaking, private content privacy can be infringed in several ways. 
First, edge servers are not trustworthy and can expose cached content to the public. Second,  malicious and unauthorized users at the edge network can access cached content during transmission or processing between end users and the EC or between different ECs.
For instance, in cache side-channel attacks~\cite{Sivaraman2021, Liang2019}, attackers attempt to access cached content by sending targeted requests, potentially allowing them to view sensitive information. For another instance, attackers can lodge cache tampering by injecting malicious content into the cache to exploit vulnerabilities in end-user systems or steal sensitive information~\cite{Qian2020, Cui2020c, Tong2022}. 

\subsubsection{\textbf{Content popularity}}
Content popularity can be defined as the relative frequency of a particular content to be requested by users.  It indicates the level of popularity of content among users. The popularity information is broadly utilized in improving caching efficiency, and caching the most popular content can effectively lower the content delivery cost. However, the popularity information is sensitive, unveiling the private preference information of users~\cite{Cui2020c, Yu2021b}.
Besides, it is possible that content popularity information can reveal sensitive information about content providers, such as their financial success and strategic direction, which should be kept confidentially~\cite{Araldo2018, Cui2020c}.

\textcolor{black}{
The popularity information is crucial for making effective edge caching decisions but is highly susceptible to leakage. 
Firstly, popularity information may be leaked during cooperative caching decision-making processes among edge caches. For instance, as the number of records owned by a single ES is limited, content providers may need to provide supplementary information~\cite{Araldo2018}. Additionally, edge caching servers may exchange popularity information to optimize caching decisions across the entire system~\cite{Yu2021b, Cui2020c}. Privacy leakage can occur because ESs might be untrusted, or the edge environment itself may be vulnerable to attacks~\cite{Cui2020c}. 
Secondly, popularity information can also be compromised through well-decided caching content. For example, through broadcasting cached content lists~\cite{Cui2020, Ni2021} or timing attacks~\cite{Sivaraman2021, Zhang2022b}, malicious entities can infer which content is more popular. This sensitive information, once exposed and tampered with, can be exploited to obtain illegal benefits, manipulate cache performance~\cite{Cui2020c, Tong2022}, or even launch cache tampering attacks~\cite{Cui2020c, Xu2020}.
Especially, as content popularity can describe specific content attributes and serve as key knowledge to improve caching efficiency, we consider it a unique type of information that intersects both content privacy and knowledge privacy, as shown in Fig.~\ref{fig: Private data in edge cache}.
}

\subsection{Knowledge Privacy}
\textcolor{black}{
Knowledge privacy refers to the insights, patterns, and parameters derived from datasets processed by machine learning models, typically owned by ECs or other service providers. Unlike user privacy and content privacy, which focus on data directly linked to users or content, knowledge privacy involves higher-level abstractions extracted from these data sources. In edge caching, service providers are particularly interested in the knowledge extracted from original datasets, as it is valuable for improving caching performance. For example, by leveraging prediction models based on this extracted knowledge, providers can make effective caching decisions in dynamic scenarios~\cite{Muller2017, Yang2019}, leading to significant improvements in edge caching performance~\cite{Ma2017b, Dhar2011, Zhang2022, Zhang2022a, Zhang2022b}.
Learning-based methods offer a feasible framework for making effective edge caching decisions, but they also pose risks of private information leakage during model training and prediction phases. Therefore, it is crucial to carefully consider and mitigate these privacy risks when employing learning-based methods for edge caching.
}

\section{Overview of Attack and Defence Methods}\label{sec: issues}

\begin{figure}[tb]
\centering
\includegraphics[width=0.8\linewidth]{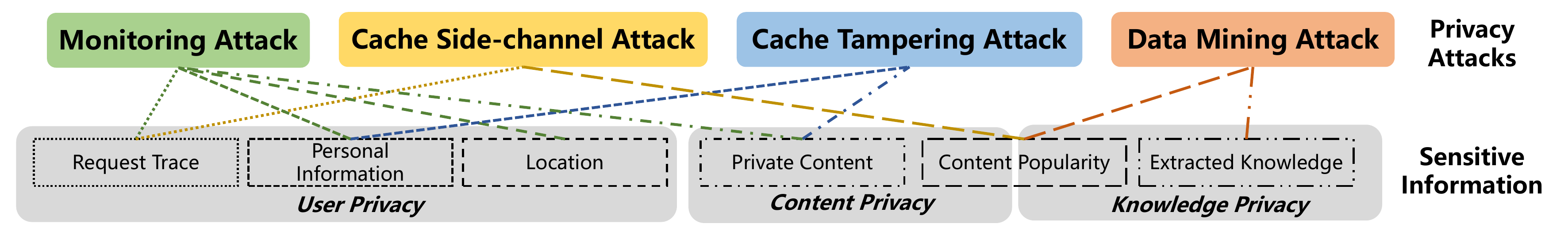}
\vspace{-3mm}
\Description{The possible privacy attacks on different sensitive information and the corresponding defence methods for enhancing privacy in edge caching systems.}
\caption{\textcolor{black}{The possible privacy attacks on different sensitive information in edge caching systems.}}
\label{fig: relation-attack-information}
\vspace{-4mm}
\end{figure}
\begin{figure}[tb]
\centering
\includegraphics[width=0.8\linewidth]{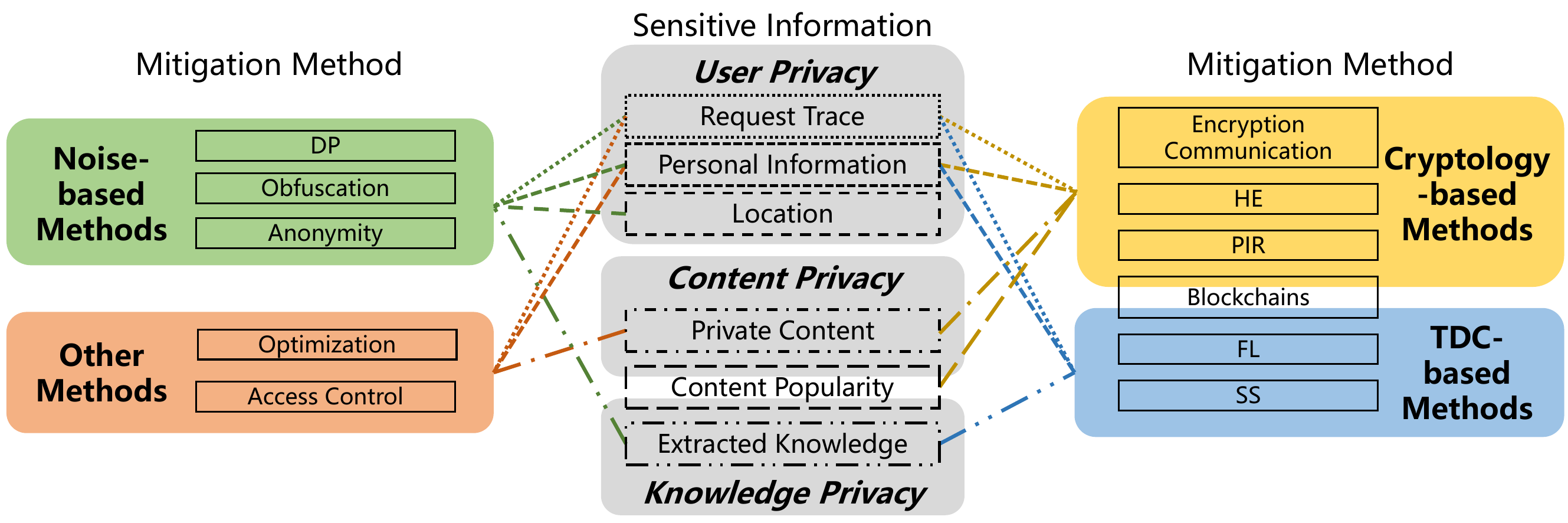}
\vspace{-4mm}
\Description{The possible privacy attacks on different sensitive data and the corresponding defence methods for enhancing privacy in edge caching systems.}
\caption{\textcolor{black}{The corresponding defence methods for enhancing different private information in edge caching systems.}}
\label{fig: relation-protection-information}
\vspace{-4mm}
\end{figure}

\textcolor{black}{
This section is divided into two parts: an overview of attack methods targeting each type of sensitive information in edge caching systems and a summary of defense methods against each type of attack. In Figs.~\ref{fig: relation-attack-information}-\ref{fig: relation-protection-information}, we present a relational map that illustrates the connections between potential privacy attacks, defense methods, and sensitive information in edge caching systems. In the remainder of this section, we briefly discuss each type of attack and defense methods as depicted in Fig.~\ref{fig: relation-attack-information} and Fig.~\ref{fig: relation-protection-information}, respectively.}

\subsection{Privacy Attack in Edge Caching Systems} 

There are mainly four types of privacy attacks in edge caching systems, which are \textit{monitoring attacks}, \textit{data mining attacks}, \textit{cache side-channel attacks} and \textit{cache tampering attacks}. 
We introduce these attacks with potential risk entities in this subsection.

\subsubsection{\textbf{Monitoring attack}}
Monitoring attacks, also known as eavesdropping attacks, can be divided into two main categories:
(1) The first is sniffing attacks on network communications, i.e., an adversary sniffs on network traffic through the edge caching node to read or intercept private information in network packets~\cite{Zhang2022b}. 
For example, the EC can monitor user requests during the caching service process. In other words, the edge caching operator can monitor users' requests intended to responding end users' requests and improve the caching efficiency. 
Through subsequent data analysis, edge caching managers can improve the caching efficiency and reduce the transmission delay of the requested content. However, a user request may contain private information, such as personal content preference~\cite{Cui2020, Yuan2016a, Schlegel2022}, location~\cite{Zhang2022a}, content popularity~\cite{Cui2020}, and other personal information~\cite{Kong2019, Xue2019, Xue2018}. 
Therefore, edge caching systems should take both caching efficiency and privacy preservation into account. 
Entities that can implement sniffing attacks in network communications include edge caching managers (e.g., content providers~\cite{Cui2020}, location service providers~\cite{Zhang2022a}, Internet services providers or based station~\cite{Yuan2016a},  edge devices~\cite{ Zhou2019, Cui2020, Xu2019, Schlegel2022, Tong2022}), malicious end devices~\cite{Xue2019, Xue2018,  Cui2020, Nikolaou2016}, and external adversaries~\cite{Zhang2022b}.
(2) The second type of monitoring attack is supervisory attacks on cached content, i.e.,  attackers conduct improper monitoring, replacement, pollution, and other privacy attack activities on cached content. By leveraging the illegal cache access, adversaries can obtain private data or information such as content popularity~\cite{Araldo2018, Cui2020c, Andreoletti2019a}, user preferences~\cite{Qian2020}, and other private information~\cite{Cui2020c, Tong2022}. If the cached content is not protected prudently, the user's privacy can be seriously compromised by edge caches, which are often deployed by honest but curious third parties (e.g., Internet service providers (ISPs)~\cite{Andreoletti2019a, Araldo2018}, edge servers~\cite{Cui2020c, Qian2020, Tong2022}, and end devices~\cite{Cui2020c, Qian2020}). 


\subsubsection{\textbf{Data mining attacks}}
Data mining attacks usually occur when an edge caching entity applies a learning-based caching algorithm to explore sensitive data for making caching decisions. 
Due to the high dynamics and complicated access patterns driven by users' interest~\cite{Muller2017, Yang2019},  designing an intelligent edge caching algorithm is essential to improve the caching performance. 
Commonly, learning-based methods make caching decisions by exploiting historical information to train a prediction model.  It is necessary to feed the model training  with  private and sensitive data related to users, and thus users may be reluctant to share.
Since edge caching decisions are generated by learning algorithms, edge caching becomes a trade-off problem between caching performance and privacy protection level.
As a consequence, learning-based methods in edge computing-assisted caching are usually vulnerable to two types of privacy risks: (1) \textit{exploratory}, in which adversaries investigate vulnerabilities (such as the training dataset, model parameters, and gradient data) without changing the training process, and 2) \textit{causative}, in which attackers manipulate and inject misleading training datasets to alter the machine learning model's training process~\cite{Tourani2018}. Additionally, previous research has shown that model parameters~\cite{Shokri2017} and gradients~\cite{Abadi2016, Zhao2020} of the machine learning model can be utilized to recover original sensitive and private information.
Learning-based methods provide a practical framework for making edge caching decisions but are susceptible to privacy risks that can compromise user privacy. 
The potential adversaries to launch data mining attacks include edge caching managers (e.g., content providers~\cite{Qiao2022, Cui2020}, Internet services providers~\cite{Wang2020}, edge devices~\cite{Qiao2022, Wang2022, Liu2022}) and malicious end devices~\cite{Wang2019}.


\subsubsection{\textbf{Cache side-channel attacks}}  
In cache side-channel attacks, attackers can learn privacy information about users and cached content by observing and measuring activities relevant to edge caches such as response time, power consumption, and return faults~\cite{Sivaraman2021, Liang2019, Wu2016, Acs2019}.
Through the edge caching service, users can conveniently upload their content to edge servers or download requested content from edge servers. 
Due to the open accessibility of edge caches (ECs)~\cite{Ni2021}, adversaries can easily access content cached by edge servers.  Adversaries can target a particular victim user by identifying content requested by the victim. The attacker may know the victim's content consumption habits or other specific characteristics to distinguish the victim from other users.
One of the main types of cache side-channel attacks is \textit{cache-timing attacks}, which allows attackers to determine whether specific content has been cached by comparing response times.
Previous works such as~\cite{Liang2019, Sivaraman2021, Acs2019} have explored cache-timing attacks in edge caching systems. 
An attacker can conduct the precise timing measurement to distinguish cache hits from misses, which can identify what content is cached at the ES. A cache hit means that a nearby user has requested the content 
(or has a high caching value), while a cache miss means that the content has not been requested (or has been ejected from the cache). A knowledgeable attacker can further determine whether the request is served by the provider or by a router somewhere along the provider's path~\cite{Liang2019}.
The main risk entities to launch cache side-channel attacks include malicious end devices~\cite{Liang2019, Acs2019, Wu2016} and external adversaries~\cite{Sivaraman2021, Zhang2022b}.

\subsubsection{\textbf{Cache tampering attacks}}
A cache tampering attack is a form of cyber-attack in which an adversary aims to alter content stored at an EC to gain unauthorized access, introduce illicit content and disrupt the caching system's regular operation. Within an edge network, a caching server offers a temporary storage area, holding frequently accessed content to expedite distribution. However, cache tampering attacks can transpire when an attacker modifies content cached in the ES or deceives the user to gain unauthorized content. The main risk entities to implement cache tampering attacks include edge servers~\cite{Tong2022, Qian2020}, malicious end devices~\cite{Qian2020, Cui2020c} and external adversary~\cite{Qian2020, Cui2020c}.

A typical instance of cache tampering attacks is \textit{cache poisoning}, where an attacker manipulates a Content Delivery Network (CDN) or edge server's cache to store and deliver malicious content or information~\cite{Zhang2022b,Cui2020c,Tong2022}.
For example, an attacker can exploit the vulnerability of the caching system by requesting a legitimate image with a specially crafted HTTP header. This header may contain malicious code that tricks the cache into storing a different image the attacker controls rather than the legitimate one. The next time when a user requests the original image, it will instead receive the attacker's image, which could contain harmful content such as malware or phishing links.

A variant of the cache tampering attack is the \textit{cache deception attack}, wherein an adversary gains access to private information by misleading and influencing a privileged user~\cite{gil2017web,Mirheidari2020,Mirheidari2022}. This process consists of two primary steps~\cite{gil2017web}. Initially, the attacker prompts the privileged user to request sensitive content and cache it in the ES. Subsequently, the adversary submits an identical request to the EC and retrieves the sensitive content.
For example, in named data networking~\cite{Lei2020}, an attacker creates a URL request targeting a victim user's private content by attaching a tag of a widely-used image. The victim is then enticed to make that request using its privilege~\cite{Mirheidari2020}. 
Upon retrieval, the cloud server disregards the invalid suffix and returns legitimate privacy content. The caching node retains the privacy content as the popular image's content. In this manner, the attacker can make the same request to access the identical privacy content in the EC, enabling them to acquire private content they are not authorized to access, potentially resulting in the victim's private content being leaked~\cite{Mirheidari2020,Mirheidari2022}.
The above kinds of cache tampering attacks give rise to unbearable privacy risks for users in edge caching systems.

\subsection{Mitigation Methods to Preserve 
  Privacy in Edge Caching Systems}

In the following subsection, we will provide a concise introduction to a range of methods that can effectively mitigate privacy leakage in content caching systems, which can be mainly classified into four types of methods: (1) \textit{noise-based methods}, (2) \textit{cryptology-based methods}, (3) \textit{trusted distributed computing}, and (4) other approaches. The specific solutions corresponding to each privacy mitigation approach are detailed in Section~\ref{sec: user privacy}-\ref{sec: knowledge privacy}. For easy reference, we also present a classification matrix for the solutions introduced in this survey based on countermeasures and privacy data in the realm of edge caching in Table~\ref{Tab: countermeasures} in Appendix.

\subsubsection{\textbf{Noise-based methods}}\
Noise-based methods represent the most prevalent approaches for preserving privacy within edge caching systems. These methods introduce disturbances to the real and genuine information before its exposure and interaction, effectively safeguarding privacy. Within the domain of edge caching, three specific types of methods are commonly employed: \textit{differential privacy (DP)}, \textit{confusion}, and \textit{anonymization}.

{\bf Differential privacy (DP)} is a data-sharing technique that allows data owners to share only some statistical characteristics of a database while withholding individual-specific information~\cite{Sivaraman2021,Acs2019}. 
    There are two ways to add noise in the DP mechanism. The traditional one is to add noise to the public database at the time of data release. However, the data collection agency is not always reliable, and thus local differential privacy (LDP) mechanism is also leveraged by data owners to distort original data before submitting private data. 
    The use of DP in edge caching systems can introduce distortion to the actual user or content information during the collection or release of sensitive data.
    DP is introduced to protect request traces~\cite{Zhang2018,Wang2019,Zhou2019,Zhang2022a,Sivaraman2021} personal information~\cite{Zhu2021,Zeng2020}, and machine learning models~\cite{Yu2021b} in edge caching systems.  
    
    
{\bf Confusion} mainly has two ways to enhance privacy in edge caching. The first one is cache obfuscation (such as proactive cache~\cite{Qian2020,Nikolaou2016}, off-path cache~\cite{Wu2016}, and request hit delay~\cite{Liang2019}), which can be used to protect users' requests when retrieving the content from monitoring or timing attacks in an untrusted or semi-trusted network environment. The second one is spatial confusion~\cite{Amini2011, GUYi-mingBAIGuang-weiSHENHang, Zhang2019b}, which is to protect the location information when users enjoy location-based services. 
    For instance, many pseudo requests for Points of Interest (PoIs) can be attached to the genuine request when retrieving content from the EC.
    
    {\bf Anonymous} methods are the last category of privacy risk mitigation measures.
    Anonymity is the act of not being named or using an alias, as opposed to the act of having a real identity~\cite{Cui2020b}. In particular, a set of public data satisfies $K$-anonymity if the information of any entity cannot be distinguished from at least $K-1$ other entities. $K$-anonymity method is often used to enhance geographical~\cite{Hu2018, Nisha2022, Yang2016} and personal privacy identity information~\cite{Sen2018, Cui2020b}  in edge caching systems. Besides, the anonymity group technology is also used in protecting users' identity information~\cite{Zhang2022b, Xue2019, Xue2018, Nguyen2023}.

\subsubsection{\textbf{Trusted distributed computing-based methods}}
Trusted distributed computing (TDC) methods encompass three primary mitigation frameworks—\textit{federated learning}, \textit{secret sharing}, and \textit{blockchain technology}—to safeguard privacy in the context of edge caching.

{\bf Federated learning (FL)} is a distributed machine learning technique that trains a learning-based algorithm across multiple decentralized devices or edge servers locally holding data samples without exposure~\cite{BrendanMcMahan2017}. The FL framework is one of the most essential methods to preserve private data during the machine learning process.
It is common that the FL framework~\cite{Yu2018,Wang2019a,Wang2020,Liu2022,Yu2020,Yu2020a,Li2020a} trains learning models by exposing model parameters or gradients. Instead, traditional machine learning methods need to collect raw data for the learning process. 
However, model parameters or gradients are also private assets of users since attackers can infer and recover users' private information from exposed model information. In addition, model information may have significant economic benefits, which will compromise the self-interest of model owners if they are exposed directly. A number of works~\cite{Yu2021b,Wang2022,Cui2022,Chen2022} have contributed to upgrading the FL framework by injecting noise or other interference to model information prior to exposure. 

{\bf Secret sharing (SS)}, also known as secret splitting, is a kind of secure multi-party computation and storage method in which each party gets a part of the secret, called a \textit{secret share}. 
The secretly shared information cannot be recovered unless a sufficient number of secret shares can be collected. A single share cannot restore the original secret. 
\textcolor{black}{
For example, the $(t,n)$-threshold scheme is the most straightforward secret-sharing scheme. In this scheme, there are a total of $n$ players, each receiving only one secret share. The secret can be recovered if at least $t$ players cooperate, where $t$ is the safety threshold parameter. 
In edge caching scenarios, secret data may include private content generated and stored by users and historical data required for edge caching decisions (e.g., request traces~\cite{Acs2019}, user preferences~\cite{Schlegel2022}, and content popularity information~\cite{Andreoletti2019}). While introducing SS may increase computational load, it significantly raises the cost for attackers attempting to obtain private information from the edge, reducing the risk of data breaches. Moreover, it enhances the fault tolerance of distributed caching systems.}
    
{\bf Blockchain} is a technical solution that does not rely on third parties to carry out network data storage, verification, transmission, and communication through its own distributed nodes. 
The blockchain mechanism can automate these four steps: (1) When a new blockchain transaction occurs, all participants can competitively record that transaction as a data block. (2) Following the rule of consensus, most participants on the blockchain network must vote for a valid recorded transaction. Depending on the type of network, the consensus mechanism of agreement can vary but is typically established at the start of the network. (3) Once participants have reached a consensus, transactions on the blockchain are written into blocks appended to a cryptographic hash that links blocks together as a chain. (4) The blockchain system finally updates and broadcasts a copy of the latest ledger to all participants. Blockchain can be used to enhance the protection of user preferences~\cite{Qian2020},  personal information~\cite{Lei2020,Dai2020,Vu2019, LiuJi2020}, and machine learning data~\cite{Cui2022} in edge caching systems. 


\subsubsection{\textbf{Cryptology-based methods}}

Cryptology-based methods, as a vital category of mitigation approaches, play a significant role in preserving privacy within edge caching systems. These methods employ cryptographic techniques to safeguard sensitive content or information, ensuring confidentiality, integrity, and authentication. Within the realm of edge caching, three specific types of methods are leveraged: \textit{encryption communication}, \textit{homomorphic encryption (HE)}, and \textit{private information retrieval (PIR)}.

    {\bf Encryption communication} is divided into two steps to protect the security and privacy of communication data. The first step is to encrypt communication data as follows. The sender encrypts the content by an encryption algorithm and the receiver's public key to obtain the ciphertext. The receiver, once getting the ciphertext, conducts decryption through the decryption algorithm and the private key to recover the original data. 
    Encryption communication is commonly used to protect the security of user request records and other data in Internet communications. 
    There are three main approaches for encryption in edge privacy-enhanced caching systems. One is symmetric encryption, which mainly uses Data Encryption Standard (DES), Advanced Encryption Standard (AES)~\cite{Pu2019, Yuan2016a}, or Searchable Encryption (SE)~\cite{Cui2020c}. Second, asymmetric encryption mainly includes Rivest-Shamir-Adleman (RSA)~\cite{Xu2019}, Attribute-Based Encryption (ABE)~\cite{Pu2019}, and Elliptic Curve Cryptography (ECC)~\cite{Zhang2022b, Cui2020}. 
    Finally, there are hashing algorithms~\cite{Xue2019, Xue2018, Xu2019}, which are sometimes used in blockchain~\cite{Cui2022, Lei2020}. 
    However, there are also three significant concerns with the use of cryptographic methods in edge caching systems. Firstly, due to the existence of encryption, third-party ECs often cannot directly use encrypted requests to retrieve related content, which may lead to the unavailability of ECs. Secondly, introducing encryption technology may pose computational pressure on the resource-constrained edge and end devices. Lastly, encryption communication may fail to prevent record privacy from content providers or service providers, who have the key to decrypt request information. Therefore, how to introduce cryptology-based techniques into edge caching systems is still a challenging problem.
    In addition, as a special communication encryption method, the digital signature~\cite{Kong2019, Chen2022, Jiang2020} is often used in edge caching systems to verify user identity and data reliability. 
    
    {\bf Homomorphic encryption (HE)} is a form of encryption by which each party co-computes the result of a specific objective function concerning their private data without a trusted third party (TTP). Each party cannot unveil private data from other parties even if the computation is completed. 
    In other words, it allows a participant to perform operations such as searching and multiplying encrypted data to produce correct results without decrypting it during calculation. HE can be used to protect user preferences~\cite{Cui2020} and information~\cite{Kong2019} when searching the EC.

    {\bf Private information retrieval (PIR)} is mainly used to protect a user's request record information~\cite{Tong2022, Kumar2019} in the edge caching system. When obtaining sensitive data, request records likely expose important privacy information of users. 
    PIR can help users with query needs to complete private data retrieval from the EC under the condition that the query privacy information is not leaked. In other words, the PIR technology can prevent attackers from obtaining precise query information and content items in cache retrieval or other sensitive queries. At the same time, PIR can let users obtain desired private content.

\subsubsection{\textbf{Other methods}} \textcolor{black}{Optimization-based methods and access control are two of the main approaches to enhancing the effectiveness of privacy protection in edge caching systems. In {\bf optimization-based methods}, metrics such as privacy exposure~\cite{Sivaraman2021,Andreoletti2019a} and credibility~\cite{Xu2019,Zhong2021,Cao2020} are mathematically modeled. The quantified metrics are then regarded as the objective function or constraint variables of the cache optimization problem. Finally, the optimal privacy protection decisions are deduced by solving the  optimization problem~\cite{Xu2020,Hassanpour2023,Shi2018,Hassanpour2021}.}
{\bf Access control} is an enforcing control method that allows or denies a user's access to a specific network resource, e.g., private content in the EC, based on the user's account or group. 
Without a defined authorization mechanism, access to system resources will have no restrictions, and thus illegal device operations can be easily launched. 
The EC can implement strict access control to filter out unauthorized or illegal accesses into the caching space for privacy protection. Access control methods have been applied to protect personal information~\cite{Lei2020, Cui2020c, Zhang2022b} and content privacy~\cite{Xue2019, Xue2018} in edge caching systems.
In the next section, we dive into the details of defence methods for protecting each type of sensitive information.

{\color{black}
\subsubsection{Summary}
In conclusion, privacy-preserving methods in edge caching systems can be broadly classified into four categories. \textit{Noise-based methods} are among the most prevalent techniques for safeguarding privacy. These methods are particularly effective in preventing monitoring attacks and cache side-channel attacks but may negatively impact the utility and performance of edge caching systems. For instance, DP is employed to add noises to data before or during its release, protecting user request traces, personal information, and machine learning models from privacy breaches. However, due to the limited privacy budget, strategic account~\cite{Abadi2016} or allocation mechanisms~\cite{Xiao2024} are required to mitigate the adverse influence of noises on utility. \textit{Noise-based methods} typically introduce an acceptable level of computational overhead, providing adaptive protection in real-time edge systems where capacity is limited.

\textit{Trusted distributed computing-based methods} refer to techniques that organize distributed devices for collaborative computation and storage while ensuring data privacy and security, such as federated learning, secret sharing, and blockchain technology. FL is valuable for protecting the privacy of machine learning models by enabling decentralized training without exposing raw data.  
These methods help mitigate the risk of data mining attacks, where adversaries might otherwise exploit original request traces or model parameters to infer private information.
However, \textit{trusted distributed computing-based methods},  regarded as a form of adaptive protection, also need integrate with some strict privacy-preserving methods and may introduce communication and computational complexity due to the need for synchronization and consensus across multiple nodes, which can affect scalability and deployability.

\textit{Cryptology-based methods} utilize cryptographic techniques to ensure the confidentiality and integrity of sensitive information. For example, HE allows computation on encrypted data without needing to decrypt it first, protecting user preferences during data processing. 
In addition to preventing request information from being monitored by external attackers, these methods are crucial for preventing cache tampering attacks, where attackers might alter cached content to gain unauthorized access or disrupt system operations. Despite offering stringent protection, \textit{cryptology-based methods} generally incur high computational complexity, particularly in scenarios involving encryption and homomorphic encryption, which can become bottlenecks for resource-constrained edge devices.


Lastly, \textit{other methods}, including optimization-based techniques and access control, enhance privacy by mathematical modeling and enforcing access restrictions to sensitive resources. 
The complexity of privacy-preserving methods in edge caching systems varies significantly depending on the approach.   Optimization-based methods involve solving complex mathematical models, with computational intensity heavily dependent on the formulations and solving algorithms. This complexity may be particularly high when optimizing  multiple privacy metrics simultaneously or under complex constraints. 
Therefore, selecting appropriate privacy-preserving methods requires a careful balance between the desired level of privacy and the performance cost.
For easy reference, we also present a classification matrix for the solutions introduced in this survey based on countermeasures and privacy data in Table~\ref{Tab: countermeasures} in Appendix.
}

\section{Enhancing User Privacy in Edge Caching Systems}\label{sec: user privacy}

User privacy is the most important privacy in edge caching systems, which has attracted tremendous research efforts dominating the research on privacy preservation in edge caching systems. We discuss these defence methods based on three types of user privacy, i.e., \textit{request traces}, \textit{personal information} and \textit{location}.

\subsection{Privacy of Request Traces in Edge Cache}
\begin{figure}[tb]
\centering
\includegraphics[width=0.75\linewidth]{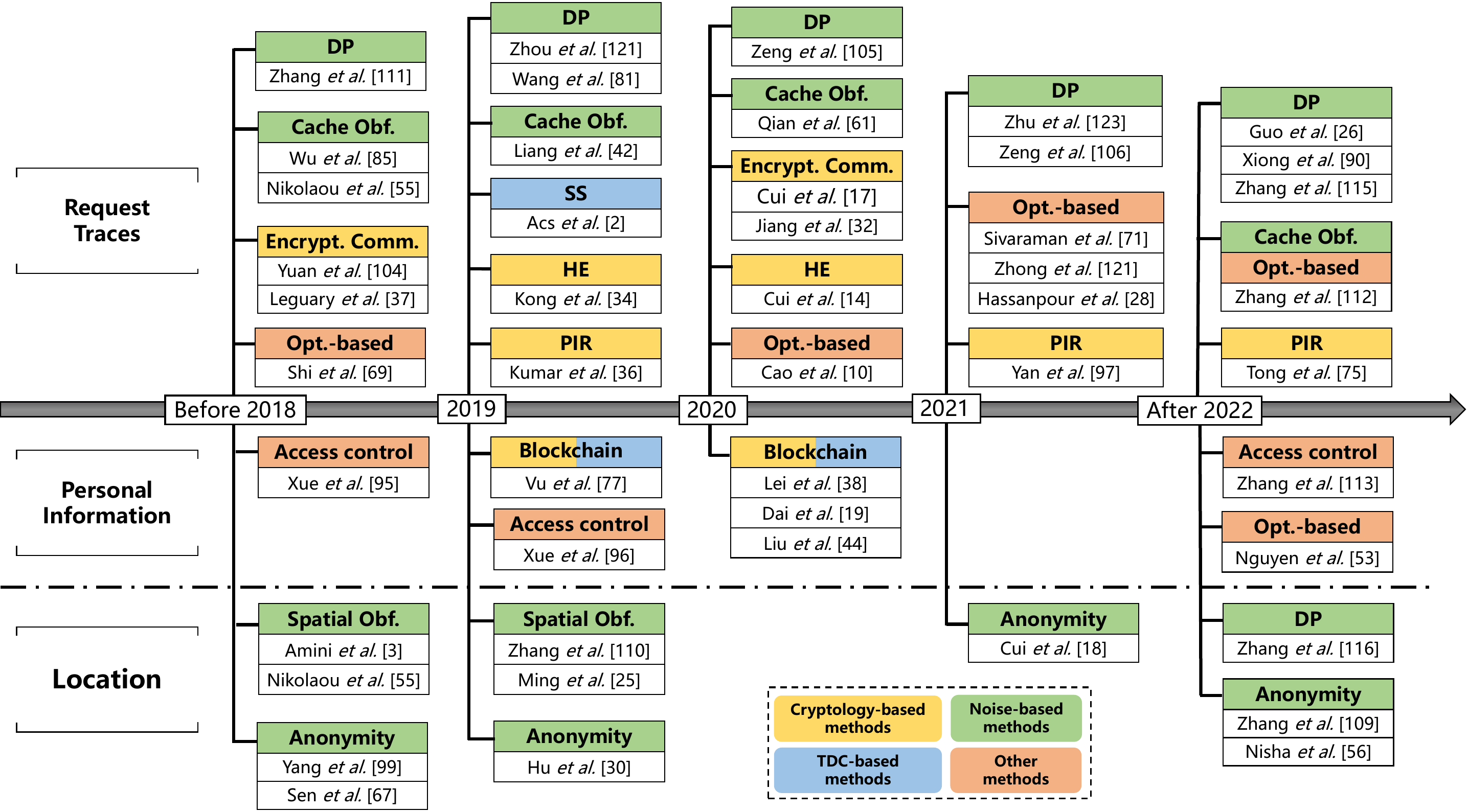}
\vspace{-3mm}
\Description{A brief timeline of solutions aimed at enhancing user privacy, including request traces, personal information and location, in the edge cache. Each solution is accompanied by its main mitigation approach.}
\caption{A brief timeline of solutions aimed at enhancing user privacy, including request traces, personal information and location, in the edge cache. Each solution is accompanied by its main mitigation approach.}
\label{fig: user privacy}
\vspace{-4mm}
\end{figure}

Request traces are the most critical privacy information in the edge cache (EC), from which adversaries can obtain user preferences~\cite{Wang2019}.
We summarize methods to protect user request records from four aspects which are \textit{noise-based methods}, \textit{cryptology-based methods}, \textit{trusted distributed computing-based methods} and other methods. A brief timeline of solutions for enhancing the privacy of request traces is presented  in Fig.~\ref{fig: user privacy}. The solutions for enhancing other user privacy, e.g., personal information and location, are also summarized in  Fig.~\ref{fig: user privacy} for the sake of brevity.


\subsubsection{\textbf{Noise-based methods}}
\textcolor{black}{The initial class of methods to protect request records are noise-based methods, which can be categorized into two main approaches. The first approach involves adding noises generated by mechanisms such as differential privacy (DP) to protect information~\cite{Zhang2018,Zhou2019,Wang2019,Zhu2021,Zeng2020, ZengYi2021, Guo2022,Wang2017}. The second approach includes cache obfuscation methods (such as proactive cache~\cite{Qian2020,Nikolaou2016}, off-path cache~\cite{Wu2016}, and request hit delay~\cite{Liang2019}) to protect users' requests from monitoring or timing attacks in untrusted or semi-trusted network environments. We will elaborate on these methods in the following sections.}

\textit{\textbf{Differential privacy (DP).}}
Content providers (CPs) often utilize edge caching nodes at the edge network and collect users' private access records to predict user preference to improve delivery efficiency. However, directly collecting users' profiles can lead to privacy breaches. Additionally, in highly dynamic scenarios, the entities of edge cache (e.g., edge nodes (ENs)~\cite{Zhou2019} and edge servers (ESs)~\cite{Zhu2021}) collect user request records in real-time and make dynamic decisions to improve the efficiency of edge caching. 
This real-time data collection process also poses a risk of privacy leakage, where DP-based methods can be employed to mitigate the risk. 

Zhou~\emph{et al.}~\cite{Zhou2019} proposed a privacy-preserving and online distributed multimedia content retrieval system. Each EN in the system is modelled as an online learner to exploit user requests with a context that includes their background information (e.g., age, gender, location, social profile, and query criteria). The ENs can collaboratively make multimedia content recommendations and cache at the edge network. When an EN needs extra context information to make a retrieval scheme, the TTP sends noisy records to ENs by deploying the DP mechanism. A trust mechanism is also proposed to identify and remove malicious ENs. 
Zhu~\emph{et al.}~\cite{Zhu2021} studied the trade-off between privacy protection and caching efficiency in edge caching systems. When a user generates a content rating vector, Gaussian noises are added to the original rating vector, and then the distorted rating vector is transmitted to the ES for privacy protection. In the global aggregation information stage, ES calculates the eigenvalues and eigenvectors of collected data based on the lightweight level calculation algorithm. Then, ES broadcasts the results to all users. 
\textcolor{black}{Xiong~\emph{et al.}~\cite{Xiong2022} presented a novel network traffic shaping framework for protecting privacy in IoT networks by integrating DP with constrained optimization. They developed a tunable DP model that shapes encrypted IoT traffic to protect against monitoring attacks, particularly eavesdropping on packet sizes and timing. This approach not only safeguards IoT traces from privacy breaches but also enhances the resilience of IoT systems against traffic analysis attacks by dynamically adapting to changing network conditions and heterogeneous user demands.}

In collaborative edge caching, managers exchange sensitive information, such as user records or preferences~\cite{Zeng2020, Zhou2019}, and routing records~\cite{Zeng2020}, to improve caching efficiency. 
However, protecting privacy often in collaborative edge caches may rely on a centralized TTP, which is challenging to obtain in practice and places more pressure on network bandwidth. Moreover, if the centralized TTP is attacked, it may pose a more serious privacy breach risk.
Zeng~\emph{et al.}~\cite{Zeng2020} proposed a distributed method to develop network caching and routing strategies for small base stations (SBSs). The scheme adds a DP noise in the routing information (i.e., the portion of the requested content served by each SBS) during the exchange process to protect the privacy of SBSs and Mobile Users (MUs). It defines an optimization problem that minimizes the global cost, which is solved by a distributed protocol. 
Guo~\emph{et al.}~\cite{Guo2022} introduced a blockchain and DP-based decentralized edge-thing system for privacy preservation and fair utilization of edge computing resources. The proposed system employed blockchain technique to deal with transactions and smart contracts' tempering issues caused by the malicious auctioneer node. Moreover, an exponential mechanism-based DP is applied to the double auction scheme to tackle the inference attack on auction results saved in the blockchain. 


Hits on the user's local cache can provide the best service experience for users. However, it is challenging for end devices that rely on a user's personal historical information to make accurate pre-fetching decisions solely.
Collaborative efforts between users are necessary, but such information exchange is risky, and the recorded history must be protected when disclosed.
Wang~\emph{et al.}~\cite{Wang2019} presented a mobile video pre-fetching strategy based on DP and distributed online learning algorithms. They formulated the pre-fetching problem as an online optimization problem considering user preferences, video popularity, and social connections. The problem is then decomposed into two sub-problems, which are solved and swapped at each terminal by a distributed method to obtain the optimal global solution. 
The DP mechanism is added in exchanging user-sensitive information during each round of iteration to protect user privacy.


\textit{\textbf{Cache obfuscation.}}
In Information-Centric Network (ICN), users can directly access desired content from edge routing nodes. However, edge routing nodes are often vulnerable to cache side-channel attacks, which can result in the exposure of requested record privacy. 
Liang~\emph{et al.}~\cite{Liang2019} designed a method to defend against timing attacks in Content-Centric Networks (CCN). According to the privacy protection degree for requested content and the honesty degree of requested nodes, evaluated by the historical information, the caching node calculates the delay in responding to requests to defend against timing attacks.
Further, Wu~\emph{et al.}~\cite{Wu2016} designed a multi-path caching strategy for ICN based on random linear network coding. The strategy encodes different video chunks into the same block for efficient content delivery. When the block is delivered along the path, it can only serve all routing nodes with related video chunk requests and keep unavailable to irrelevant nodes. It adopts a random forwarding method which increases the diversity of routing paths, thereby increasing the size of anonymity sets and the cost of inferring user privacy.

In addition, proactive caching of redundant and obfuscated content at the edge can interfere with an attacker's ability to access the user's actual request records.
Qian~\emph{et al.}~\cite{Qian2020} proposed a privacy-aware content caching architecture for cognitive Internet of vehicles (CIoV) networks with proactive caching and blockchain technology. 
In this system, roadside units (RSUs) and smart vehicles can cache content in advance, which can provide the cached content in the form of a broadcast to meet the content needs of other vehicles. 
Therefore, a vehicle only needs to obtain content from broadcast data without further requests, which can reduce user privacy exposure. 
At the same time, blockchain technology is introduced to ensure a more secure and reliable transaction mode to guarantee the reliability of the content.
Additionally, Nikolaou~\cite{Nikolaou2016} proposed two cache placement strategies for the joint caching of users. The first strategy considers the graph network structure between user terminals, and the second one focuses on the workload change of the server. However, transmitting requested videos between clients will leak privacy for both sides. The requested user proactively fetches and caches obfuscated content. At the same time, the server adds randomly obfuscated addresses when sending  feasible retrieval address lists to reduce the risk of privacy exposure.

\subsubsection{\textbf{Trusted distributed computing-based methods}}
{The second category of trusted distributed computing methods aim at safeguarding request records primarily comprises \textbf{secret sharing (SS)}, a secure multi-party computation technique, that can effectively prevent attackers from acquiring valued request records.}
Acs~\emph{et al.}~\cite{Acs2019} proposed two timing attack defence methods for the edge router cache in the ICN network. For interactive traffic-type communication, random naming and SS are used for privacy protection to prevent attackers from obtaining specific traffic information.
In view of the content distribution traffic, a method of increasing artificial delay is proposed to protect privacy, and a certain delay is added to the private content that is hit by the router cache to prevent adversaries from determining the hit status of private-sensitive content. 

\subsubsection{\textbf{Cryptology-based methods}}
Cryptology-based methods have been widely used to protect the security and privacy of user request records and other information in Internet communications. 
However, there are also three challenging problems when using cryptographic methods to protect the privacy of request records in edge caching systems. Firstly, due to the existence of encryption, third-party edge caches probably cannot directly use encrypted requests to retrieve related content, leading to the unavailability of edge caches~\cite{Yuan2016a, Leguay2017}. Secondly, introducing encryption technology may pose heavy computational pressure on the resource-constrained edge and end devices. Lastly, cryptology-based methods fail to prevent the leakage of record privacy from content providers or service providers, who have the key to decrypt requests. Therefore, how to apply cryptology-based techniques to edge caching systems is still a challenging problem.

\textit{\textbf{Encryption communication.}}
To prevent the monitoring of users' request records by Internet service providers (ISPs), efforts have been made to encrypt request records and the corresponding transmitted data using encryption algorithms while ensuring the availability of the cache within the ISP.
Yuan~\emph{et al.}~\cite{Yuan2016a} designed a system to achieve efficient encrypted video delivery in the ISP network. The content cached in the network is encrypted and distributed in the ISP network. This system can efficiently and safely locate and retrieve related content from the ISP network with a proposed encrypted content fingerprint index for a given encrypted request.


In order to improve privacy in the Content Delivery Network (CDN), Cui~\emph{et al.}~\cite{Cui2020c} proposed a novel encrypted method that combines searchable encryption (SE) and a multi-CDN strategy to achieve both content delivery performance and security in edge CDN nodes.
The work introduces the SE method to realize content security and searchability. In addition, a semantically secure algorithm is used to encrypt user requests so that the same query can correspond to different request content.
To further protect user preference privacy, a one-time nonce will also be used for secondary encryption, which will be transmitted together with the content transferred between CDN node clusters. For each request, the node must receive the nonce to search, and after the search hits, the nonce must be regenerated and re-encrypted before continuing to deliver the content.


\textit{\textbf{Homomorphic encryption (HE).}}
{
In previous works, HE has been introduced to protect the privacy of vehicles' request records in IoV while collaborating with RSUs to improve the efficiency of edge caching.
}
Cui~\emph{et al.}~\cite{Cui2020} proposed a cooperative download scheme in the IoV network, considering the security and privacy protection of request traces. This scheme uses edge computing architecture to reduce transmission delay. It uses lightweight encryption methods, such as elliptic-curve cryptography, the Tesla broadcast authentication, and additive HE, to protect user privacy and content security.
The strategy proposed in this work is composed of two phases: the \textit{non-accelerated phase} and the \textit{accelerated phase}, the details of which can be found in~\cite{Cui2020}.

Kong~\emph{et al.}~\cite{Kong2019} utilized an invertible matrix to construct multiple content requests sent by different vehicles such that the RSUs can recover each request without being associated with a specific car. Specifically, when a vehicle needs to initiate a request, it will first generate a $k*k$ random invertible matrix and send secret information required for HE to $k$ vehicle users within a unified range. Then, in the response,  a collaborative request group is randomly selected for the requested vehicle. Other vehicles in the group first generate the requested information according to the Paillier HE algorithm and send it to the RSU, returning the HE information to the requested vehicle. That vehicle completes the corresponding HE according to the returned information and the invertible matrix. Finally, it sends the encrypted request to the RSU to retrieve the private content without exposing its privacy.

\textit{\textbf{Private information retrieval (PIR).}}
By utilizing PIR methods, users are able to obtain the content they desire while preventing potential leaks of their private interests.
Kumar~\emph{et al.}~\cite{Kumar2019} were the first to introduce a PIR strategy based on encoding cache into wireless edge caching. Erasure-correcting codes are used to encode cached content, and different bit rates can be selected for videos with varying popularity to conserve backhaul bandwidth usage. Additionally, the scheme is based on general Reed-Solomon coding to safeguard user privacy from SBSs that may collude with one another.
Furthermore, ensuring the integrity of content in the edge cache is essential for maintaining a stable edge caching system. This is particularly important because edge devices owned by individuals or small organizations are susceptible to cache tampering attacks and internal hardware failures. However, verifying the integrity of the content can compromise its privacy, especially when third-party verifiers are involved. To address this issue, Tong~\emph{et al.}~\cite{Tong2022} proposed an integrity-checking protocol for edge storage based on provable data possession to verify the integrity of cached content on a single EN. The protocol employs a PIR scheme and homomorphic verifiable tags to prevent the disclosure of sensitive information (e.g., user request traces, edge download schemes, and private content) to verifiers. 



\subsubsection{\textbf{Other methods}}
Other methods, such as \textbf{Optimization-based methods}, are also introduced to enhance the privacy of request traces or user preferences in EC.
Sivaraman~\emph{et al.}~\cite{Sivaraman2021} used game theory to formulate an off-path and cooperative caching problem in the edge of ICN, where users can choose their optimal routers at the edge network to cache content. Constraints in the problem include network latency, caching cost, and the amount of exposed user privacy. 
\textcolor{black}{Two different privacy measures (i.e., mutual information and differential privacy) are used as constraints in the work.}
Finally, it is proved that a Nash equilibrium point exists in the game, which can be solved by an iterative method. 
\textcolor{black}{
In addition, to mislead adversaries eavesdropping on edge caches (EC), Hassanpour~\emph{et al.}~\cite{Hassanpour2023,Hassanpour2021} proposed caching approaches aimed at enhancing privacy and reducing communication costs in edge networks. The solution presented in~\cite{Hassanpour2021} employs an $\epsilon$-constraint optimization approach to balance the trade-off between minimizing the average delivery load and maximizing context-oriented privacy. By optimizing cache placement probabilities, the approach in~\cite{Hassanpour2023} utilizes chunk-based joint probabilistic caching (JPC) to increase adversarial errors while maintaining the desired privacy levels. Furthermore, to address the exponential growth of the feasible solution set in the JPC optimization problem, they proposed a scalable JPC strategy to solve the linear programming optimization problem efficiently.
}

Furthermore, Cao~\emph{et al.}~\cite{Cao2020} studied the reliable and efficient performance of multimedia transmission services between base stations (BS) and MUs through a two-stage joint optimization. In the first stage of optimization, a service reliability evaluation mechanism is designed to evaluate the credibility of BS to ensure the security of user privacy information. Then, the price and reliability competition among BSes and the strategic interaction of all players are modelled by the Stackelberg game~\cite{He2007}. A resource allocation problem is further proposed in the second stage to coordinate multiple MUs serving on the same BS. The potential game model is used to improve the transmission service performance. 
Additionally, Shi~\emph{et al.}~\cite{Shi2018} proposed a model for the cache placement problem in wireless edge caching, considering a multi-attacker scenario where both benign users' and attackers' locations follow a homogeneous Poisson Point Process (PPP). An optimization problem is formulated to determine the probability of each caching file, considering the average probability of successful eavesdropper attacks and transmissions at the wireless edge network. Finally, the genetic algorithm is used to maximize the secure transmission performance of the system.


\subsection{Privacy of Personal Information in Edge Cache}
Personal identity information is also sensitive in the network, which can be used by edge cache for carrying out sensitive operations such as permission control and cache admission control. However, excessive disclosure of users' personal identity information makes it convenient for malicious nodes and attackers to spam users with advertisements and recommendations and attack edge servers by polluting cached content. 


\subsubsection{\textbf{Blockchain-based methods}}

Previous works mainly employ blockchain to protect users' identity information~\cite{Lei2020, Dai2020, Vu2019, LiuJi2020}.
Specifically, Vu~\emph{et al.}~\cite{Vu2019} proposed a blockchain-based CDN (B-CDN) architecture for content delivery, which enables anonymous operations on users. The B-CDN leverages intelligent contracts to maintain the blockchain and provide CPs with users' registration and subscription functions while ensuring user privacy. Additionally, the B-CDN can reduce the cost of CP management by utilizing a public database of requested traces, which allows CPs to estimate users' preferences with virtual identities and maximize the efficiency of their caching services.

Named Data Network (NDN) is a variant of the ICN, where content can be retrieved by the content name. 
Lei~\emph{et al.}~\cite{Lei2020} introduced a blockchain-based security architecture for improving the security and privacy of NDN-based vehicular edge computing systems. This work deploys blockchain nodes in edge servers and ISP nodes, where a delegated consensus algorithm is designed to enhance the efficiency of the blockchain. A three-layer management framework and an access control strategy are proposed for key management based on blockchain verification and vehicle attributes, respectively. A resource requester needs to prove to blockchain consensus nodes that it satisfies the access condition according to the access policy of the resource owner.
Dai~\emph{et al.}~\cite{Dai2020} designed a content caching mechanism based on the permissioned blockchain technology to address the problem of privacy and security in the vehicle edge computing network. A new block validator selection method is proposed to achieve a fast and efficient blockchain consensus mechanism. In addition, this work presents a deep reinforcement learning-based vehicle content caching algorithm. 
Liu~\emph{et al.}~\cite{LiuJi2020} designed a decentralized caching framework empowered with blockchain credentials to tackle the challenges of content data verification and edge device authentication. In the designed system, it is possible to trace each transaction at an active edge network without a central manager. A cache order matching technique is devised to use the cache resources efficiently. Further, data integrity verification is done with the help of a content trading mechanism which helps data sharing among the edge devices of the edge network and ensures the efficiency of trading in the edge cashing system.

\subsubsection{\textbf{Other methods}}
The \textbf{access control} is also exploited to protect users' identity information~\cite{Nguyen2023, Zhang2022b, Xue2019, Xue2018}.
In~\cite{Xue2019, Xue2018}, Xue~\emph{et al.} proposed a secure and efficient network access framework (SEAF) for cache resources at the edge of ICN. 
The SEAF provides several security and privacy features, including content confidentiality, user privacy protection, user privilege revocation, countability, and efficiency. 
In SEAF, routers at the edge network authenticate user requests to separate access control from content provisioning. Only authenticated requests can enter the network; thus, authorized users can only access the bandwidth and cache resources inside the ICN. Meanwhile, to protect privacy, users can verify their identity to the edge router by generating a valid group signature, thereby maintaining users' anonymity to the edge router.
Zhang~\emph{et al. }~\cite{Zhang2022b} focused on the security issues of cache-based software-defined networks, using the Tesla protocol to achieve fast authentication of the cache of vehicles and fog nodes. Besides, the Pedersen commitment mechanism is used to directly authenticate vehicles and fog nodes without exposing user identity privacy. Considering the limited computing power and delay-sensitive characteristics of the IoV, the author designed a set of cryptographic mechanisms supporting batch verification.


\subsection{Privacy of Location in Edge Cache}


The location information is a kind of critical privacy of a high value,  including moving trajectory~\cite{Wu2023, Li2023, Zhang2022a}, spatial coordinates~\cite{Zhang2019b}, and other unique features~\cite{Zhang2022a}. Noise-based methods comprise the primary class of techniques employed to enhance location privacy as illustrated in Fig.~\ref{fig: user privacy}.

\subsubsection{\textbf{Noise-based methods}}
As such,  noise-based methods are mainly introduced to protect location privacy, including geographic differential privacy~\cite{Zhang2022a}, Spatial Confusion~\cite{Amini2011,GUYi-mingBAIGuang-weiSHENHang,Zhang2019b}, $k$-anonymity~\cite{Cui2020b,Nisha2022,Yang2016,Sen2018,Zhang2019b}, etc.

\textit{\textbf{Differential privacy.}}
{
With the increasing mobility of users and the constant threat of malicious attacks from third parties, there is a growing risk of privacy breaches in mobile edge caching. In order to address this issue, Zhang~\emph{et al.}~\cite{Zhang2022a} proposed a DP-based method for improving the video Quality of Experience (VQoE) for mobile users while protecting their location and preference privacy in mobile edge caching.
The proposed scheme utilizes a privacy-preserving approach for computing the location transfer model and aggregating user preferences, achieving a balance between caching service efficiency and privacy protection at mobile edge networks. Specifically, the Laplacian perturbation model is employed to protect users' location and preferences when submitting their information.
Based on the perturbed information, mobile edge caching nodes can evaluate the popularity of videos in the user's area, and Q-learning~\cite{Sutton1998} is employed to achieve cache optimization goals combined with transcoding technologies. 
}

\textit{\textbf{Spatial obfuscation.}}
Amini~\emph{et al.}~\cite{Amini2011} were one of the first to utilize devices' cache to protect users' location information, where location-based content can be periodically prefetched to devices in large geographic blocks before they are actually consumed. When content has been cached in a user's local area, the user can access it directly on their device without needing external network services. This can effectively reduce privacy exposure risks for the user.

Additionally, privacy protection can be achieved through a distributed collaborative cache that forms anonymous user groups within the vicinity.
Zhang~\emph{et al.}~\cite{Zhang2019b} proposed a multi-level caching strategy to reduce the number of users directly requesting LBS from the local service provider (LSP). In turn, users can obtain the required services from the local cache, surrounding neighbour caches, and trusted anonymizers. In this way, the interaction with untrusted LBS is reduced, and privacy exposure is mitigated. When the request is lost, it has to request the LSP by generating a stealth zone and making a request to the LSP. The anonymizer will select the optimal K-space anonymity to request content according to the prediction result (considering a user's future geographic location, the caching contribution rate of each unit, and the freshness of the content in the unit).
However, the high communication overhead and computational energy consumption of users collaborating as a group pose problems in protecting privacy. Moreover, the introduction of centralized anonymizers is vulnerable to attacks, and if breached, all users' private information may be compromised.
To address these limitations, Ming~\emph{et al.}~\cite{GUYi-mingBAIGuang-weiSHENHang} proposed a method that employs the trusted ESs to preprocess user requests and blur their location information during the snapshot query (i.e., one-shot query) of their POI. The ESs cache the requested POI for further query, thus minimizing the number of queries exposed to LBS providers and potential attackers. Additionally, in continuous queries, fuzzy prediction queries are generated and correlated with the actual query to enhance the queries' utility while interfering with attackers.

\textit{\textbf{Anonymity.}}
The utilization of cache in edge devices, such as user devices~\cite{Yang2016, Cui2020b, Nisha2022}, ESs~\cite{Sen2018} and RSUs~\cite{Hu2018}, can keep users' transparency from LBS providers by reusing the users' POI within a specific region. This approach allows users to access the cached POI directly at the edge network instead of relying on remote LBS service providers. 
Additional privacy protections (e.g., $k$-anonymity~\cite{Yang2016, Hu2018, Zhang2023}, $l$-diversity~\cite{Cui2020b}, anonymity groups~\cite{Sen2018, Nisha2022}) are exploited when resources have to be obtained from LBS providers.
As a result, the likelihood of exposing sensitive location information to the service provider is reduced.

Zhang~\emph{et al.}~\cite{Zhang2023} devised a Caching-based Dual $k$-Anonymous (CDKA) mechanism to preserve location privacy. CDKA uses double anonymity and multilevel caching to reduce communication overhead while providing location privacy. For this, an edge server is used to intervene between the user and the LBS server, and location privacy is ensured by making mobile clients and edge servers anonymous. The proposed mechanism is assessed for computational efficiency, communication overhead, and cache hit ratio.
Additionally, dealing with vehicles' high-speed movement characteristics in vehicular networks, Hu~\emph{et al.}~\cite{Hu2018} designed a privacy protection algorithm combining proactive caching and the $k$-anonymity method. When a vehicle user requests a specific POI, it needs to send $k-1$ obfuscated requests simultaneously. Besides, the corresponding request content will be obtained through multiple passing RSUs to protect the user's location information, including factual geographic and POI.


Moreover, at the edge of the wireless network, Sen~\emph{et al.}~\cite{Sen2018} proposed a double cache strategy to deploy a pair of caches for each region. Cache $A$ records previous request results of users in the region, and cache $B$ caches all user requests and maintains cooperation between users. When querying private content, a user queries cache $A$ first. If it is not hit, the request will be redirected to cache $B$ for conversion. Finally, the converted request is sent to other users within the region to request LBS together, and the received results may be maintained in cache $A$ for further queries.

To further prevent users' location and personal information from being accessed by untrustworthy EC and malicious users,
Nisha~\emph{et al.}~\cite{Nisha2022} proposed a caching scheme called Group Collaboration Scheme (GCS) to request POI combining with spatial obfuscation. In this scheme, users who need to find POI in a specific area will modify the requested area according to the proposed random area obfuscation algorithm and then register with the group authenticator to obtain virtual group identity information and cooperative anonymous user groups. The collaboration is one-time, and the anonymous group changes as the user moves. Users with request requirements will cooperate with nearby users to query whether the cache of other users in the anonymous group meets the request requirements. If the request POI is unavailable in the user group, the required content will be requested in the name of the anonymous group. 

\subsubsection{\textbf{Trusted distributed computing}}
To enhance the Quality of Service (QoS), CPs collaborate with ISPs to deploy edge caching resources as close to the users as possible. ISPs can support edge cache by placing Virtual Servers (VSes) at the network's edge and assigning them to CPs. However, CPs only possess the request records of users, while ISPs only have access to their geographic location information. In the caching process, CPs do not want to disclose all the requested information to the ISPs, and vice versa. 
To deal with this challenge, Andreoletti~\emph{et al.}~\cite{Andreoletti2018} proposed a secure multi-party computation protocol to facilitate cooperation between ISPs and CPs without requiring either party to disclose sensitive information. The protocol enables ISPs to obtain the number of requests for specific video content in a given area at a low computational cost. Once the ISP has this information, it can deploy VSes efficiently, and the CP can use these VSes to place the edge cache, thereby minimizing the number of hops for content delivery and reducing communication delays.

\textcolor{black}{Despite the comprehensive introduction of major solutions, our discussion is not exhaustive. Thus, we provide a supplementary introduction Table~\ref{Tab: user privacy} in Appendix, briefly introducing  other solutions to protect user privacy in edge caching systems that have  not been discussed in detail in Section~\ref{sec: user privacy}. }

\section{Enhancing Content Privacy in Edge Caching Systems}\label{sec: content privacy}
\begin{figure}[t]
\includegraphics[width=0.82\linewidth]{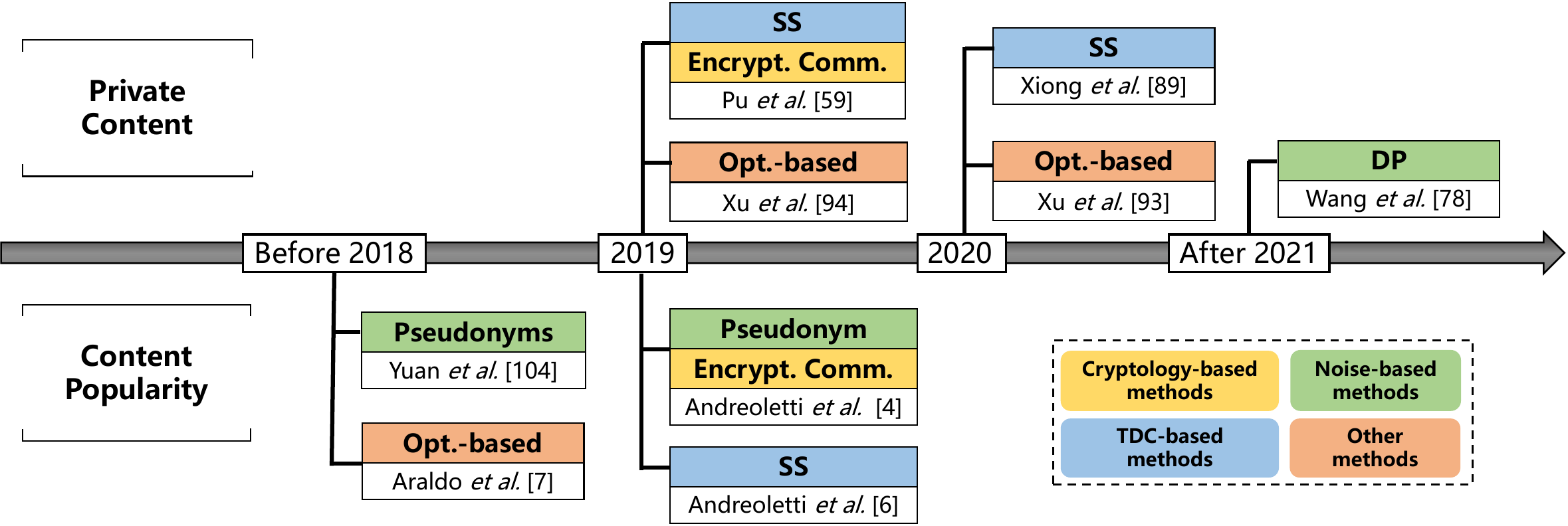}
\Description{A brief timeline of solutions for enhancing content privacy, including private content data and content popularity.}
\vspace{-3mm}
\caption{A brief timeline of solutions for enhancing content privacy, including private content data and content popularity.}
\label{fig: content privacy}
\vspace{-4mm}
\end{figure}

In this section, we move on to discuss defence methods that can preserve the second type of sensitive information, i.e., content privacy, in edge caching systems. 
We present a timeline, as depicted in Fig.~\ref{fig: content privacy}, summarizing the methods employed to safeguard content privacy, encompassing private content data and content popularity.

\subsection{Privacy of Content Data in Edge Cache}

Other than caching content for CPs, edge servers are also able to cache private content generated by users. 
However, due to the presence of incompletely trusted ESs~\cite{Pu2019, Xu2019} or malicious and unauthorized users~\cite{Xu2019, Xu2020} at the edge network, stored content in EC may face privacy leakage risks.

\textbf{DP-based methods}
are used to upload local data in the network cache while preserving its privacy. For example, Wang~\emph{et al.}~\cite{WangHu2022} proposed a Differential Privacy-Preserving Peep Learning Caching Framework (DP-DLCF) to deal with the privacy leakage problem of private content in edge caching networks. The privacy budget is utilized adaptively to strike a trade-off between the privacy and accuracy of the  prediction. In the proposed technique, users upload their data after perturbing it with a randomized response technique based on LDP to preserve the privacy of their local data. Next, the neighboring BS accumulates the uploaded data and transfers it to the deep model for training. Moreover, the prediction accuracy of the model training is improved by the bootstrap aggregation algorithm.

\textbf{Crytology-based methods} can also be leveraged in protecting the private content in the edge cache.
Pu~\emph{et al.}~\cite{Pu2019} proposed a secure and privacy-aware content-sharing strategy to protect sharing data stored and delivered by incompletely trusted ESs. To ensure the secure sharing of content, the content generator first encrypts the content using the Ciphertext-Policy Attribute-Based Encryption (CP-ABE) algorithm and calculates its signature based on its private key. Additionally, by utilizing the public key cached at the nearest ES, the generator performs secondary encryption of the content to the nearest ES.
When the ES receives the encrypted content from the content generator, it will first decrypt the content with its private key and check the security of the content. According to the secret sharing (SS) scheme, ES randomly divides the content into $n$ parts and distributes the content parts to other $n-1$ ESs to store the content. The proposed scheme can effectively ensure the integrity and recovery ability of the content in case any edge cache node becomes offline.
\textcolor{black}{
The SS method was also integrated by Xiong~\emph{et al.}~\cite{Xiong2020} to design an edge-assisted privacy-preserving data-sharing framework for autonomous vehicles. This approach encrypts raw data into two ciphertexts, which are processed by two edge servers. Additionally, a privacy-preserving convolutional neural network (P-CNN) was developed to ensure that the classification results are identical to those of the original CNN model, without any data leakage. The framework effectively addresses threats such as potential data leakage and unauthorized access to private content by malicious vehicles or edge servers.
}

\textbf{
Optimization-based methods} are introduced to enhance the privacy of content caching in edge servers.
To prevent private content from leaking to the unreliable edges and make optimal caching decisions for MUs, Xu~\emph{et al.}~\cite{Xu2019} used the multi-leader and multi-follower Stackelberg game to model a multi-link cache scenario at the mobile edge network. 
In the scenario, Edge Computing small base stations (ECSBS) act as leaders and, firstly, set pricing strategies in a non-cooperative game. Then, a trust mechanism is proposed to evaluate the reliability of each ECSBS, which consists of two parts: \textit{direct trust degree} and \textit{indirect trust degree}. Based on the caching reliability and pricing offered by ECSBS, MUs can make their optimal caching decisions as followers.
Additionally, Xu~\emph{et al.}~\cite{Xu2020} proposed a Stackelberg game model to encourage edge cache devices (ECDs) to provide secure caching services in both static and dynamic scenarios. The model takes into account the selfish and open nature of ECDs and employs a zero-payment mechanism to penalize ECDs that provide poor services. The optimal strategies for the CP and ECDs in a static game are analyzed, proving the existence of a unique equilibrium in the Stackelberg game. Besides, in dynamic games with incomplete information, the Q-learning algorithm is used to solve the problem. 

\subsection{Privacy of Content Popularity in Edge Cache}
Content popularity, which can be used as the key knowledge to improve caching efficiency, is business-critical information for the CPs and edge caching managers (e.g., ISP). 
Due to the limited number of records in the service scope of edge cache (e.g., serving a specific geographical location range or a particular network level), edge caching suppliers may require content providers and other edge caching entities to provide the critical content popularity information so that they can judiciously make caching decisions so as to shrink bandwidth consumption of the core network. 

{
Andreoletti~\emph{et al.}~\cite{Andreoletti2019a} improved the solution proposed in~\cite{Yuan2016a} by allowing CPs to encrypt content and associate them with pseudonyms to prevent privacy leakage to edge caching managers. ISPs only count the occurrences of these pseudonyms to infer content popularity without examining the original content. The authors introduced the mathematical definition of privacy and studied the trade-off relationship between privacy and hit rate, retrieval latency, and traffic load metrics.
Additionally, Andreoletti~\emph{et al.}~\cite{Andreoletti2019} proposed a protocol for spatial partitioning of ISP caches based on the popularity of different CPs' content, which aims to improve the quality of service (QoS) of edge caching services while protecting CPs' privacy of popularity information. The protocol employs the Shamir secret sharing scheme for CPs to share the popularity information between the ISP and the regulator authority, which guarantees a fair subdivision of the cache storage and the preservation of privacy. The ISP can calculate the caching space requirement for each CP using the secret information, thus protecting CPs' privacy.

Similarly, Araldo~\emph{et al.}~\cite{Araldo2018} proposed a caching space partitioning method that protects the popularity information of CPs while ensuring the efficiency of edge caching. The method divides the ISP's caching space into multiple slices and assigns each slice to different CPs using the stochastic dynamic cache partitioning algorithm. The algorithm takes an initial slice allocation as input and iteratively optimizes the slice allocation scheme by testing the Cache Miss rate of the allocation scheme in each round. However, unlike the partitioning method proposed by Andreoletti~\emph{et al.}~\cite{Andreoletti2019}, this method does not depend on the private information of CPs' popularity. Additionally, this architecture also supports a transparent cache of encrypted content deployed at the edge of the ISP network.
}

\section{Enhancing Knowledge Privacy in Edge Caching Systems} \label{sec: knowledge privacy}

In this section, we discuss defence methods that can preserve privacy for the last type of privacy, i.e., knowledge privacy, in edge caching systems. 
All edge caching service providers have the motivation to extract knowledge for improving caching performance, which gives rise to the  trade-off between caching performance and privacy protection. Due to the high dynamics and complicated access patterns driven by users' interest~\cite{Muller2017, Yang2019}, it is essential to come up with intelligent edge caching algorithms to improve the caching performance. 
Machine learning-based methods provide a feasible framework to extract user access patterns by exploiting collected datasets related to users, which may contain sensitive information. 
For example, video request access patterns are driven by users' interest in different locations~\cite{Ma2017b, Dhar2011}. Users may keep dynamic moving~\cite{Dai2020}, and their interests evolve over time~\cite{Zhang2022}. Thus, it is necessary to make edge caching decisions based on features which can be extracted from localized and private user information by machine learning methods.

\textbf{Federated learning (FL)} as a distributed machine learning framework is the  most popular method to preserve knowledge privacy. FL trains a learning-based algorithm across multiple decentralized devices or edge servers holding local data samples without exposing them.
Additionally, we provide a comprehensive summary of the FL-based methods employed to safeguard knowledge privacy, presented in a timeline illustrated in Fig.~\ref{fig: Knowledge privacy}. Besides, Table~\ref{tab: solutions for knowledge privacy} in Appendix offers a detailed classification of these solutions based on the combination of methods used.

\subsection{Enhacing Knowledge Privacy with FL Frameworks}
The most common approach is to use an FL framework to train prediction models.
Unlike traditional machine learning methods, FL does not collect raw data for model training~\cite{Yu2018, Yu2020, Yu2020a}.  This framework encourages models to be trained on local data, and all training works upload model parameters or gradients rather than sensitive raw data.
Yu~\emph{et al.}~\cite{Yu2018} were probably the first to propose a learning-based proactive content caching method following the FL framework. This work proposes a hybrid filtering method based on the autoencoder to calculate the user-content similarity and predict the content of a user's interest. 
Yu~\emph{et al.}~\cite{Yu2020} also designed an FL-based proactive caching method for vehicular networks. Considering the high mobility of vehicles and dynamic content popularity in vehicular networks, RSUs integrate the mobility-aware cache replacement policy to make proactive caching decisions. 
Following the FL framework, the above three works enable users to train machine learning models (e.g., autoencoder model) with their private datasets, locally and distributively, and upload trained models to the corresponding parameter server for aggregation.

Reinforcement learning can be realized in the FL framework to solve the complex dynamic control problem and mitigate the privacy leakage problem in edge caching systems~\cite{Wang2019a, Li2020a, Wang2020, Liu2022, Abadi2016, Qiao2022, Xiao2018a}
to improve the caching performance and privacy protection simultaneously.
Wang~\emph{et al.}~\cite{Wang2019a} proposed an ``In-EDGE AI" system with deep reinforcement learning in FL. It delegates the reinforcement learning training task to the device side to protect the private dataset and brings more intelligence to edge systems.
Liu~\emph{et al.}~\cite{Liu2022} 
proposed a privacy-preserving distributed deep deterministic policy gradient scheme to make caching decisions for EC. 
At the same time, to preserve user privacy, the model only predicts content popularity by avoiding mining sensitive historical information. The model training process is completed by FL in order to prevent users from leaking privacy to ESes. 
Qiao~\emph{et al.}~\cite{Qiao2022} proposed an FL-based proactive content caching scheme to shorten content retrieval latency and protect users' private datasets. Firstly, the edge computing architecture reduces energy consumption and transmission overhead. The problems of client selection and local iteration round selection in the FL process are modeled as an MDP, which is solved by the deep reinforcement learning algorithm. The solution can alleviate the non-independent and independent distributed (Non-IID) data distribution problem and limited resources for end users.

In vehicular networks, privacy-preserving edge caching nodes, such as at RSUs, can also be effectively achieved by combining FL and DRL frameworks. However, the high mobility of vehicles introduces additional challenges to edge caching efficiency and privacy security. To tackle these challenges, Wu~\emph{et al.}~\cite{Wu2023} designed an asynchronous federated learning model to evaluate regional content popularity, taking into account vehicle movement speed, RSU coverage, and network channel conditions. They modified the selection of training vehicles and the aggregation function's weight, assigning different weights to vehicles with varying dwell times and channel conditions. They proposed a joint content placement strategy based on dueling DRL to overcome the caching efficiency degradation caused by high vehicle mobility. This strategy further reduces content transmission delay while ensuring user data privacy and RSU joint caching efficiency in edge vehicle computing scenarios.
Li~\emph{et al.}~\cite{Li2023} tackled the privacy and long-term training delay issues in high-precision map caching in intelligent connected vehicles (ICV) by formulating a framework called \textit{federated deep reinforcement learning} (F-DRL). F-DRL is an MDP-based edge cooperative caching technique in which Dueling-Deep-Q-Network (Dueling-DQN) is employed to optimize the adaptive edge caching scheme with an improved FL approach to preserve the privacy of ICV. For FL, resource provision and member vehicle selection are made using joint optimization to minimize the delays in training and load in the edge cache.

\begin{figure}[tb]
\centering
\includegraphics[width=0.95\linewidth]{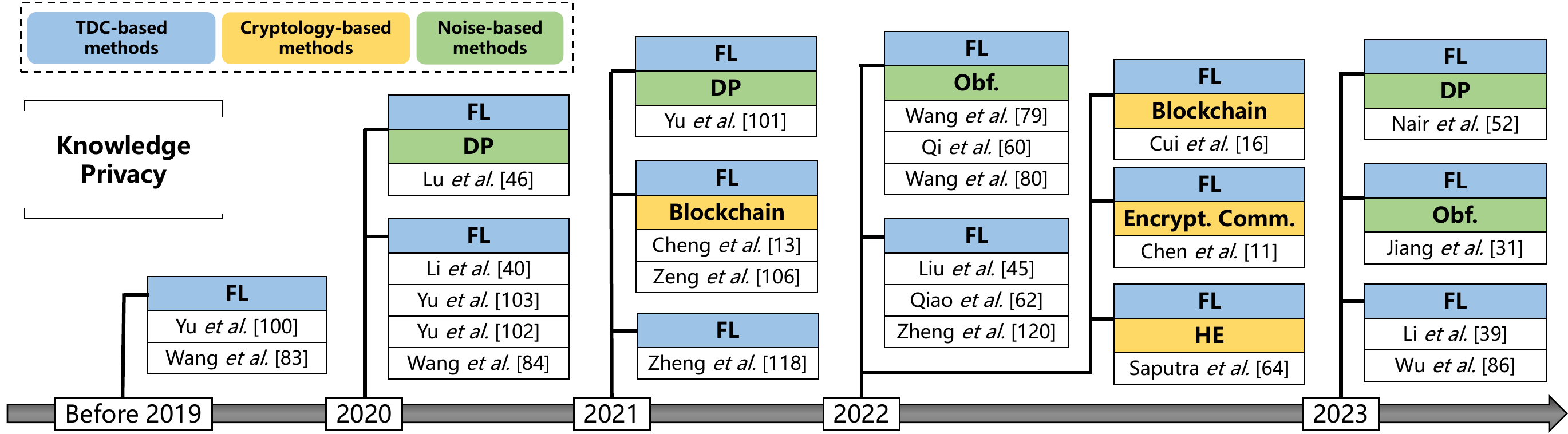}
\Description{A brief timeline of solutions for enhancing knowledge privacy.}
\vspace{-3mm}
\caption{A brief timeline of solutions for enhancing knowledge privacy.}
\label{fig: Knowledge privacy}
\vspace{-4mm}
\end{figure}
\subsection{Combining FL with Other Methods}

Other than requiring sensitive data to train machine learning models, the edge caching system may also need private information to make edge caching decisions. Therefore, some works~\cite{Qi2020, Cheng2021, Zheng2022, Wang2022a} have introduced additional privacy protection methods into the FL framework to enhance data privacy during the model training process.
Zheng~\emph{et al.}~\cite{Zheng2022} proposed a privacy-preserving FL model to predict popularity in an unsupervised manner. The prediction method introduces two concepts: local and global popularity, considering both efficiency and privacy. Local popularity can be evaluated by historical information by the Long Short-Term Memory (LSTM) model on users. In contrast, global popularity can only be predicted by the information at the current moment, which will be erased immediately at the next moment. 
FL is applied to perform offline training and online popularity evaluation with distributed information to avoid exposing privacy.
Wang~\emph{et al.}~\cite{Wang2022a} proposed a private FL-based caching scheme, which utilizes an FL framework and a pseudo rating matrix to collect statistical characteristics of user groups. With this distorted information, the server can predict the popularity of content and make caching decisions. The scheme also protects the privacy of individual users from being accessed by servers and other users.
Saputra~\emph{et al.}~\cite{Saputra2022} introduced the HE method into the FL framework to protect the privacy of MUs with constrained computing resources. The scheme allows MUs to upload encrypted training data to ESs, which can perform additional training processes. The portions of the encrypted decision problem are modeled as a multi-objective profit maximization problem considering both privacy and training costs. The optimization problem is proved to be a concave function that can be solved by the interior point method. At the same time, the training data cached at the EN or the cloud node is HE based on the Brakerski Fan Vercauteren (BFV) method.

\subsection{Combining  Noise-based FL with Other Methods}
More parameters or gradients (representing knowledge extracted from user-related data) can expose user privacy because attackers can probably infer and restore user information from exposed model information. In addition, model parameters may have a huge economic value, and directly uploading model parameters will compromise the self-interest of model owners. Therefore, there are works dedicated to upgrading the FL framework by adding noises~\cite{LuYun2020, Yu2021b, Jiang2023, NAIR2023} or other interference~\cite{Wang2022, Chen2022, Cui2022} to model information prior to exposure. 

The FL framework has further employed DP-based noise to safeguard the parameters or gradients in previous works.
Lu~\emph{et al.}~\cite{LuYun2020} designed a differentially private asynchronous federated learning scheme to share resources in vehicular networks. The proposed scheme uses LDP to perturb the local model parameters with noise drawn from the Gaussian distribution. Moreover, a distributed random update method is used to preserve the privacy of the global ML model during the update process.
Yu~\emph{et al.}~\cite{Yu2021b} proposed an FL framework based on privacy protection so that the user dataset is always kept locally. Further, the LDP mechanism is added while exchanging model parameters for aggregation to protect user privacy. In addition, this work proposes a hierarchical joint caching mechanism to combine the characteristics of local caching and global caching. A weighted aggregation method is used to solve the data imbalance problem. 
Jiang~\emph{et al.}~\cite{Jiang2023} developed a privacy-preserving FL framework for industrial data processing. This framework works by compressing adaptive gradients in the first place during model training at the edge terminal. Afterwards, hybrid DP is applied to optimize the FL framework, and the privacy-preserved gradients are transferred in the industrial environment.

Furthermore, some efforts try to enhance privacy protection in edge caching systems by integrating the Generative Adversarial Network (GAN) technique with FL.
Wang~\emph{et al.}~\cite{Wang2022} combined FL and Wasserstein Generative Adversarial Network  (WGAN) to improve further the efficiency of model training and accuracy of the popularity prediction model. With the fake data generated by WGAN, the privacy of users' real preferences can be enhanced.  Besides, gradient clipping and model parameter restriction are applied at the training time to protect model privacy and security.

Privacy preservation in vehicular edge computing is demanded since new attack types are developed continuously. To cope with the situation, Chen~\emph{et al.}~\cite{Chen2022} proposed a novel edge computing approach that utilizes unmanned aerial vehicle swarms as edge computing nodes to aggregate model parameters and caches the model parameters, thereby reducing the communication cost of the core network and protecting users' dataset. To enhance the security and privacy protection of cached model parameters, the authors designed a comprehensive protocol for model aggregation, storage, and transmission, which can effectively prevent potential security threats, such as poisoning attacks, man-in-the-middle attacks, and eavesdropping attacks.
Meanwhile, to defend against pollution attacks, the cosine similarity between local parameters and its edge aggregation parameters is calculated to exclude parameters uploaded by malicious nodes. Then, parameters are re-aggregated, and the aggregated parameters are sent to the cloud servers for the final process. Schnorr signature is also added before uploading the aggregated model parameters to ensure the reliability of the parameters.

In the Internet of Things (IoT) realm, edge computing architectures can expedite data processing, while edge caching can accelerate file delivery speeds for IoT devices. To ensure the reliability and privacy of data in IoT networks, Cui~\emph{et al.}~\cite{Cui2022} proposed a blockchain system comprising four contracts to predict content popularity, cache, and deliver sensitive content.
Meanwhile, to improve the security and throughput of the system, the Proof of Stake (PoS) consensus mechanism based on reputation is modified and applied to reach consensus more efficiently. Besides, the FL algorithm based on compressed gradients is used to protect the privacy information of ESs and reduce communication overhead. The K-means algorithm filters important gradients that must be uploaded accurately. 
These gradients are then quantified using a clustering-based quantization algorithm to reduce the amount of data uploaded. 
Meanwhile, an averaged gradient value is uploaded to the server for other gradients with a small value. 
Blockchain technique is also used to verify uploaded data.
Recently, the Internet of Medical Things (IoMT) is becoming popular. However, it is also prone to privacy threats like other edge computing-based approaches. To tackle these challenges for IoMT-based big data analytics, Nair~\emph{et al.}~\cite{NAIR2023} proposed an edge computing-based FL scheme called \textit{Fed\_select}, which ensures privacy and provides load reduction at the central FL server by introducing an edge server. To ensure privacy, \textit{Fed\_select} performs user anonymity at the edge server by employing hybrid encryption techniques with client and attribute selection performed at the edge server. Moreover, DP with Laplace noise is applied to the shared gradients to make them private during transfer.

\textcolor{black}{Due to limited space, we briefly present  an overview of additional solutions for safeguarding knowledge privacy in edge caching systems in Table~\ref{Tab: knowledge privacy} in Appendix, which covers methods not fully discussed in Section~\ref{sec: knowledge privacy}.}

\section{Open Challenges and Future Research Directions}\label{sec: Future Directions}

\begin{figure}[tb]
\centering
\includegraphics[width=0.70\linewidth]{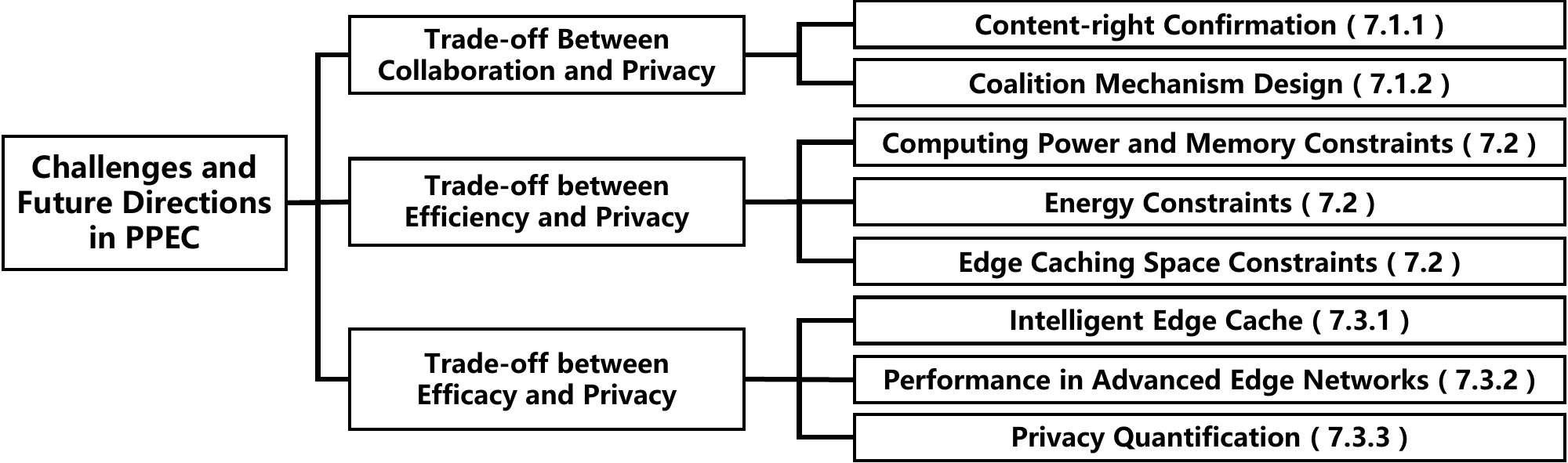}
\vspace{-4mm}
\Description{A outline for challenges and open issues section.}
\caption{A outline for challenges and open issues section.}
\label{fig: future work}
\vspace{-4mm}
\end{figure}

\textcolor{black}{In this section, we discuss open challenges and future research directions worth exploring in privacy-preserving edge caching (PPEC). As shown in Fig.~\ref{fig: future work}, we elaborate on the challenges and open issues from three major perspectives in PPEC: collaboration, efficiency, and efficacy.}

\subsection{Trade-off Between Collaboration and Privacy in PPEC}

Due to the large scale of network applications, 
it is common to deploy multiple edge servers for collaboratively caching content. To enable collaborations between edge servers, critical information such as cached content or other private information will be exchanged between edge servers, which can expose user privacy and raise privacy concerns. 
We outline two open privacy concerns when multiple edge servers share sensitive information. 

\subsubsection{\textbf{Content-right confirmation}}

Digital content can be easily copied and distributed, which is a double-edged sword making content-right confirmation difficult.
For instance, when social media content cached on a particular edge server is accessed by other edge servers, the edge server completely loses control of cached social media content because other edge servers can easily copy and redistribute this social media content~\cite{Araldo2018, Cui2020c, Yuan2016a}. 
Content-right confirmation is essential for content owners to maintain availability and accountability when using edge caching services to preserve content privacy. 
On the one hand, with content-right confirmation, it is easy to determine content ownership. Privacy strategies can be implemented to ensure that only authorized parties can access the content cached in an edge server. 
On the other hand, content-right confirmation is the basis of content accountability. With content-right confirmation, the content right can be authenticated when the content is used once, with the right changed accordingly.

\textcolor{black}{\textit{\textbf{Challenges.}}}
However, realizing content-right confirmation in edge caching systems is non-trivial due to several challenges. First, the distributed nature of edge caching systems makes it difficult to maintain a centralized and trusted authority for content ownership verification. 
Second, edge devices' dynamic and heterogeneous nature introduces complexity in implementing content-right confirmation mechanisms, which must be scalable and adaptable to different devices.
Furthermore, the use of encryption and privacy-enhancing technologies in edge caching systems further complicates content-right confirmation. While these technologies are essential for protecting the privacy of cached content, they may also prevent content owners from verifying the use of their content in the cache.

\textcolor{black}{\textit{\textbf{Future directions.}}} The challenges of realizing content-right confirmation in edge caching systems call for future work in several directions. Firstly, new verification mechanisms are needed that can handle the distributed nature, dynamics, and heterogeneity of edge devices. To overcome these challenges, several approaches have been proposed in the literature, such as blockchain-based solutions~\cite{Lei2020, Vu2019}. However, these mechanisms are short in scalability and the ability to adapt to different edge devices. 
Additionally, privacy-preserving verification methods that can coexist with encryption and other privacy-enhancing techniques should be explored. One possible solution is to leverage secure multi-party computation~\cite{Andreoletti2018} to enable verification while preserving the privacy of cached content. Finally, standardization efforts are needed to ensure interoperability between different edge caching systems and content providers. For example, the trust management mechanisms~\cite{Zhong2021, Xu2019} are proposed to enable content-right confirmation in ISP and D2D edge caching. Thus, promoting the adoption of content-right confirmation mechanisms and facilitating the collaboration between different stakeholders in different edge caching scenarios need to be further discussed.

\subsubsection{\textbf{Coalition mechanism design}}

Collaborative edge caching is essential for enhancing QoS. However, many edge caches are deployed on leased nodes provided by profit-oriented third-party providers, which are often decentralized~\cite{Cui2020c}, unreliable~\cite{Leguay2017} or self-interested~\cite{Cui2020c}. For instance, edge caching routers can be unreliable~\cite{Leguay2017, Cui2020c} in CDN, while RSUs and vehicles can be semi-trust~\cite{Qian2020, Zhang2022b, Hu2018, Jiang2020} or self-interested~\cite{Dai2020, Cui2020, Kong2019} in the edge IoV caching network. Similarly, in social media networks,  edge servers can be self-interested~\cite{Xu2019, Xu2020}.
To enable privacy-preserving applications and technology cooperation among  edge caching systems, coalition mechanisms are required. These mechanisms involve designing an incentive and allocation model that encourages participants in edge caching systems to join the coalition and maximize their benefits through a reasonable selection. 
Additionally, punishment mechanisms should also be considered when there are untrustworthy or dishonest nodes in the system.
Hence, incentive and allocation mechanisms can be designed in a thoughtful manner to foster participation and cooperation among edge caching systems.

\textcolor{black}{\textit{\textbf{Challenges.}}}  However, designing coalition mechanisms for PPEC is complicated  because it is necessary to balance several conflicting objectives. On the one hand, the mechanisms should encourage participants to contribute their resources to the coalition, ensuring that the costs and benefits of participants are distributed fairly~\cite{Andreoletti2018}. 
On the other hand, they must incentivize participants to prioritize the interests of the coalition over their individual interests and punish illegal strategies~\cite{Xu2020}, which is challenging when  participants are profit-oriented with conflicting goals~\cite{Xu2019}. 

\textcolor{black}{\textit{\textbf{Future directions.}}} Game theory is a powerful mathematical framework for investigating decision-making processes, and interactions among rational individuals or entities in coalition mechanisms. Its applications can optimize edge caching, capturing interactions between content providers, network operators and end users. Various game-theoretic models such as the non-cooperative game~\cite{Sivaraman2021}, Stackelberg game~\cite{Xu2019, Xu2020, Cao2020}, coalition games and potential games~\cite{Cao2020} can analyze the interaction among participants in edge caching systems. For instance, in a Stackelberg game, one player acts as the leader while the others follow. In the context of edge caching, the content provider can be modelled as the leader, while network caching operators are the followers~\cite{Xu2020}. 
 Similarly, the edge cache can act as the leader, followed by end users~\cite{Xu2019}. 
Nevertheless, game theory-based approaches with complete information~\cite{Xu2019, Xu2020} can not be directly applied in privacy-preserving scenarios since the information is likely incomplete to players in edge caching systems.  Besides, it is impractical to assume that every player is benign at an open-access edge network. There may exist semi-honest and even malicious players. 
Therefore, it is necessary to conduct further investigations into the coalition mechanisms when analyzing the complex interactions between different kinds of players in  collaborative PPEC.

\subsection{Limited Capacity for Running Privacy-enhancing Caching Algorithms}

\textcolor{black}{\textit{\textbf{Challenges.}}}  Edge devices are becoming increasingly crucial in edge caching networks. However, these devices are typically limited in processing power, memory, caching space, and energy capacity, which present challenges for running privacy-enhancing algorithms: 
    (1) Limited computing power and memory pose a significant challenge on implementing complex privacy-preserving algorithms on edge devices~\cite{Ni2021}. 
    To address this challenge, the development of lightweight privacy-preserving algorithms is desired to protect user privacy without compromising caching performance. 
    Lightweight homomorphic encryption~\cite{Cui2020}, identity authentication~\cite{Xue2019, Xue2018, Zhang2022b, Tong2022}, and differential privacy~\cite{Sivaraman2021, Acs2019} are prospective approaches that can reconcile privacy protection and computational efficiency. Additionally, it is vital to ensure that the developed algorithms are robust and productive, meeting the needs of edge devices. 
    (2) Energy constraints: Edge devices such as autonomous vehicles and smartphones are often battery-powered~\cite{Mao2017}, which can limit the ability of caching~\cite{Saputra2022} and communication~\cite{Cui2022}, and hence lower the performance of privacy-preserving algorithms~\cite{Qiao2022, Saputra2022}. To address this challenge, energy-efficient caching management techniques should be developed to minimize the energy consumption of PPEC approaches. Techniques such as data compression~\cite{Cui2022} and optimization models~\cite{Qiao2022} can be adopted to reduce the consumption of caching and communication to minimize energy usage. For instance, an energy-aware client selection and communication method for FL was proposed in~\cite{Qiao2022} that reduced energy consumption by up to 50\% compared to traditional FL methods when protecting the privacy of data sources. 
    (3) Cache space is another significant constraint for edge caching systems due to the limited storage capacity compared to the vast amount of content that can be cached. However, research has shown that only a small fraction of content is popular, while the majority of users concentrate their access on popular content, implying a long tail distribution of content popularity~\cite{Ma2017b, Wang2019}. Therefore, it is crucial to determine which content should be cached based on popularity and user preferences while considering privacy concerns. PPEC approaches need to trade-off between privacy protection and caching performance. 

\textcolor{black}{\textit{\textbf{Future directions.}}} Implementing privacy-preserving algorithms can complicate the system, which adversely impacts caching performance~\cite{Liang2019}. Conversely, simplifying the system may increase the risk of privacy breaches. Challenges associated with PPEC include balancing the complexity of protection algorithms with caching performance~\cite{Xue2019, Xue2018, Zhu2021} or limited resources~\cite{Sivaraman2021, Tong2022, Cui2022}, and ensuring user privacy while enabling efficient content distribution~\cite{Cui2020, Yuan2016a}. 
To sum up, an in-depth understanding of the limitations of edge devices and designing practical and feasible solutions are vital in enhancing PPEC.

\subsection{Trade-off between Efficacy and Privacy in PPEC}

{
\subsubsection{\textcolor{black}{\textbf{Privacy-enhancing intelligent edge cache}}}
Machine learning-based methods have become a powerful tool for optimizing edge caching performance and developing intelligent caching algorithms~\cite{Qiao2022, Liu2022}. However, there has been some controversy regarding privacy violations in edge caching. Privacy concerns arise when edge caching providers analyze and manage content in their cache since the storage spaces of edge cache are limited, and the content scale is growing rapidly~\cite{Wang2019}. To provide intelligent edge caching, providers may be curious about the content stored in their cache (e.g., popular content~\cite{Araldo2018, Cui2020, Cui2020c}) and the confidential information about consumers (e.g., request record~\cite{Wu2016, Yuan2016a, Nikolaou2016}, identifiable information~\cite{Nguyen2023, Zhang2022b, Xue2019, Xue2018}). Providers may use monitoring and inference attacks to compromise consumers' privacy to improve caching efficiency and gain economic benefits. 
\textcolor{black}{Additionally, the rise of generative AI applications can further complicate this landscape. Generative AI can be leveraged to design more sophisticated caching algorithms that model user behavior and more accurately anticipate content popularity~\cite{Tang2022}. Furthermore, by caching pretrained foundation models (PFMs) at the edge network~\cite{Xu2023}, various multimedia enhancement techniques, such as super-resolution, can be deployed to effectively improve user QoE and reduce transport delays. 
However, these improvements may also come with increased privacy risks, as these models might require more granular user data for fine-tuning at edge servers, potentially exposing sensitive information.
}
Therefore, developing effective privacy-preserving mechanisms in intelligent edge caching algorithms is crucial to address these problems.

\textcolor{black}{\textit{\textbf{Challenges.}}}  However, the open-edge network provides an ideal entrance or interface for attackers to obtain private data or knowledge from machine learning methods designed for edge caching systems. 
Therefore, reconciling privacy and efficiency in intelligent caching methods at the open-edge network is challenging for several reasons. 
Firstly, the diversity of user requirements and content in machine learning-based edge networks can be more significant than traditional caching systems~\cite{Zhou2019}, which makes it difficult to apply traditional privacy-enhancing mechanisms directly. 
Secondly, the edge network is usually open, and multi-access~\cite{Xu2019}, implying that it is difficult to control the access of the cache so as to preserve privacy. 
Finally, the semi-trust or unreliable third-party caching service providers exacerbate the challenge to the design of protection method~\cite{Xue2019, Xue2018, Qiao2022, Araldo2018, Liu2022}.
\textcolor{black}{For instance, the introduction of generative AI models could intensify these challenges, as these models may require ongoing access to real-time data for training content-specific generative models~\cite{Xu2023}, thereby creating new vectors for privacy breaches.}
Therefore, designing intelligent caching methods that can well balance privacy and efficiency is challenging.


\textit{\textbf{Future directions.}}} The FL framework is one of the essential methods to preserve private data in the machine learning process~\cite{Wang2020, Wang2019a}.
Based on the FL framework, different parties may distributedly predict the critical information, e.g., content population~\cite{Qiao2022, Liu2022, Zheng2022} or user location migratory pattern~\cite{Wu2023}, for intelligent caching decisions at the edge network. However, some parties may be dishonest and malicious. In particular, malicious users in FL may bring poisoned data to affect the overall computing of the global model. For example, dishonest parties may back-infer their partners' model by collecting their gradients to infer private information~\cite{Cui2022, Yu2021b}. These attacks can lead to the disclosure of critical privacy information or destroy caching performance. 
Additionally, some adversaries even deliberately provide incorrect model parameters during lateral FL to disrupt the overall computation and impact model performance~\cite{Wang2022}. 
The security of private computing in the FL framework is also a challenging open topic for edge caching.
\textcolor{black}{
Furthermore, generative AI could introduce new methods for obfuscating or anonymizing data before it is cached~\cite{Wang2022}, thereby adding an additional layer of privacy protection. Similarly, with federated learning or other distributed learning frameworks, a lightweight generative model can be fine-tuned or trained with local data, stored, and run on edge devices, providing personalized and customized AIGC services in real time while maintaining user privacy~\cite{Xu2024}.
}

\subsubsection{\textbf{Reconcile efficacy and privacy in advanced edge networks}}
{\color{black}
Traditional edge caching algorithms often rely on pre-determined and suboptimal caching policies, which may not be effective in dynamic network environments where conditions~\cite{Zhang2022d} and user behavior~\cite{Ma2017b} can change rapidly over time. In modern network systems, resource access patterns are highly dynamic and complex~\cite{Zhang2022,Zhao2024}. Additionally, factors such as user interests, geographical locations, and IoT device connectivity in edge scenarios are also highly dynamic, making one-time trained models less adaptable to these evolving conditions~\cite{Chen2022a}. Some existing works propose high-cost training of machine learning models, which may not be feasible for resource-constrained edge or terminal devices~\cite{Cui2023}. Furthermore, advanced network infrastructures, such as Named Data Networks (NDN) and Information-Centric Networks (ICN), are becoming more prevalent in edge networks, introducing new challenges for privacy preservation.
}

\textcolor{black}{\textit{\textbf{Challenges.}}}
Several open challenges exist in designing privacy-preserving algorithms for advanced network systems. One challenge is to balance the privacy protection strength and the accuracy of the model prediction. The decision-making process in online scenarios is already highly challenging, and the introduction of privacy protection methods, such as noise perturbation, can further compromise the algorithm's performance or even make it unusable. Another challenge is to develop privacy-preserving algorithms that are computationally efficient and can be easily deployed in dynamic network environments. 
\textcolor{black}{Moreover, advanced network architectures inherently focus on data rather than specific endpoints, leading to new vulnerabilities. For instance, in NDN and ICN, content is named and cached throughout the network, which could increase privacy risks if sensitive data is cached without proper safeguards~\cite{Sivaraman2021,Xue2019,Cui2020c}. The ability to cache and retrieve data based on content names rather than IP addresses can expose more granular user preferences and access patterns, making it easier for adversaries to infer sensitive information. Additionally, these architectures may complicate the implementation of privacy-preserving caching strategies, as they require more sophisticated mechanisms to control data access and ensure data integrity across distributed caches.}

\textcolor{black}{\textit{\textbf{Future directions.}} Online learning algorithms, such as reinforcement learning~\cite{Xiao2018a} and continuous learning, has become increasingly popular for solving complex problems in various fields, including edge caching in dynamic network environments~\cite{Krishnendu2022, Zhang2022d, Chen2022a, Cui2023}.} However, online learning algorithms can also pose a risk to user privacy when collecting and processing sensitive user data. 
Therefore, how to safely use the latest historical information to make efficient online caching decisions is a problem worthy of discussion. There are little efforts to address the privacy concerns associated with online learning algorithms. The FL may be a possible framework to allow multiple parties to process data jointly without revealing their raw datasets in dynamic scenarios~\cite{Krishnendu2022}. Additionally, differential privacy techniques can be used to add random noises to the data in advanced caching systems to obscure individual information~\cite{Zhou2023}.

\subsubsection{\textbf{Privacy quantification for PPEC}}
Privacy quantification is a critical aspect of PPEC systems as it allows for the measurement and assessment of privacy protection levels provided by these systems~\cite{Sivaraman2021, Yan2021, Wu2016, Acs2019}. However, most current work on privacy-enhanced intelligent edge caching lacks specific privacy metrics. Rather than developing clear and effective privacy metrics, researchers often combine existing privacy protection schemes and claim that their works can protect privacy. Unfortunately, without clear quantification of the privacy protection effect, it fails to identify weaknesses for improving privacy protection~\cite{Acs2019, Yan2021}. 

\textcolor{black}{\textit{\textbf{Challenges.}}}  One of the primary challenges in privacy quantification is developing an accurate and consistent metric for measuring privacy protection levels. It is a complicated task to propose a universal method to measure different types of data leakage in the edge cache. Therefore, an appropriate privacy quantification method with a formalized definition should be established to guide the design of PPEC systems. For instance, in intelligent caching algorithms based on reinforcement learning models, a good privacy exposure quantitative index can guide the model's reward design and help the agent make better caching decisions~\cite{Xu2020}. Various privacy metrics have been proposed in the literature, such as the information-theoretic converse bound~\cite{Yan2021} and the size of the anonymity set~\cite{Wu2016}. However, each metric has its limitations and may not be suitable for general PPEC systems. 
Evaluating the privacy protection degree in dynamic network environments is another challenge in PPEC. Edge caching systems operate in a constantly changing environment, and various factors can impact privacy protection levels, which makes it challenging to determine an accurate and consistent privacy metric that can be applied in a dynamic environment. 

\textcolor{black}{\textit{\textbf{Future directions.}}} To address these challenges, researchers can explore the use of online algorithms~\cite{Pang2022, Li2021} to predict privacy protection levels in real time based on network traffic patterns and user behaviour. 
This approach can help to dynamically adjust privacy protection levels in response to changes in the network environment and improve the effectiveness of PPEC.
{\color{black}
Future work for designing privacy metrics (similar to the privacy budget in differential privacy~\cite{Sivaraman2021, Acs2019} and mutual information in information theory~\cite{Sivaraman2021}) is desired.
An innovative definition of privacy measurement applicable in intelligent edge caching scenarios~\cite{Zhang2022a} should be designed to guide PPEC. Finally, exploring the integration of blockchain technology in PPEC could offer new avenues for ensuring data integrity and transparency~\cite{Qian2020,Cui2022}. Blockchain can be used to create an immutable record of data access and modifications, thus providing a reliable audit trail that enhances trust and accountability in edge caching systems.}

\section{Conclusion}\label{sec: Conclusion}
Edge caching has shown significant potential for improving network performance and resource utilization, but privacy concerns must be considered when deploying edge caches. \textcolor{black}{This article has analyzed and summarized the most prominent privacy issues in edge caching systems from a \textit{sensitive information} perspective, based on which a comprehensive classification has been proposed.} The recent countermeasures for alleviating the exposed threats of different private information have been retrospectively reviewed. The article concludes with lessons learned and highlights open challenges for future research in the PPEC. Further investigations are needed to ensure the privacy and performance of edge caching while also reconciling the trade-off between privacy protection and caching performance optimization.

\begin{acks}
This work was supported in part by Shenzhen Science and Technology Program under Grant KJZD20230923113901004, in part by the Science and Technology Planning Project of Guangdong Province under Grant 2023A0505020006, in part by Science and Technology Development Fund, Macau SAR under Grant 0008/2022/AGJ, in part by the National Natural Science Foundation of China under Grant U2001209 and Grant 62072486. (Corresponding author: Di Wu.)
\end{acks}

\bibliographystyle{ACM-Reference-Format.bst}
\bibliography{ref}


\begin{thebibliography}{124}


\ifx \showCODEN    \undefined \def \showCODEN     #1{\unskip}     \fi
\ifx \showDOI      \undefined \def \showDOI       #1{#1}\fi
\ifx \showISBNx    \undefined \def \showISBNx     #1{\unskip}     \fi
\ifx \showISBNxiii \undefined \def \showISBNxiii  #1{\unskip}     \fi
\ifx \showISSN     \undefined \def \showISSN      #1{\unskip}     \fi
\ifx \showLCCN     \undefined \def \showLCCN      #1{\unskip}     \fi
\ifx \shownote     \undefined \def \shownote      #1{#1}          \fi
\ifx \showarticletitle \undefined \def \showarticletitle #1{#1}   \fi
\ifx \showURL      \undefined \def \showURL       {\relax}        \fi
\providecommand\bibfield[2]{#2}
\providecommand\bibinfo[2]{#2}
\providecommand\natexlab[1]{#1}
\providecommand\showeprint[2][]{arXiv:#2}

\bibitem[\protect\citeauthoryear{Abadi, McMahan, Chu, Mironov, Zhang, et~al\mbox{.}}{Abadi et~al\mbox{.}}{2016}]%
        {Abadi2016}
\bibfield{author}{\bibinfo{person}{Mart{\'{i}}n Abadi}, \bibinfo{person}{H.~Brendan McMahan}, {et~al\mbox{.}}} \bibinfo{year}{2016}\natexlab{}.
\newblock \showarticletitle{{Deep learning with differential privacy}}. In \bibinfo{booktitle}{\emph{Proceedings of the ACM Conference on Computer and Communications Security (CCS)}}. \bibinfo{publisher}{ACM}, \bibinfo{pages}{308--318}.
\newblock
\showISBNx{9781450341394}
\showISSN{15437221}


\bibitem[\protect\citeauthoryear{Acs, Conti, Gasti, Ghali, Tsudik, et~al\mbox{.}}{Acs et~al\mbox{.}}{2019}]%
        {Acs2019}
\bibfield{author}{\bibinfo{person}{Gergely Acs}, \bibinfo{person}{Mauro Conti}, {et~al\mbox{.}}} \bibinfo{year}{2019}\natexlab{}.
\newblock \showarticletitle{{Privacy-Aware Caching in Information-Centric Networking}}.
\newblock \bibinfo{journal}{\emph{IEEE Transactions on Dependable and Secure Computing}} \bibinfo{volume}{16}, \bibinfo{number}{2} (\bibinfo{date}{Mar} \bibinfo{year}{2019}), \bibinfo{pages}{313--328}.
\newblock
\showISSN{19410018}


\bibitem[\protect\citeauthoryear{Amini, Lindqvist, Hong, Lin, Toch, et~al\mbox{.}}{Amini et~al\mbox{.}}{2011}]%
        {Amini2011}
\bibfield{author}{\bibinfo{person}{Shahriyar Amini}, \bibinfo{person}{Janne Lindqvist}, {et~al\mbox{.}}} \bibinfo{year}{2011}\natexlab{}.
\newblock \showarticletitle{{Cach{\'{e}}: Caching location-enhanced content to improve user privacy}}. In \bibinfo{booktitle}{\emph{Proceedings of the 9th International Conference on Mobile Systems, Applications, and Services (MobiSys)}}. \bibinfo{publisher}{ACM}, \bibinfo{pages}{197--209}.
\newblock
\showISBNx{9781450306430}


\bibitem[\protect\citeauthoryear{Andreoletti, Ayoub, Giordano, Verticale, and Tornatore}{Andreoletti et~al\mbox{.}}{2019a}]%
        {Andreoletti2019a}
\bibfield{author}{\bibinfo{person}{Davide Andreoletti}, \bibinfo{person}{Omran Ayoub}, {et~al\mbox{.}}} \bibinfo{year}{2019}\natexlab{a}.
\newblock \showarticletitle{{Privacy-preserving caching in ISP networks}}. In \bibinfo{booktitle}{\emph{IEEE 20th International Conference on High Performance Switching and Routing (HPSR)}}. \bibinfo{publisher}{IEEE}, \bibinfo{pages}{1--6}.
\newblock
\showISBNx{9781728116860}
\showISSN{23255609}


\bibitem[\protect\citeauthoryear{Andreoletti, Giordano, Verticale, and Tornatore}{Andreoletti et~al\mbox{.}}{2018}]%
        {Andreoletti2018}
\bibfield{author}{\bibinfo{person}{Davide Andreoletti}, \bibinfo{person}{Silvia Giordano}, {et~al\mbox{.}}} \bibinfo{year}{2018}\natexlab{}.
\newblock \showarticletitle{{Discovering the Geographic Distribution of Live Videos' Users: A Privacy-Preserving Approach}}. In \bibinfo{booktitle}{\emph{IEEE Global Communications Conference (GLOBECOM)}}. \bibinfo{publisher}{IEEE}, \bibinfo{pages}{1--6}.
\newblock
\showISBNx{9781538647271}


\bibitem[\protect\citeauthoryear{Andreoletti, Rottondi, Giordano, Verticale, and Tornatore}{Andreoletti et~al\mbox{.}}{2019b}]%
        {Andreoletti2019}
\bibfield{author}{\bibinfo{person}{Davide Andreoletti}, \bibinfo{person}{Cristina Rottondi}, {et~al\mbox{.}}} \bibinfo{year}{2019}\natexlab{b}.
\newblock \showarticletitle{{An Open Privacy-Preserving and Scalable Protocol for a Network-Neutrality Compliant Caching}}. In \bibinfo{booktitle}{\emph{IEEE International Conference on Communications (ICC)}}. \bibinfo{publisher}{IEEE}, \bibinfo{pages}{1--6}.
\newblock
\showISBNx{9781538680889}
\showISSN{15503607}


\bibitem[\protect\citeauthoryear{Araldo, Dan, and Rossi}{Araldo et~al\mbox{.}}{2018}]%
        {Araldo2018}
\bibfield{author}{\bibinfo{person}{Andrea Araldo}, \bibinfo{person}{Gyorgy Dan}, {et~al\mbox{.}}} \bibinfo{year}{2018}\natexlab{}.
\newblock \showarticletitle{{Caching Encrypted Content Via Stochastic Cache Partitioning}}.
\newblock \bibinfo{journal}{\emph{IEEE/ACM Transactions on Networking}} \bibinfo{volume}{26}, \bibinfo{number}{1} (\bibinfo{date}{Jan} \bibinfo{year}{2018}), \bibinfo{pages}{548--561}.
\newblock
\showISSN{10636692}


\bibitem[\protect\citeauthoryear{Bellavista, Foschini, and Mora}{Bellavista et~al\mbox{.}}{2021}]%
        {Paolo2021}
\bibfield{author}{\bibinfo{person}{Paolo Bellavista}, \bibinfo{person}{Luca Foschini}, {et~al\mbox{.}}} \bibinfo{year}{2021}\natexlab{}.
\newblock \showarticletitle{Decentralised Learning in Federated Deployment Environments: A System-Level Survey}.
\newblock \bibinfo{journal}{\emph{Comput. Surveys}} \bibinfo{volume}{54}, \bibinfo{number}{1}, Article \bibinfo{articleno}{15} (\bibinfo{date}{Feb} \bibinfo{year}{2021}), \bibinfo{numpages}{38}~pages.
\newblock
\showISSN{0360-0300}


\bibitem[\protect\citeauthoryear{{Brendan McMahan}, Moore, Ramage, Hampson, and {Ag{\"{u}}era y Arcas}}{{Brendan McMahan} et~al\mbox{.}}{2017}]%
        {BrendanMcMahan2017}
\bibfield{author}{\bibinfo{person}{H. {Brendan McMahan}}, \bibinfo{person}{Eider Moore}, {et~al\mbox{.}}} \bibinfo{year}{2017}\natexlab{}.
\newblock \showarticletitle{{Communication-efficient learning of deep networks from decentralized data}}. In \bibinfo{booktitle}{\emph{Proceedings of the 20th International Conference on Artificial Intelligence and Statistics}}. \bibinfo{publisher}{PMLR}, \bibinfo{pages}{1273--1282}.
\newblock
\showISSN{2640-3498}


\bibitem[\protect\citeauthoryear{Cao, Xu, Du, Li, Xiao, et~al\mbox{.}}{Cao et~al\mbox{.}}{2020}]%
        {Cao2020}
\bibfield{author}{\bibinfo{person}{Tengfei Cao}, \bibinfo{person}{Changqiao Xu}, {et~al\mbox{.}}} \bibinfo{year}{2020}\natexlab{}.
\newblock \showarticletitle{{Reliable and Efficient Multimedia Service Optimization for Edge Computing-Based 5G Networks: Game Theoretic Approaches}}.
\newblock \bibinfo{journal}{\emph{IEEE Transactions on Network and Service Management}} \bibinfo{volume}{17}, \bibinfo{number}{3} (\bibinfo{date}{Sep} \bibinfo{year}{2020}), \bibinfo{pages}{1610--1625}.
\newblock
\showISSN{19324537}


\bibitem[\protect\citeauthoryear{Chen, Chen, Hu, and Zhang}{Chen et~al\mbox{.}}{2022a}]%
        {Chen2022}
\bibfield{author}{\bibinfo{person}{Qi Chen}, \bibinfo{person}{Bing Chen}, {et~al\mbox{.}}} \bibinfo{year}{2022}\natexlab{a}.
\newblock \showarticletitle{{Edge-based Protection Against Malicious Poisoning for Distributed Federated Learning}}. In \bibinfo{booktitle}{\emph{IEEE 25th International Conference on Computer Supported Cooperative Work in Design (CSCWD)}}. \bibinfo{publisher}{IEEE}, \bibinfo{pages}{459--464}.
\newblock
\showISBNx{9781665405270}


\bibitem[\protect\citeauthoryear{Chen, Zhang, Ma, Zhang, Qian, et~al\mbox{.}}{Chen et~al\mbox{.}}{2022b}]%
        {Chen2022a}
\bibfield{author}{\bibinfo{person}{Zhiqi Chen}, \bibinfo{person}{Sheng Zhang}, {et~al\mbox{.}}} \bibinfo{year}{2022}\natexlab{b}.
\newblock \showarticletitle{{An Online Approach for DNN Model Caching and Processor Allocation in Edge Computing}}. In \bibinfo{booktitle}{\emph{IEEE/ACM 30th International Symposium on Quality of Service (IWQoS)}}. \bibinfo{publisher}{IEEE}, \bibinfo{pages}{1--10}.
\newblock
\showISBNx{9781665468244}


\bibitem[\protect\citeauthoryear{Cheng, Sun, Liu, Xia, Sun, et~al\mbox{.}}{Cheng et~al\mbox{.}}{2021}]%
        {Cheng2021}
\bibfield{author}{\bibinfo{person}{Runze Cheng}, \bibinfo{person}{Yao Sun}, {et~al\mbox{.}}} \bibinfo{year}{2021}\natexlab{}.
\newblock \showarticletitle{{A Privacy-preserved D2D Caching Scheme Underpinned by Blockchain-enabled Federated Learning}}. In \bibinfo{booktitle}{\emph{IEEE Global Communications Conference (GLOBECOM)}}. \bibinfo{publisher}{IEEE}, \bibinfo{pages}{1--6}.
\newblock
\showISBNx{9781728181042}


\bibitem[\protect\citeauthoryear{Cui, Wei, Zhong, Zhang, Xu, et~al\mbox{.}}{Cui et~al\mbox{.}}{2020c}]%
        {Cui2020}
\bibfield{author}{\bibinfo{person}{Jie Cui}, \bibinfo{person}{Lu Wei}, {et~al\mbox{.}}} \bibinfo{year}{2020}\natexlab{c}.
\newblock \showarticletitle{{Edge computing in VANETs-An efficient and privacy-preserving cooperative downloading scheme}}.
\newblock \bibinfo{journal}{\emph{IEEE Journal on Selected Areas in Communications}} \bibinfo{volume}{38}, \bibinfo{number}{6} (\bibinfo{date}{Apr} \bibinfo{year}{2020}), \bibinfo{pages}{1191--1204}.
\newblock
\showISSN{15580008}


\bibitem[\protect\citeauthoryear{Cui, Ni, Zhou, Wang, Zhang, et~al\mbox{.}}{Cui et~al\mbox{.}}{2023}]%
        {Cui2023}
\bibfield{author}{\bibinfo{person}{Laizhong Cui}, \bibinfo{person}{Erchao Ni}, {et~al\mbox{.}}} \bibinfo{year}{2023}\natexlab{}.
\newblock \showarticletitle{{Towards Real-Time Video Caching at Edge Servers: A Cost-Aware Deep Q-Learning Solution}}.
\newblock \bibinfo{journal}{\emph{IEEE Transactions on Multimedia}}  \bibinfo{volume}{25} (\bibinfo{date}{Nov} \bibinfo{year}{2023}), \bibinfo{pages}{302--314}.
\newblock
\showISSN{19410077}


\bibitem[\protect\citeauthoryear{Cui, Su, Ming, Chen, Yang, et~al\mbox{.}}{Cui et~al\mbox{.}}{2022}]%
        {Cui2022}
\bibfield{author}{\bibinfo{person}{Laizhong Cui}, \bibinfo{person}{Xiaoxin Su}, {et~al\mbox{.}}} \bibinfo{year}{2022}\natexlab{}.
\newblock \showarticletitle{{CREAT: Blockchain-Assisted Compression Algorithm of Federated Learning for Content Caching in Edge Computing}}.
\newblock \bibinfo{journal}{\emph{IEEE Internet of Things Journal}} \bibinfo{volume}{9}, \bibinfo{number}{16} (\bibinfo{date}{Aug} \bibinfo{year}{2022}), \bibinfo{pages}{14151--14161}.
\newblock
\showISSN{23274662}


\bibitem[\protect\citeauthoryear{Cui, Asghar, and Russello}{Cui et~al\mbox{.}}{2020a}]%
        {Cui2020c}
\bibfield{author}{\bibinfo{person}{Shujie Cui}, \bibinfo{person}{Muhammad~Rizwan Asghar}, {et~al\mbox{.}}} \bibinfo{year}{2020}\natexlab{a}.
\newblock \showarticletitle{{Multi-CDN: Towards Privacy in Content Delivery Networks}}.
\newblock \bibinfo{journal}{\emph{IEEE Transactions on Dependable and Secure Computing}} \bibinfo{volume}{17}, \bibinfo{number}{5} (\bibinfo{date}{Sep} \bibinfo{year}{2020}), \bibinfo{pages}{984--999}.
\newblock
\showISSN{19410018}


\bibitem[\protect\citeauthoryear{Cui, Gao, Li, Shi, Zhang, et~al\mbox{.}}{Cui et~al\mbox{.}}{2020b}]%
        {Cui2020b}
\bibfield{author}{\bibinfo{person}{Yuanbo Cui}, \bibinfo{person}{Fei Gao}, {et~al\mbox{.}}} \bibinfo{year}{2020}\natexlab{b}.
\newblock \showarticletitle{{Cache-based privacy preserving solution for location and content protection in location-based services}}.
\newblock \bibinfo{journal}{\emph{Sensors}} \bibinfo{volume}{20}, \bibinfo{number}{16} (\bibinfo{date}{Aug} \bibinfo{year}{2020}), \bibinfo{pages}{4651}.
\newblock
\showISSN{14248220}


\bibitem[\protect\citeauthoryear{Dai, Xu, Zhang, Maharjan, and Zhang}{Dai et~al\mbox{.}}{2020}]%
        {Dai2020}
\bibfield{author}{\bibinfo{person}{Yueyue Dai}, \bibinfo{person}{Du Xu}, {et~al\mbox{.}}} \bibinfo{year}{2020}\natexlab{}.
\newblock \showarticletitle{{Deep Reinforcement Learning and Permissioned Blockchain for Content Caching in Vehicular Edge Computing and Networks}}.
\newblock \bibinfo{journal}{\emph{IEEE Transactions on Vehicular Technology}} \bibinfo{volume}{69}, \bibinfo{number}{4} (\bibinfo{date}{Apr} \bibinfo{year}{2020}), \bibinfo{pages}{4312--4324}.
\newblock
\showISSN{19399359}


\bibitem[\protect\citeauthoryear{Dhar and Varshney}{Dhar and Varshney}{2011}]%
        {Dhar2011}
\bibfield{author}{\bibinfo{person}{Subhankar Dhar} {and} \bibinfo{person}{Upkar Varshney}.} \bibinfo{year}{2011}\natexlab{}.
\newblock \showarticletitle{{Challenges and business models for mobile location-based services and advertising}}.
\newblock \bibinfo{journal}{\emph{Commun. ACM}} \bibinfo{volume}{54}, \bibinfo{number}{5} (\bibinfo{date}{May} \bibinfo{year}{2011}), \bibinfo{pages}{121--129}.
\newblock
\showISSN{00010782}


\bibitem[\protect\citeauthoryear{Famaey, Iterbeke, Wauters, and {De Turck}}{Famaey et~al\mbox{.}}{2013}]%
        {Famaey2013}
\bibfield{author}{\bibinfo{person}{Jeroen Famaey}, \bibinfo{person}{Fr{\'{e}}d{\'{e}}ric Iterbeke}, {et~al\mbox{.}}} \bibinfo{year}{2013}\natexlab{}.
\newblock \showarticletitle{{Towards a predictive cache replacement strategy for multimedia content}}.
\newblock \bibinfo{journal}{\emph{Journal of Network and Computer Applications}} \bibinfo{volume}{36}, \bibinfo{number}{1} (\bibinfo{date}{Jan} \bibinfo{year}{2013}), \bibinfo{pages}{219--227}.
\newblock
\showISSN{10848045}


\bibitem[\protect\citeauthoryear{Feng, Jiang, Niyato, Zheng, and You}{Feng et~al\mbox{.}}{2019}]%
        {Feng2019}
\bibfield{author}{\bibinfo{person}{Haojie Feng}, \bibinfo{person}{Yanxiang Jiang}, {et~al\mbox{.}}} \bibinfo{year}{2019}\natexlab{}.
\newblock \showarticletitle{{Content popularity prediction via deep learning in cache-enabled fog radio access networks}}. In \bibinfo{booktitle}{\emph{Proceedings of the 38th GLOBECOM}}. \bibinfo{publisher}{IEEE}, \bibinfo{pages}{1--6}.
\newblock
\showISBNx{9781728109626}


\bibitem[\protect\citeauthoryear{Fotia, Delicato, and Fortino}{Fotia et~al\mbox{.}}{2023}]%
        {Lidia2023}
\bibfield{author}{\bibinfo{person}{Lidia Fotia}, \bibinfo{person}{Fl\'{a}via Delicato}, {et~al\mbox{.}}} \bibinfo{year}{2023}\natexlab{}.
\newblock \showarticletitle{Trust in Edge-based Internet of Things Architectures: State of the Art and Research Challenges}.
\newblock \bibinfo{journal}{\emph{Comput. Surveys}} \bibinfo{volume}{55}, \bibinfo{number}{9}, Article \bibinfo{articleno}{182} (\bibinfo{date}{Jan} \bibinfo{year}{2023}), \bibinfo{numpages}{34}~pages.
\newblock
\showISSN{0360-0300}


\bibitem[\protect\citeauthoryear{Gil}{Gil}{2017}]%
        {gil2017web}
\bibfield{author}{\bibinfo{person}{Omer Gil}.} \bibinfo{year}{2017}\natexlab{}.
\newblock \bibinfo{booktitle}{\emph{Web cache deception attack}}.
\newblock \bibinfo{type}{{T}echnical {R}eport}. \bibinfo{institution}{Black Hat USA}.
\newblock


\bibitem[\protect\citeauthoryear{{GU Yi-ming, BAI Guang-wei, SHEN Hang}}{{GU Yi-ming, BAI Guang-wei, SHEN Hang}}{2019}]%
        {GUYi-mingBAIGuang-weiSHENHang}
\bibfield{author}{\bibinfo{person}{HU~Yu-jia. {GU Yi-ming, BAI Guang-wei, SHEN Hang}}.} \bibinfo{year}{2019}\natexlab{}.
\newblock \showarticletitle{{Pre-cache Based Privacy Protection Mechanism in Continuous LBS Queries}}.
\newblock \bibinfo{journal}{\emph{Computer Science}} \bibinfo{volume}{46}, \bibinfo{number}{5} (\bibinfo{date}{May} \bibinfo{year}{2019}), \bibinfo{pages}{122--128}.
\newblock
\showISSN{1002-137X}


\bibitem[\protect\citeauthoryear{Guo, Ding, Wang, and Jia}{Guo et~al\mbox{.}}{2022}]%
        {Guo2022}
\bibfield{author}{\bibinfo{person}{Jianxiong Guo}, \bibinfo{person}{Xingjian Ding}, {et~al\mbox{.}}} \bibinfo{year}{2022}\natexlab{}.
\newblock \showarticletitle{{Combinatorial resources auction in decentralized edge-thing systems using blockchain and differential privacy}}.
\newblock \bibinfo{journal}{\emph{Information Sciences}}  \bibinfo{volume}{607} (\bibinfo{date}{Aug} \bibinfo{year}{2022}), \bibinfo{pages}{211--229}.
\newblock
\showISSN{0020-0255}


\bibitem[\protect\citeauthoryear{Hassanpour, Diyanat, Khonsari, Shariatpanahi, and Dadlani}{Hassanpour et~al\mbox{.}}{2021}]%
        {Hassanpour2021}
\bibfield{author}{\bibinfo{person}{Seyedeh~Bahereh Hassanpour}, \bibinfo{person}{Abolfazl Diyanat}, {et~al\mbox{.}}} \bibinfo{year}{2021}\natexlab{}.
\newblock \showarticletitle{Context-Aware Privacy Preservation in Network Caching: An Information Theoretic Approach}.
\newblock \bibinfo{journal}{\emph{IEEE Communications Letters}} \bibinfo{volume}{25}, \bibinfo{number}{1} (\bibinfo{date}{Jan} \bibinfo{year}{2021}), \bibinfo{pages}{54--58}.
\newblock


\bibitem[\protect\citeauthoryear{Hassanpour, Khonsari, Moradian, and Shariatpanahi}{Hassanpour et~al\mbox{.}}{2023}]%
        {Hassanpour2023}
\bibfield{author}{\bibinfo{person}{Seyedeh~Bahereh Hassanpour}, \bibinfo{person}{Ahmad Khonsari}, {et~al\mbox{.}}} \bibinfo{year}{2023}\natexlab{}.
\newblock \showarticletitle{{Privacy-preserving edge caching: A probabilistic approach}}.
\newblock \bibinfo{journal}{\emph{Computer Networks}}  \bibinfo{volume}{226} (\bibinfo{date}{May} \bibinfo{year}{2023}), \bibinfo{pages}{109654}.
\newblock
\showISSN{13891286}


\bibitem[\protect\citeauthoryear{He, Prasad, Sethi, and Gutierrez}{He et~al\mbox{.}}{2007}]%
        {He2007}
\bibfield{author}{\bibinfo{person}{Xiuli He}, \bibinfo{person}{Ashutosh Prasad}, {et~al\mbox{.}}} \bibinfo{year}{2007}\natexlab{}.
\newblock \showarticletitle{{A survey of Stackelberg differential game models in supply and marketing channels}}.
\newblock \bibinfo{journal}{\emph{Journal of Systems Science and Systems Engineering}} \bibinfo{volume}{16}, \bibinfo{number}{4} (\bibinfo{date}{Nov} \bibinfo{year}{2007}), \bibinfo{pages}{385--413}.
\newblock
\showISSN{10043756}


\bibitem[\protect\citeauthoryear{Hu, Qian, Chen, Hossain, and Muhammad}{Hu et~al\mbox{.}}{2018}]%
        {Hu2018}
\bibfield{author}{\bibinfo{person}{Long Hu}, \bibinfo{person}{Yongfeng Qian}, {et~al\mbox{.}}} \bibinfo{year}{2018}\natexlab{}.
\newblock \showarticletitle{{Proactive Cache-Based Location Privacy Preserving for Vehicle Networks}}.
\newblock \bibinfo{journal}{\emph{IEEE Wireless Communications}} \bibinfo{volume}{25}, \bibinfo{number}{6} (\bibinfo{date}{Dec} \bibinfo{year}{2018}), \bibinfo{pages}{77--83}.
\newblock
\showISSN{15580687}


\bibitem[\protect\citeauthoryear{Jiang, Li, Wang, and Song}{Jiang et~al\mbox{.}}{2023}]%
        {Jiang2023}
\bibfield{author}{\bibinfo{person}{Bin Jiang}, \bibinfo{person}{Jianqiang Li}, {et~al\mbox{.}}} \bibinfo{year}{2023}\natexlab{}.
\newblock \showarticletitle{{Privacy-Preserving Federated Learning for Industrial Edge Computing via Hybrid Differential Privacy and Adaptive Compression}}.
\newblock \bibinfo{journal}{\emph{IEEE Transactions on Industrial Informatics}} \bibinfo{volume}{19}, \bibinfo{number}{2} (\bibinfo{date}{Feb} \bibinfo{year}{2023}), \bibinfo{pages}{1136--1144}.
\newblock


\bibitem[\protect\citeauthoryear{Jiang, Liu, Huang, Wu, and Zhou}{Jiang et~al\mbox{.}}{2020}]%
        {Jiang2020}
\bibfield{author}{\bibinfo{person}{Shunrong Jiang}, \bibinfo{person}{Jianqing Liu}, {et~al\mbox{.}}} \bibinfo{year}{2020}\natexlab{}.
\newblock \showarticletitle{{Vehicular Edge Computing Meets Cache: An Access Control Scheme for Content Delivery}}. In \bibinfo{booktitle}{\emph{IEEE International Conference on Communications (ICC)}}. \bibinfo{publisher}{IEEE}, \bibinfo{pages}{1--6}.
\newblock
\showISBNx{9781728150895}
\showISSN{15503607}


\bibitem[\protect\citeauthoryear{Jiang, Feng, and Qin}{Jiang et~al\mbox{.}}{2017}]%
        {Jiang2017}
\bibfield{author}{\bibinfo{person}{Wei Jiang}, \bibinfo{person}{Gang Feng}, {et~al\mbox{.}}} \bibinfo{year}{2017}\natexlab{}.
\newblock \showarticletitle{{Optimal Cooperative Content Caching and Delivery Policy for Heterogeneous Cellular Networks}}.
\newblock \bibinfo{journal}{\emph{IEEE Transactions on Mobile Computing}} \bibinfo{volume}{16}, \bibinfo{number}{5} (\bibinfo{date}{May} \bibinfo{year}{2017}), \bibinfo{pages}{1382--1393}.
\newblock
\showISSN{15361233}


\bibitem[\protect\citeauthoryear{Kong, Lu, Ma, and Bao}{Kong et~al\mbox{.}}{2019}]%
        {Kong2019}
\bibfield{author}{\bibinfo{person}{Qinglei Kong}, \bibinfo{person}{Rongxing Lu}, {et~al\mbox{.}}} \bibinfo{year}{2019}\natexlab{}.
\newblock \showarticletitle{{A Privacy-Preserving and Verifiable Querying Scheme in Vehicular Fog Data Dissemination}}.
\newblock \bibinfo{journal}{\emph{IEEE Transactions on Vehicular Technology}} \bibinfo{volume}{68}, \bibinfo{number}{2} (\bibinfo{date}{Feb} \bibinfo{year}{2019}), \bibinfo{pages}{1877--1887}.
\newblock
\showISSN{00189545}


\bibitem[\protect\citeauthoryear{Krishnendu, Bharath, Garg, Bhatia, and Ratnarajah}{Krishnendu et~al\mbox{.}}{2022}]%
        {Krishnendu2022}
\bibfield{author}{\bibinfo{person}{S. Krishnendu}, \bibinfo{person}{B.~N. Bharath}, {et~al\mbox{.}}} \bibinfo{year}{2022}\natexlab{}.
\newblock \showarticletitle{{Learning to Cache: Federated Caching in a Cellular Network with Correlated Demands}}.
\newblock \bibinfo{journal}{\emph{IEEE Transactions on Communications}} \bibinfo{volume}{70}, \bibinfo{number}{3} (\bibinfo{date}{Mar} \bibinfo{year}{2022}), \bibinfo{pages}{1653--1665}.
\newblock


\bibitem[\protect\citeauthoryear{Kumar, {Graell Amat}, Rosnes, and Senigagliesi}{Kumar et~al\mbox{.}}{2019}]%
        {Kumar2019}
\bibfield{author}{\bibinfo{person}{Siddhartha Kumar}, \bibinfo{person}{Alexandre~I. {Graell Amat}}, {et~al\mbox{.}}} \bibinfo{year}{2019}\natexlab{}.
\newblock \showarticletitle{{Private information retrieval from a cellular network with caching at the edge}}.
\newblock \bibinfo{journal}{\emph{IEEE Transactions on Communications}} \bibinfo{volume}{67}, \bibinfo{number}{7} (\bibinfo{date}{Jul} \bibinfo{year}{2019}), \bibinfo{pages}{4900--4912}.
\newblock
\showISSN{15580857}


\bibitem[\protect\citeauthoryear{Leguay, Paschos, Quaglia, and Smyth}{Leguay et~al\mbox{.}}{2017}]%
        {Leguay2017}
\bibfield{author}{\bibinfo{person}{Jeremie Leguay}, \bibinfo{person}{Georgios~S. Paschos}, {et~al\mbox{.}}} \bibinfo{year}{2017}\natexlab{}.
\newblock \showarticletitle{{CryptoCache: Network caching with confidentiality}}. In \bibinfo{booktitle}{\emph{IEEE International Conference on Communications (ICC)}}. \bibinfo{publisher}{IEEE}, \bibinfo{pages}{1--6}.
\newblock
\showISBNx{9781467389990}
\showISSN{15503607}


\bibitem[\protect\citeauthoryear{Lei, Fang, Zhang, Lou, Du, et~al\mbox{.}}{Lei et~al\mbox{.}}{2020}]%
        {Lei2020}
\bibfield{author}{\bibinfo{person}{Kai Lei}, \bibinfo{person}{Junjie Fang}, {et~al\mbox{.}}} \bibinfo{year}{2020}\natexlab{}.
\newblock \showarticletitle{{Blockchain-Based Cache Poisoning Security Protection and Privacy-Aware Access Control in NDN Vehicular Edge Computing Networks}}.
\newblock \bibinfo{journal}{\emph{Journal of Grid Computing}} \bibinfo{volume}{18}, \bibinfo{number}{4} (\bibinfo{date}{Aug} \bibinfo{year}{2020}), \bibinfo{pages}{593--613}.
\newblock
\showISSN{1572-9184}


\bibitem[\protect\citeauthoryear{Li, Zhang, and Luo}{Li et~al\mbox{.}}{2023}]%
        {Li2023}
\bibfield{author}{\bibinfo{person}{Chunlin Li}, \bibinfo{person}{Yong Zhang}, {et~al\mbox{.}}} \bibinfo{year}{2023}\natexlab{}.
\newblock \showarticletitle{{A Federated Learning-Based Edge Caching Approach for Mobile Edge Computing-Enabled Intelligent Connected Vehicles}}.
\newblock \bibinfo{journal}{\emph{IEEE Transactions on Intelligent Transportation Systems}} \bibinfo{volume}{24}, \bibinfo{number}{3} (\bibinfo{date}{Nov} \bibinfo{year}{2023}), \bibinfo{pages}{3360--3369}.
\newblock


\bibitem[\protect\citeauthoryear{Li, Zhao, Wang, Wang, Leung, et~al\mbox{.}}{Li et~al\mbox{.}}{2020}]%
        {Li2020a}
\bibfield{author}{\bibinfo{person}{Ruibin Li}, \bibinfo{person}{Yiwei Zhao}, {et~al\mbox{.}}} \bibinfo{year}{2020}\natexlab{}.
\newblock \showarticletitle{{Edge Caching Replacement Optimization for D2D Wireless Networks via Weighted Distributed DQN}}. In \bibinfo{booktitle}{\emph{IEEE Wireless Communications and Networking Conference (WCNC)}}. \bibinfo{publisher}{IEEE}, \bibinfo{pages}{1--6}.
\newblock
\showISBNx{9781728131061}
\showISSN{15253511}


\bibitem[\protect\citeauthoryear{Li, Xiang, Zhou, and Peng}{Li et~al\mbox{.}}{2021}]%
        {Li2021}
\bibfield{author}{\bibinfo{person}{Weiting Li}, \bibinfo{person}{Liyao Xiang}, {et~al\mbox{.}}} \bibinfo{year}{2021}\natexlab{}.
\newblock \showarticletitle{{Privacy budgeting for growing machine learning datasets}}. In \bibinfo{booktitle}{\emph{IEEE Conference on Computer Communications (INFOCOM)}}. \bibinfo{publisher}{IEEE}, \bibinfo{pages}{1--10}.
\newblock
\showISBNx{9780738112817}
\showISSN{0743166X}


\bibitem[\protect\citeauthoryear{Liang and Liu}{Liang and Liu}{2019}]%
        {Liang2019}
\bibfield{author}{\bibinfo{person}{Jie Liang} {and} \bibinfo{person}{Yinlong Liu}.} \bibinfo{year}{2019}\natexlab{}.
\newblock \showarticletitle{{A Cache Privacy Protection Strategy Based on Content Privacy and User Security Classification in CCN}}. In \bibinfo{booktitle}{\emph{IEEE Wireless Communications and Networking Conference (WCNC)}}. \bibinfo{publisher}{IEEE}, \bibinfo{pages}{1--6}.
\newblock
\showISBNx{9781538676462}
\showISSN{15253511}


\bibitem[\protect\citeauthoryear{Liu, Chen, Yang, and Molisch}{Liu et~al\mbox{.}}{2016}]%
        {Liu2016}
\bibfield{author}{\bibinfo{person}{Dong Liu}, \bibinfo{person}{Binqiang Chen}, {et~al\mbox{.}}} \bibinfo{year}{2016}\natexlab{}.
\newblock \showarticletitle{{Caching at the wireless edge: Design aspects, challenges, and future directions}}.
\newblock \bibinfo{journal}{\emph{IEEE Communications Magazine}} \bibinfo{volume}{54}, \bibinfo{number}{9} (\bibinfo{date}{Sep} \bibinfo{year}{2016}), \bibinfo{pages}{22--28}.
\newblock
\showISSN{01636804}


\bibitem[\protect\citeauthoryear{Liu, Guo, Shi, Feng, and Wang}{Liu et~al\mbox{.}}{2020}]%
        {LiuJi2020}
\bibfield{author}{\bibinfo{person}{Jiadi Liu}, \bibinfo{person}{Songtao Guo}, {et~al\mbox{.}}} \bibinfo{year}{2020}\natexlab{}.
\newblock \showarticletitle{{Decentralized Caching Framework Toward Edge Network Based on Blockchain}}.
\newblock \bibinfo{journal}{\emph{IEEE Internet of Things Journal}} \bibinfo{volume}{7}, \bibinfo{number}{9} (\bibinfo{date}{Jun} \bibinfo{year}{2020}), \bibinfo{pages}{9158--9174}.
\newblock


\bibitem[\protect\citeauthoryear{Liu, Zheng, Huang, and Quek}{Liu et~al\mbox{.}}{2022}]%
        {Liu2022}
\bibfield{author}{\bibinfo{person}{Shengheng Liu}, \bibinfo{person}{Chong Zheng}, {et~al\mbox{.}}} \bibinfo{year}{2022}\natexlab{}.
\newblock \showarticletitle{{Distributed Reinforcement Learning for Privacy-Preserving Dynamic Edge Caching}}.
\newblock \bibinfo{journal}{\emph{IEEE Journal on Selected Areas in Communications}} \bibinfo{volume}{40}, \bibinfo{number}{3} (\bibinfo{date}{Mar} \bibinfo{year}{2022}), \bibinfo{pages}{749--760}.
\newblock
\showISSN{15580008}


\bibitem[\protect\citeauthoryear{Lu, Huang, Dai, Maharjan, and Zhang}{Lu et~al\mbox{.}}{2020}]%
        {LuYun2020}
\bibfield{author}{\bibinfo{person}{Yunlong Lu}, \bibinfo{person}{Xiaohong Huang}, {et~al\mbox{.}}} \bibinfo{year}{2020}\natexlab{}.
\newblock \showarticletitle{{Differentially Private Asynchronous Federated Learning for Mobile Edge Computing in Urban Informatics}}.
\newblock \bibinfo{journal}{\emph{IEEE Transactions on Industrial Informatics}} \bibinfo{volume}{16}, \bibinfo{number}{3} (\bibinfo{date}{Mar} \bibinfo{year}{2020}), \bibinfo{pages}{2134--2143}.
\newblock


\bibitem[\protect\citeauthoryear{Ma, Wang, Zhang, Ye, Chen, et~al\mbox{.}}{Ma et~al\mbox{.}}{2017}]%
        {Ma2017b}
\bibfield{author}{\bibinfo{person}{Ge Ma}, \bibinfo{person}{Zhi Wang}, {et~al\mbox{.}}} \bibinfo{year}{2017}\natexlab{}.
\newblock \showarticletitle{{Understanding Performance of Edge Content Caching for Mobile Video Streaming}}.
\newblock \bibinfo{journal}{\emph{IEEE Journal on Selected Areas in Communications}} \bibinfo{volume}{35}, \bibinfo{number}{5} (\bibinfo{date}{May} \bibinfo{year}{2017}), \bibinfo{pages}{1076--1089}.
\newblock
\showISSN{07338716}


\bibitem[\protect\citeauthoryear{Mao, You, Zhang, Huang, and Letaief}{Mao et~al\mbox{.}}{2017}]%
        {Mao2017}
\bibfield{author}{\bibinfo{person}{Yuyi Mao}, \bibinfo{person}{Changsheng You}, {et~al\mbox{.}}} \bibinfo{year}{2017}\natexlab{}.
\newblock \showarticletitle{{A Survey on Mobile Edge Computing: The Communication Perspective}}.
\newblock \bibinfo{journal}{\emph{IEEE Communications Surveys and Tutorials}} \bibinfo{volume}{19}, \bibinfo{number}{4} (\bibinfo{date}{Oct} \bibinfo{year}{2017}), \bibinfo{pages}{2322--2358}.
\newblock
\showISSN{1553877X}


\bibitem[\protect\citeauthoryear{Mirheidari, Arshad, Onarlioglu, Crispo, Kirda, et~al\mbox{.}}{Mirheidari et~al\mbox{.}}{2020}]%
        {Mirheidari2020}
\bibfield{author}{\bibinfo{person}{Seyed~Ali Mirheidari}, \bibinfo{person}{Sajjad Arshad}, {et~al\mbox{.}}} \bibinfo{year}{2020}\natexlab{}.
\newblock \showarticletitle{Cached and Confused: Web Cache Deception in the Wild}. In \bibinfo{booktitle}{\emph{29th USENIX Security Symposium (USENIX Security)}}. \bibinfo{publisher}{USENIX Association}, \bibinfo{pages}{665--682}.
\newblock
\showISBNx{978-1-939133-17-5}


\bibitem[\protect\citeauthoryear{Mirheidari, Golinelli, Onarlioglu, Kirda, and Crispo}{Mirheidari et~al\mbox{.}}{2022}]%
        {Mirheidari2022}
\bibfield{author}{\bibinfo{person}{Seyed~Ali Mirheidari}, \bibinfo{person}{Matteo Golinelli}, {et~al\mbox{.}}} \bibinfo{year}{2022}\natexlab{}.
\newblock \showarticletitle{Web Cache Deception Escalates}. In \bibinfo{booktitle}{\emph{31st USENIX Security Symposium (USENIX Security)}}. \bibinfo{publisher}{USENIX Association}, \bibinfo{pages}{179--196}.
\newblock
\showISBNx{978-1-939133-31-1}


\bibitem[\protect\citeauthoryear{Muller, Atan, {Van Der Schaar}, and Klein}{Muller et~al\mbox{.}}{2017}]%
        {Muller2017}
\bibfield{author}{\bibinfo{person}{Sabrina Muller}, \bibinfo{person}{Onur Atan}, {et~al\mbox{.}}} \bibinfo{year}{2017}\natexlab{}.
\newblock \showarticletitle{{Context-Aware Proactive Content Caching with Service Differentiation in Wireless Networks}}.
\newblock \bibinfo{journal}{\emph{IEEE Transactions on Wireless Communications}} \bibinfo{volume}{16}, \bibinfo{number}{2} (\bibinfo{date}{Feb} \bibinfo{year}{2017}), \bibinfo{pages}{1024--1036}.
\newblock
\showISSN{15361276}


\bibitem[\protect\citeauthoryear{Nair, Sahoo, and Raj}{Nair et~al\mbox{.}}{2023}]%
        {NAIR2023}
\bibfield{author}{\bibinfo{person}{Akarsh~K Nair}, \bibinfo{person}{Jayakrushna Sahoo}, {et~al\mbox{.}}} \bibinfo{year}{2023}\natexlab{}.
\newblock \showarticletitle{{Privacy preserving Federated Learning framework for IoMT based big data analysis using edge computing}}.
\newblock \bibinfo{journal}{\emph{Computer Standards \& Interfaces}}  \bibinfo{volume}{86} (\bibinfo{date}{Jan} \bibinfo{year}{2023}), \bibinfo{pages}{103720}.
\newblock
\showISSN{0920-5489}


\bibitem[\protect\citeauthoryear{Nguyen, Cheng, Nguyen, and Nedich}{Nguyen et~al\mbox{.}}{2023}]%
        {Nguyen2023}
\bibfield{author}{\bibinfo{person}{Duong Thuy~Anh Nguyen}, \bibinfo{person}{Jiaming Cheng}, {et~al\mbox{.}}} \bibinfo{year}{2023}\natexlab{}.
\newblock \bibinfo{title}{CrowdCache: A Decentralized Game-Theoretic Framework for Mobile Edge Content Sharing}.
\newblock
\newblock
\showeprint[arxiv]{cs.GT/2304.13246}
\urldef\tempurl%
\url{https://arxiv.org/abs/2304.13246}
\showURL{%
\tempurl}


\bibitem[\protect\citeauthoryear{Ni, Zhang, and Vasilakos}{Ni et~al\mbox{.}}{2021}]%
        {Ni2021}
\bibfield{author}{\bibinfo{person}{Jianbing Ni}, \bibinfo{person}{Kuan Zhang}, {et~al\mbox{.}}} \bibinfo{year}{2021}\natexlab{}.
\newblock \showarticletitle{{Security and Privacy for Mobile Edge Caching: Challenges and Solutions}}.
\newblock \bibinfo{journal}{\emph{IEEE Wireless Communications}} \bibinfo{volume}{28}, \bibinfo{number}{3} (\bibinfo{date}{Jun} \bibinfo{year}{2021}), \bibinfo{pages}{77--83}.
\newblock
\showISSN{15580687}


\bibitem[\protect\citeauthoryear{Nikolaou, {Van Renesse}, and Schiper}{Nikolaou et~al\mbox{.}}{2016}]%
        {Nikolaou2016}
\bibfield{author}{\bibinfo{person}{Stavros Nikolaou}, \bibinfo{person}{Robbert {Van Renesse}}, {et~al\mbox{.}}} \bibinfo{year}{2016}\natexlab{}.
\newblock \showarticletitle{{Proactive Cache Placement on Cooperative Client Caches for Online Social Networks}}.
\newblock \bibinfo{journal}{\emph{IEEE Transactions on Parallel and Distributed Systems}} \bibinfo{volume}{27}, \bibinfo{number}{4} (\bibinfo{date}{Apr} \bibinfo{year}{2016}), \bibinfo{pages}{1174--1186}.
\newblock
\showISSN{10459219}


\bibitem[\protect\citeauthoryear{Nisha, Natgunanathan, Gao, and Xiang}{Nisha et~al\mbox{.}}{2022}]%
        {Nisha2022}
\bibfield{author}{\bibinfo{person}{Nisha Nisha}, \bibinfo{person}{Iynkaran Natgunanathan}, {et~al\mbox{.}}} \bibinfo{year}{2022}\natexlab{}.
\newblock \showarticletitle{{A novel privacy protection scheme for location-based services using collaborative caching}}.
\newblock \bibinfo{journal}{\emph{Computer Networks}}  \bibinfo{volume}{213} (\bibinfo{date}{Aug} \bibinfo{year}{2022}), \bibinfo{pages}{109107}.
\newblock
\showISSN{1389-1286}


\bibitem[\protect\citeauthoryear{Otoum, Gottimukkala, Kumar, and Nayak}{Otoum et~al\mbox{.}}{2024}]%
        {Yazan2024}
\bibfield{author}{\bibinfo{person}{Yazan Otoum}, \bibinfo{person}{Navya Gottimukkala}, {et~al\mbox{.}}} \bibinfo{year}{2024}\natexlab{}.
\newblock \showarticletitle{Machine Learning in Metaverse Security: Current Solutions and Future Challenges}.
\newblock \bibinfo{journal}{\emph{Comput. Surveys}} \bibinfo{volume}{56}, \bibinfo{number}{8}, Article \bibinfo{articleno}{215} (\bibinfo{date}{Apr} \bibinfo{year}{2024}), \bibinfo{numpages}{36}~pages.
\newblock
\showISSN{0360-0300}


\bibitem[\protect\citeauthoryear{Pang, Wang, Li, Zhou, Ren, et~al\mbox{.}}{Pang et~al\mbox{.}}{2022}]%
        {Pang2022}
\bibfield{author}{\bibinfo{person}{Xiaoyi Pang}, \bibinfo{person}{Zhibo Wang}, {et~al\mbox{.}}} \bibinfo{year}{2022}\natexlab{}.
\newblock \showarticletitle{{Towards Online Privacy-preserving Computation Offloading in Mobile Edge Computing}}. In \bibinfo{booktitle}{\emph{IEEE Conference on Computer Communications (INFOCOM)}}. \bibinfo{publisher}{IEEE}, \bibinfo{pages}{1179--1188}.
\newblock
\showISBNx{9781665458221}
\showISSN{0743166X}


\bibitem[\protect\citeauthoryear{Pu, Wang, Yang, Luo, Hu, et~al\mbox{.}}{Pu et~al\mbox{.}}{2019}]%
        {Pu2019}
\bibfield{author}{\bibinfo{person}{Yuwen Pu}, \bibinfo{person}{Ying Wang}, {et~al\mbox{.}}} \bibinfo{year}{2019}\natexlab{}.
\newblock \showarticletitle{{An Efficient and Recoverable Data Sharing Mechanism for Edge Storage}}. In \bibinfo{booktitle}{\emph{International Conference on Wireless Algorithms, Systems, and Applications (WASA)}}. \bibinfo{publisher}{Springer Verlag}, \bibinfo{pages}{247--259}.
\newblock
\showISBNx{9783030235963}
\showISSN{16113349}


\bibitem[\protect\citeauthoryear{Qi and Yang}{Qi and Yang}{2020}]%
        {Qi2020}
\bibfield{author}{\bibinfo{person}{Kaiqiang Qi} {and} \bibinfo{person}{Chenyang Yang}.} \bibinfo{year}{2020}\natexlab{}.
\newblock \showarticletitle{{Popularity Prediction with Federated Learning for Proactive Caching at Wireless Edge}}. In \bibinfo{booktitle}{\emph{IEEE Wireless Communications and Networking Conference (WCNC)}}. \bibinfo{publisher}{IEEE}, \bibinfo{pages}{1--6}.
\newblock
\showISBNx{9781728131061}
\showISSN{15253511}


\bibitem[\protect\citeauthoryear{Qian, Jiang, Hu, Hossain, Alrashoud, et~al\mbox{.}}{Qian et~al\mbox{.}}{2020}]%
        {Qian2020}
\bibfield{author}{\bibinfo{person}{Yongfeng Qian}, \bibinfo{person}{Yingying Jiang}, {et~al\mbox{.}}} \bibinfo{year}{2020}\natexlab{}.
\newblock \showarticletitle{{Blockchain-based privacy-aware content caching in cognitive internet of vehicles}}.
\newblock \bibinfo{journal}{\emph{IEEE Network}} \bibinfo{volume}{34}, \bibinfo{number}{2} (\bibinfo{date}{Mar} \bibinfo{year}{2020}), \bibinfo{pages}{46--51}.
\newblock
\showISSN{1558156X}


\bibitem[\protect\citeauthoryear{Qiao, Guo, Liu, Long, Zhou, et~al\mbox{.}}{Qiao et~al\mbox{.}}{2022}]%
        {Qiao2022}
\bibfield{author}{\bibinfo{person}{Dewen Qiao}, \bibinfo{person}{Songtao Guo}, {et~al\mbox{.}}} \bibinfo{year}{2022}\natexlab{}.
\newblock \showarticletitle{{Adaptive Federated Deep Reinforcement Learning for Proactive Content Caching in Edge Computing}}.
\newblock \bibinfo{journal}{\emph{IEEE Transactions on Parallel and Distributed Systems}} \bibinfo{volume}{33}, \bibinfo{number}{12} (\bibinfo{date}{Dec} \bibinfo{year}{2022}), \bibinfo{pages}{4767--4782}.
\newblock
\showISSN{15582183}


\bibitem[\protect\citeauthoryear{Ren, Zhang, He, Zhang, and Li}{Ren et~al\mbox{.}}{2019}]%
        {Ren2019}
\bibfield{author}{\bibinfo{person}{Ju Ren}, \bibinfo{person}{Deyu Zhang}, {et~al\mbox{.}}} \bibinfo{year}{2019}\natexlab{}.
\newblock \showarticletitle{A Survey on End-Edge-Cloud Orchestrated Network Computing Paradigms: Transparent Computing, Mobile Edge Computing, Fog Computing, and Cloudlet}.
\newblock \bibinfo{journal}{\emph{Comput. Surveys}} \bibinfo{volume}{52}, \bibinfo{number}{6}, Article \bibinfo{articleno}{125} (\bibinfo{date}{Oct} \bibinfo{year}{2019}), \bibinfo{numpages}{36}~pages.
\newblock
\showISSN{0360-0300}


\bibitem[\protect\citeauthoryear{Saputra, Nguyen, Hoang, and Dutkiewicz}{Saputra et~al\mbox{.}}{2022}]%
        {Saputra2022}
\bibfield{author}{\bibinfo{person}{Yuris~Mulya Saputra}, \bibinfo{person}{Diep~N. Nguyen}, {et~al\mbox{.}}} \bibinfo{year}{2022}\natexlab{}.
\newblock \showarticletitle{{In-Network Caching and Learning Optimization for Federated Learning in Mobile Edge Networks}}. In \bibinfo{booktitle}{\emph{IEEE International Conference on Communications (ICC)}}. \bibinfo{publisher}{IEEE}, \bibinfo{pages}{1653--1658}.
\newblock
\showISBNx{9781538683477}
\showISSN{15503607}


\bibitem[\protect\citeauthoryear{Sarwar, Yongchareon, Yu, and Ur~Rehman}{Sarwar et~al\mbox{.}}{2021}]%
        {Kinza2021}
\bibfield{author}{\bibinfo{person}{Kinza Sarwar}, \bibinfo{person}{Sira Yongchareon}, {et~al\mbox{.}}} \bibinfo{year}{2021}\natexlab{}.
\newblock \showarticletitle{A Survey on Privacy Preservation in Fog-Enabled Internet of Things}.
\newblock \bibinfo{journal}{\emph{Comput. Surveys}} \bibinfo{volume}{55}, \bibinfo{number}{1}, Article \bibinfo{articleno}{2} (\bibinfo{date}{Nov} \bibinfo{year}{2021}), \bibinfo{numpages}{39}~pages.
\newblock
\showISSN{0360-0300}


\bibitem[\protect\citeauthoryear{Schlegel, Kumar, Rosnes, and {Graell I Amat}}{Schlegel et~al\mbox{.}}{2022}]%
        {Schlegel2022}
\bibfield{author}{\bibinfo{person}{Reent Schlegel}, \bibinfo{person}{Siddhartha Kumar}, {et~al\mbox{.}}} \bibinfo{year}{2022}\natexlab{}.
\newblock \showarticletitle{{Privacy-Preserving Coded Mobile Edge Computing for Low-Latency Distributed Inference}}.
\newblock \bibinfo{journal}{\emph{IEEE Journal on Selected Areas in Communications}} \bibinfo{volume}{40}, \bibinfo{number}{3} (\bibinfo{date}{Mar} \bibinfo{year}{2022}), \bibinfo{pages}{788--799}.
\newblock
\showISSN{15580008}


\bibitem[\protect\citeauthoryear{Sen, Eassa, Yamin, and Jambi}{Sen et~al\mbox{.}}{2018}]%
        {Sen2018}
\bibfield{author}{\bibinfo{person}{Adnan~A.Abi Sen}, \bibinfo{person}{Fathy~B. Eassa}, {et~al\mbox{.}}} \bibinfo{year}{2018}\natexlab{}.
\newblock \showarticletitle{{Double Cache Approach with Wireless Technology for Preserving User Privacy}}.
\newblock \bibinfo{journal}{\emph{Wireless Communications and Mobile Computing}}  \bibinfo{volume}{2018} (\bibinfo{date}{Jan} \bibinfo{year}{2018}), \bibinfo{pages}{1--11}.
\newblock
\showISSN{15308677}


\bibitem[\protect\citeauthoryear{Shafiq, Liu, and Khakpour}{Shafiq et~al\mbox{.}}{2014}]%
        {Shafiq2014}
\bibfield{author}{\bibinfo{person}{M.~Zubair Shafiq}, \bibinfo{person}{Alex~X. Liu}, {et~al\mbox{.}}} \bibinfo{year}{2014}\natexlab{}.
\newblock \showarticletitle{{Revisiting caching in content delivery networks}}. In \bibinfo{booktitle}{\emph{Proceedings of the ACM SIGMETRICS 2014}}. \bibinfo{publisher}{ACM}, \bibinfo{pages}{567--568}.
\newblock
\showISBNx{9781450327893}
\showISSN{0163-5999}


\bibitem[\protect\citeauthoryear{Shi, Fan, Liu, Na, and Liu}{Shi et~al\mbox{.}}{2018}]%
        {Shi2018}
\bibfield{author}{\bibinfo{person}{Fang Shi}, \bibinfo{person}{Lisheng Fan}, {et~al\mbox{.}}} \bibinfo{year}{2018}\natexlab{}.
\newblock \showarticletitle{{Probabilistic Caching Placement in the Presence of Multiple Eavesdroppers}}.
\newblock \bibinfo{journal}{\emph{Wireless Communications and Mobile Computing}}  \bibinfo{volume}{2018} (\bibinfo{date}{May} \bibinfo{year}{2018}), \bibinfo{pages}{1--10}.
\newblock
\showISSN{15308677}


\bibitem[\protect\citeauthoryear{Shokri, Stronati, Song, and Shmatikov}{Shokri et~al\mbox{.}}{2017}]%
        {Shokri2017}
\bibfield{author}{\bibinfo{person}{Reza Shokri}, \bibinfo{person}{Marco Stronati}, {et~al\mbox{.}}} \bibinfo{year}{2017}\natexlab{}.
\newblock \showarticletitle{{Membership Inference Attacks Against Machine Learning Models}}. In \bibinfo{booktitle}{\emph{IEEE Symposium on Security and Privacy (S\&P)}}. \bibinfo{publisher}{IEEE}, \bibinfo{pages}{3--18}.
\newblock
\showISBNx{9781509055326}
\showISSN{10816011}


\bibitem[\protect\citeauthoryear{Sivaraman and Sikdar}{Sivaraman and Sikdar}{2021}]%
        {Sivaraman2021}
\bibfield{author}{\bibinfo{person}{Vignesh Sivaraman} {and} \bibinfo{person}{Biplab Sikdar}.} \bibinfo{year}{2021}\natexlab{}.
\newblock \showarticletitle{{A Defense Mechanism against Timing Attacks on User Privacy in ICN}}.
\newblock \bibinfo{journal}{\emph{IEEE/ACM Transactions on Networking}} \bibinfo{volume}{29}, \bibinfo{number}{6} (\bibinfo{date}{Dec} \bibinfo{year}{2021}), \bibinfo{pages}{2709--2722}.
\newblock
\showISSN{15582566}


\bibitem[\protect\citeauthoryear{{Splunk, Inc.}}{{Splunk, Inc.}}{2023}]%
        {splunk2023internet}
\bibfield{author}{\bibinfo{person}{{Splunk, Inc.}}} \bibinfo{year}{2023}\natexlab{}.
\newblock \bibinfo{title}{Internet Trends in 2024: Mary Meeker, Stats, \& Predictions}.
\newblock
\newblock
\urldef\tempurl%
\url{https://www.splunk.com/en_us/blog/learn/internet-trends.html}
\showURL{%
\tempurl}
\newblock
\shownote{Accessed: 2024-07-21.}


\bibitem[\protect\citeauthoryear{Sutton and Barto}{Sutton and Barto}{1998}]%
        {Sutton1998}
\bibfield{author}{\bibinfo{person}{R.S. Sutton} {and} \bibinfo{person}{A.G. Barto}.} \bibinfo{year}{1998}\natexlab{}.
\newblock \showarticletitle{{Reinforcement Learning: An Introduction}}.
\newblock \bibinfo{journal}{\emph{IEEE Transactions on Neural Networks}} \bibinfo{volume}{9}, \bibinfo{number}{5} (\bibinfo{year}{1998}), \bibinfo{pages}{1054--1054}.
\newblock
\showISSN{1045-9227}


\bibitem[\protect\citeauthoryear{Tang, Li, Ma, Gao, Wang, et~al\mbox{.}}{Tang et~al\mbox{.}}{2022}]%
        {Tang2022}
\bibfield{author}{\bibinfo{person}{Shisong Tang}, \bibinfo{person}{Qing Li}, {et~al\mbox{.}}} \bibinfo{year}{2022}\natexlab{}.
\newblock \showarticletitle{Knowledge-based Temporal Fusion Network for Interpretable Online Video Popularity Prediction}. In \bibinfo{booktitle}{\emph{Proceedings of the ACM Web Conference (WWW)}}. \bibinfo{publisher}{ACM}, \bibinfo{pages}{2879–2887}.
\newblock
\showISBNx{9781450390965}


\bibitem[\protect\citeauthoryear{Tong, Chen, Jiang, Xu, Li, et~al\mbox{.}}{Tong et~al\mbox{.}}{2023}]%
        {Tong2022}
\bibfield{author}{\bibinfo{person}{Wei Tong}, \bibinfo{person}{Wenjie Chen}, {et~al\mbox{.}}} \bibinfo{year}{2023}\natexlab{}.
\newblock \showarticletitle{{Privacy-Preserving Data Integrity Verification for Secure Mobile Edge Storage}}.
\newblock \bibinfo{journal}{\emph{IEEE Transactions on Mobile Computing}} \bibinfo{volume}{22}, \bibinfo{number}{9} (\bibinfo{date}{Mar} \bibinfo{year}{2023}), \bibinfo{pages}{5463--5478}.
\newblock
\showISSN{15580660}


\bibitem[\protect\citeauthoryear{Tourani, Misra, Mick, and Panwar}{Tourani et~al\mbox{.}}{2018}]%
        {Tourani2018}
\bibfield{author}{\bibinfo{person}{Reza Tourani}, \bibinfo{person}{Satyajayant Misra}, {et~al\mbox{.}}} \bibinfo{year}{2018}\natexlab{}.
\newblock \showarticletitle{{Security, Privacy, and Access Control in Information-Centric Networking: A Survey}}.
\newblock \bibinfo{journal}{\emph{IEEE Communications Surveys and Tutorials}} \bibinfo{volume}{20}, \bibinfo{number}{1} (\bibinfo{date}{Jan} \bibinfo{year}{2018}), \bibinfo{pages}{556--600}.
\newblock
\showISSN{1553877X}


\bibitem[\protect\citeauthoryear{Vu, Chatzinotas, and Ottersten}{Vu et~al\mbox{.}}{2019}]%
        {Vu2019}
\bibfield{author}{\bibinfo{person}{Thang~X. Vu}, \bibinfo{person}{Symeon Chatzinotas}, {et~al\mbox{.}}} \bibinfo{year}{2019}\natexlab{}.
\newblock \showarticletitle{{Blockchain-based Content Delivery Networks: Content Transparency Meets User Privacy}}. In \bibinfo{booktitle}{\emph{IEEE Wireless Communications and Networking Conference (WCNC)}}. \bibinfo{publisher}{IEEE}, \bibinfo{pages}{1--6}.
\newblock
\showISBNx{9781538676462}
\showISSN{15253511}


\bibitem[\protect\citeauthoryear{Wang, Zhang, Xia, and Wu}{Wang et~al\mbox{.}}{2022b}]%
        {WangHu2022}
\bibfield{author}{\bibinfo{person}{Huanhuan Wang}, \bibinfo{person}{Xiao Zhang}, {et~al\mbox{.}}} \bibinfo{year}{2022}\natexlab{b}.
\newblock \showarticletitle{{A differential privacy-preserving deep learning caching framework for heterogeneous communication network systems}}.
\newblock \bibinfo{journal}{\emph{International Journal of Intelligent Systems}} \bibinfo{volume}{37}, \bibinfo{number}{12} (\bibinfo{date}{Aug} \bibinfo{year}{2022}), \bibinfo{pages}{11142--11166}.
\newblock


\bibitem[\protect\citeauthoryear{Wang and Deng}{Wang and Deng}{2022}]%
        {Wang2022a}
\bibfield{author}{\bibinfo{person}{Kailun Wang} {and} \bibinfo{person}{Na Deng}.} \bibinfo{year}{2022}\natexlab{}.
\newblock \showarticletitle{{A Privacy-Protected Popularity Prediction Scheme for Content Caching Based on Federated Learning}}.
\newblock \bibinfo{journal}{\emph{IEEE Transactions on Vehicular Technology}} \bibinfo{volume}{71}, \bibinfo{number}{9} (\bibinfo{date}{Jun} \bibinfo{year}{2022}), \bibinfo{pages}{10191--10196}.
\newblock
\showISSN{19399359}


\bibitem[\protect\citeauthoryear{Wang, Deng, and Li}{Wang et~al\mbox{.}}{2022a}]%
        {Wang2022}
\bibfield{author}{\bibinfo{person}{Kailun Wang}, \bibinfo{person}{Na Deng}, {et~al\mbox{.}}} \bibinfo{year}{2022}\natexlab{a}.
\newblock \showarticletitle{{An Efficient Content Popularity Prediction of Privacy Preserving Based on Federated Learning and Wasserstein GAN}}.
\newblock \bibinfo{journal}{\emph{IEEE Internet of Things Journal}} \bibinfo{volume}{10}, \bibinfo{number}{5} (\bibinfo{date}{May} \bibinfo{year}{2022}), \bibinfo{pages}{3786--3798}.
\newblock
\showISSN{23274662}


\bibitem[\protect\citeauthoryear{Wang, Xu, Chen, Hao, Zhong, et~al\mbox{.}}{Wang et~al\mbox{.}}{2019b}]%
        {Wang2019}
\bibfield{author}{\bibinfo{person}{Mu Wang}, \bibinfo{person}{Changqiao Xu}, {et~al\mbox{.}}} \bibinfo{year}{2019}\natexlab{b}.
\newblock \showarticletitle{{Differential privacy oriented distributed online learning for mobile social video prefetching}}.
\newblock \bibinfo{journal}{\emph{IEEE Transactions on Multimedia}} \bibinfo{volume}{21}, \bibinfo{number}{3} (\bibinfo{date}{Jan} \bibinfo{year}{2019}), \bibinfo{pages}{636--651}.
\newblock
\showISSN{15209210}


\bibitem[\protect\citeauthoryear{Wang, Blocki, Li, and Jha}{Wang et~al\mbox{.}}{2017}]%
        {Wang2017}
\bibfield{author}{\bibinfo{person}{Tianhao Wang}, \bibinfo{person}{Jeremiah Blocki}, {et~al\mbox{.}}} \bibinfo{year}{2017}\natexlab{}.
\newblock \showarticletitle{{Locally differentially private protocols for frequency estimation}}. In \bibinfo{booktitle}{\emph{Proceedings of the 26th USENIX Security Symposium}}. \bibinfo{publisher}{ACM}, \bibinfo{pages}{729--745}.
\newblock
\showISBNx{9781931971409}


\bibitem[\protect\citeauthoryear{Wang, Han, Wang, Zhao, Chen, et~al\mbox{.}}{Wang et~al\mbox{.}}{2019a}]%
        {Wang2019a}
\bibfield{author}{\bibinfo{person}{Xiaofei Wang}, \bibinfo{person}{Yiwen Han}, {et~al\mbox{.}}} \bibinfo{year}{2019}\natexlab{a}.
\newblock \showarticletitle{{In-edge AI: Intelligentizing mobile edge computing, caching and communication by federated learning}}.
\newblock \bibinfo{journal}{\emph{IEEE Network}} \bibinfo{volume}{33}, \bibinfo{number}{5} (\bibinfo{date}{Sep} \bibinfo{year}{2019}), \bibinfo{pages}{156--165}.
\newblock
\showISSN{1558156X}


\bibitem[\protect\citeauthoryear{Wang, Wang, Li, Leung, and Taleb}{Wang et~al\mbox{.}}{2020}]%
        {Wang2020}
\bibfield{author}{\bibinfo{person}{Xiaofei Wang}, \bibinfo{person}{Chenyang Wang}, {et~al\mbox{.}}} \bibinfo{year}{2020}\natexlab{}.
\newblock \showarticletitle{{Federated Deep Reinforcement Learning for Internet of Things with Decentralized Cooperative Edge Caching}}.
\newblock \bibinfo{journal}{\emph{IEEE Internet of Things Journal}} \bibinfo{volume}{7}, \bibinfo{number}{10} (\bibinfo{date}{Oct} \bibinfo{year}{2020}), \bibinfo{pages}{9441--9455}.
\newblock
\showISSN{23274662}


\bibitem[\protect\citeauthoryear{Wu, Li, Tyson, Uhlig, Kaafar, et~al\mbox{.}}{Wu et~al\mbox{.}}{2016}]%
        {Wu2016}
\bibfield{author}{\bibinfo{person}{Qinghua Wu}, \bibinfo{person}{Zhenyu Li}, {et~al\mbox{.}}} \bibinfo{year}{2016}\natexlab{}.
\newblock \showarticletitle{{Privacy-Aware Multipath Video Caching for Content-Centric Networks}}.
\newblock \bibinfo{journal}{\emph{IEEE Journal on Selected Areas in Communications}} \bibinfo{volume}{34}, \bibinfo{number}{8} (\bibinfo{date}{Aug} \bibinfo{year}{2016}), \bibinfo{pages}{2219--2230}.
\newblock
\showISSN{07338716}


\bibitem[\protect\citeauthoryear{Wu, Zhao, Fan, Fan, and Wang}{Wu et~al\mbox{.}}{2023}]%
        {Wu2023}
\bibfield{author}{\bibinfo{person}{Qiong Wu}, \bibinfo{person}{Yu Zhao}, {et~al\mbox{.}}} \bibinfo{year}{2023}\natexlab{}.
\newblock \showarticletitle{{Mobility-Aware Cooperative Caching in Vehicular Edge Computing Based on Asynchronous Federated and Deep Reinforcement Learning}}.
\newblock \bibinfo{journal}{\emph{IEEE Journal on Selected Topics in Signal Processing}} \bibinfo{volume}{17}, \bibinfo{number}{1} (\bibinfo{date}{Jan} \bibinfo{year}{2023}), \bibinfo{pages}{66--81}.
\newblock
\showISSN{19410484}


\bibitem[\protect\citeauthoryear{Xiao, Wan, Dai, Du, Chen, et~al\mbox{.}}{Xiao et~al\mbox{.}}{2018}]%
        {Xiao2018a}
\bibfield{author}{\bibinfo{person}{Liang Xiao}, \bibinfo{person}{Xiaoyue Wan}, {et~al\mbox{.}}} \bibinfo{year}{2018}\natexlab{}.
\newblock \showarticletitle{{Security in Mobile Edge Caching with Reinforcement Learning}}.
\newblock \bibinfo{journal}{\emph{IEEE Wireless Communications}} \bibinfo{volume}{25}, \bibinfo{number}{3} (\bibinfo{date}{Jun} \bibinfo{year}{2018}), \bibinfo{pages}{116--122}.
\newblock
\showISSN{15361284}


\bibitem[\protect\citeauthoryear{Xiao, Zhang, Wu, Hu, Zhou, et~al\mbox{.}}{Xiao et~al\mbox{.}}{2024}]%
        {Xiao2024}
\bibfield{author}{\bibinfo{person}{Linchang Xiao}, \bibinfo{person}{Xianzhi Zhang}, {et~al\mbox{.}}} \bibinfo{year}{2024}\natexlab{}.
\newblock \showarticletitle{History-Aware Privacy Budget Allocation for Model Training on Evolving Data-Sharing Platforms}.
\newblock \bibinfo{journal}{\emph{IEEE Transactions on Services Computing}}  \bibinfo{volume}{Early Access} (\bibinfo{year}{2024}), \bibinfo{pages}{1--15}.
\newblock


\bibitem[\protect\citeauthoryear{Xiong, Bi, Zhao, Guo, and Yang}{Xiong et~al\mbox{.}}{2020}]%
        {Xiong2020}
\bibfield{author}{\bibinfo{person}{Jinbo Xiong}, \bibinfo{person}{Renwan Bi}, {et~al\mbox{.}}} \bibinfo{year}{2020}\natexlab{}.
\newblock \showarticletitle{Edge-Assisted Privacy-Preserving Raw Data Sharing Framework for Connected Autonomous Vehicles}.
\newblock \bibinfo{journal}{\emph{IEEE Wireless Communications}} \bibinfo{volume}{27}, \bibinfo{number}{3} (\bibinfo{year}{2020}), \bibinfo{pages}{24--30}.
\newblock


\bibitem[\protect\citeauthoryear{Xiong, Sarwate, and Mandayam}{Xiong et~al\mbox{.}}{2022}]%
        {Xiong2022}
\bibfield{author}{\bibinfo{person}{Sijie Xiong}, \bibinfo{person}{Anand~D. Sarwate}, {et~al\mbox{.}}} \bibinfo{year}{2022}\natexlab{}.
\newblock \showarticletitle{Network Traffic Shaping for Enhancing Privacy in IoT Systems}.
\newblock \bibinfo{journal}{\emph{IEEE/ACM Transactions on Networking}} \bibinfo{volume}{30}, \bibinfo{number}{3} (\bibinfo{year}{2022}), \bibinfo{pages}{1162--1177}.
\newblock


\bibitem[\protect\citeauthoryear{Xu, Du, Niyato, Kang, Xiong, et~al\mbox{.}}{Xu et~al\mbox{.}}{2024}]%
        {Xu2024}
\bibfield{author}{\bibinfo{person}{Minrui Xu}, \bibinfo{person}{Hongyang Du}, {et~al\mbox{.}}} \bibinfo{year}{2024}\natexlab{}.
\newblock \showarticletitle{Unleashing the Power of Edge-Cloud Generative AI in Mobile Networks: A Survey of AIGC Services}.
\newblock \bibinfo{journal}{\emph{IEEE Communications Surveys \& Tutorials}} \bibinfo{volume}{26}, \bibinfo{number}{2} (\bibinfo{year}{2024}), \bibinfo{pages}{1127--1170}.
\newblock


\bibitem[\protect\citeauthoryear{Xu, Niyato, Zhang, Kang, Xiong, et~al\mbox{.}}{Xu et~al\mbox{.}}{2023}]%
        {Xu2023}
\bibfield{author}{\bibinfo{person}{Minrui Xu}, \bibinfo{person}{Dusit Niyato}, {et~al\mbox{.}}} \bibinfo{year}{2023}\natexlab{}.
\newblock \showarticletitle{Joint Foundation Model Caching and Inference of Generative AI Services for Edge Intelligence}. In \bibinfo{booktitle}{\emph{IEEE Global Communications Conference (GLOBECOM)}}. \bibinfo{publisher}{IEEE}, \bibinfo{pages}{3548--3553}.
\newblock


\bibitem[\protect\citeauthoryear{Xu, Su, and Lu}{Xu et~al\mbox{.}}{2020}]%
        {Xu2020}
\bibfield{author}{\bibinfo{person}{Qichao Xu}, \bibinfo{person}{Zhou Su}, {et~al\mbox{.}}} \bibinfo{year}{2020}\natexlab{}.
\newblock \showarticletitle{{Game Theory and Reinforcement Learning Based Secure Edge Caching in Mobile Social Networks}}.
\newblock \bibinfo{journal}{\emph{IEEE Transactions on Information Forensics and Security}}  \bibinfo{volume}{15} (\bibinfo{date}{Mar} \bibinfo{year}{2020}), \bibinfo{pages}{3415--3429}.
\newblock


\bibitem[\protect\citeauthoryear{Xu, Su, Zheng, Luo, Dong, et~al\mbox{.}}{Xu et~al\mbox{.}}{2019}]%
        {Xu2019}
\bibfield{author}{\bibinfo{person}{Qichao Xu}, \bibinfo{person}{Zhou Su}, {et~al\mbox{.}}} \bibinfo{year}{2019}\natexlab{}.
\newblock \showarticletitle{{Game theoretical secure caching scheme in multihoming edge computing-enabled heterogeneous networks}}.
\newblock \bibinfo{journal}{\emph{IEEE Internet of Things Journal}} \bibinfo{volume}{6}, \bibinfo{number}{3} (\bibinfo{date}{Jun} \bibinfo{year}{2019}), \bibinfo{pages}{4536--4546}.
\newblock
\showISSN{23274662}


\bibitem[\protect\citeauthoryear{Xue, He, Zhang, Xia, Wei, et~al\mbox{.}}{Xue et~al\mbox{.}}{2019}]%
        {Xue2019}
\bibfield{author}{\bibinfo{person}{Kaiping Xue}, \bibinfo{person}{Peixuan He}, {et~al\mbox{.}}} \bibinfo{year}{2019}\natexlab{}.
\newblock \showarticletitle{{A Secure, Efficient, and Accountable Edge-Based Access Control Framework for Information Centric Networks}}.
\newblock \bibinfo{journal}{\emph{IEEE/ACM Transactions on Networking}} \bibinfo{volume}{27}, \bibinfo{number}{3} (\bibinfo{date}{Jun} \bibinfo{year}{2019}), \bibinfo{pages}{1220--1233}.
\newblock
\showISSN{15582566}


\bibitem[\protect\citeauthoryear{Xue, Zhang, Xia, Wei, Yue, et~al\mbox{.}}{Xue et~al\mbox{.}}{2018}]%
        {Xue2018}
\bibfield{author}{\bibinfo{person}{Kaiping Xue}, \bibinfo{person}{Xiang Zhang}, {et~al\mbox{.}}} \bibinfo{year}{2018}\natexlab{}.
\newblock \showarticletitle{{SEAF: A Secure, Efficient and Accountable Access Control Framework for Information Centric Networking}}. In \bibinfo{booktitle}{\emph{IEEE Conference on Computer Communications (INFOCOM)}}. \bibinfo{publisher}{IEEE}, \bibinfo{pages}{2213--2221}.
\newblock
\showISBNx{9781538641286}
\showISSN{0743166X}


\bibitem[\protect\citeauthoryear{Yan and Tuninetti}{Yan and Tuninetti}{2021}]%
        {Yan2021}
\bibfield{author}{\bibinfo{person}{Qifa Yan} {and} \bibinfo{person}{Daniela Tuninetti}.} \bibinfo{year}{2021}\natexlab{}.
\newblock \showarticletitle{{Fundamental Limits of Caching for Demand Privacy Against Colluding Users}}.
\newblock \bibinfo{journal}{\emph{IEEE Journal on Selected Areas in Information Theory}} \bibinfo{volume}{2}, \bibinfo{number}{1} (\bibinfo{date}{Jan} \bibinfo{year}{2021}), \bibinfo{pages}{192--207}.
\newblock


\bibitem[\protect\citeauthoryear{Yang, Zhang, Zhang, Yu, Zhang, et~al\mbox{.}}{Yang et~al\mbox{.}}{2019}]%
        {Yang2019}
\bibfield{author}{\bibinfo{person}{Peng Yang}, \bibinfo{person}{Ning Zhang}, {et~al\mbox{.}}} \bibinfo{year}{2019}\natexlab{}.
\newblock \showarticletitle{{Content Popularity Prediction Towards Location-Aware Mobile Edge Caching}}.
\newblock \bibinfo{journal}{\emph{IEEE Transactions on Multimedia}} \bibinfo{volume}{21}, \bibinfo{number}{4} (\bibinfo{date}{Apr} \bibinfo{year}{2019}), \bibinfo{pages}{915--929}.
\newblock
\showISSN{15209210}


\bibitem[\protect\citeauthoryear{Yang and Kong}{Yang and Kong}{2016}]%
        {Yang2016}
\bibfield{author}{\bibinfo{person}{Qiuwei Yang} {and} \bibinfo{person}{Pan Kong}.} \bibinfo{year}{2016}\natexlab{}.
\newblock \showarticletitle{{RuleCache: A mobility pattern based multi-level cache approach for location privacy protection}}. In \bibinfo{booktitle}{\emph{IEEE 22nd International Conference on Parallel and Distributed Systems (ICPADS)}}. \bibinfo{publisher}{IEEE}, \bibinfo{pages}{448--455}.
\newblock
\showISBNx{9781509044573}
\showISSN{15219097}


\bibitem[\protect\citeauthoryear{Yu, Hu, Min, Lu, Zhao, et~al\mbox{.}}{Yu et~al\mbox{.}}{2018}]%
        {Yu2018}
\bibfield{author}{\bibinfo{person}{Zhengxin Yu}, \bibinfo{person}{Jia Hu}, {et~al\mbox{.}}} \bibinfo{year}{2018}\natexlab{}.
\newblock \showarticletitle{{Federated Learning Based Proactive Content Caching in Edge Computing}}. In \bibinfo{booktitle}{\emph{IEEE Global Communications Conference (GLOBECOM)}}. \bibinfo{publisher}{IEEE}, \bibinfo{pages}{1--6}.
\newblock
\showISBNx{9781538647271}


\bibitem[\protect\citeauthoryear{Yu, Hu, Min, Wang, Miao, et~al\mbox{.}}{Yu et~al\mbox{.}}{2022}]%
        {Yu2021b}
\bibfield{author}{\bibinfo{person}{Zhengxin Yu}, \bibinfo{person}{Jia Hu}, {et~al\mbox{.}}} \bibinfo{year}{2022}\natexlab{}.
\newblock \showarticletitle{{Privacy-Preserving Federated Deep Learning for Cooperative Hierarchical Caching in Fog Computing}}.
\newblock \bibinfo{journal}{\emph{IEEE Internet of Things Journal}} \bibinfo{volume}{9}, \bibinfo{number}{22} (\bibinfo{date}{May} \bibinfo{year}{2022}), \bibinfo{pages}{22246--22255}.
\newblock
\showISSN{23274662}


\bibitem[\protect\citeauthoryear{Yu, Hu, Min, Xu, and Mills}{Yu et~al\mbox{.}}{2020}]%
        {Yu2020a}
\bibfield{author}{\bibinfo{person}{Zhengxin Yu}, \bibinfo{person}{Jia Hu}, {et~al\mbox{.}}} \bibinfo{year}{2020}\natexlab{}.
\newblock \showarticletitle{{Proactive content caching for internet-of-vehicles based on peer-to-peer federated learning}}. In \bibinfo{booktitle}{\emph{IEEE 26th International Conference on Parallel and Distributed Systems (ICPADS)}}. \bibinfo{publisher}{IEEE}, \bibinfo{pages}{601--608}.
\newblock
\showISBNx{9781728190747}
\showISSN{15219097}


\bibitem[\protect\citeauthoryear{Yu, Hu, Min, Zhao, Miao, et~al\mbox{.}}{Yu et~al\mbox{.}}{2021}]%
        {Yu2020}
\bibfield{author}{\bibinfo{person}{Zhengxin Yu}, \bibinfo{person}{Jia Hu}, {et~al\mbox{.}}} \bibinfo{year}{2021}\natexlab{}.
\newblock \showarticletitle{{Mobility-Aware Proactive Edge Caching for Connected Vehicles Using Federated Learning}}.
\newblock \bibinfo{journal}{\emph{IEEE Transactions on Intelligent Transportation Systems}} \bibinfo{volume}{22}, \bibinfo{number}{8} (\bibinfo{date}{Aug} \bibinfo{year}{2021}), \bibinfo{pages}{5341--5351}.
\newblock
\showISSN{15580016}


\bibitem[\protect\citeauthoryear{Yuan, Wang, Wang, Chu, Wang, et~al\mbox{.}}{Yuan et~al\mbox{.}}{2016}]%
        {Yuan2016a}
\bibfield{author}{\bibinfo{person}{Xingliang Yuan}, \bibinfo{person}{Xinyu Wang}, {et~al\mbox{.}}} \bibinfo{year}{2016}\natexlab{}.
\newblock \showarticletitle{{Enabling secure and efficient video delivery through encrypted in-network caching}}.
\newblock \bibinfo{journal}{\emph{IEEE Journal on Selected Areas in Communications}} \bibinfo{volume}{34}, \bibinfo{number}{8} (\bibinfo{date}{Aug} \bibinfo{year}{2016}), \bibinfo{pages}{2077--2090}.
\newblock
\showISSN{15580008}


\bibitem[\protect\citeauthoryear{Zeng, Huang, Liu, and Yang}{Zeng et~al\mbox{.}}{2020}]%
        {Zeng2020}
\bibfield{author}{\bibinfo{person}{Yiming Zeng}, \bibinfo{person}{Yaodong Huang}, {et~al\mbox{.}}} \bibinfo{year}{2020}\natexlab{}.
\newblock \showarticletitle{{Privacy-preserving distributed edge caching for mobile data offloading in 5G networks}}. In \bibinfo{booktitle}{\emph{IEEE 40th International Conference on Distributed Computing Systems (ICDCS)}}. \bibinfo{publisher}{IEEE}, \bibinfo{pages}{541--551}.
\newblock
\showISBNx{9781728170022}


\bibitem[\protect\citeauthoryear{Zeng, Huang, Liu, Liu, and Yang}{Zeng et~al\mbox{.}}{2021}]%
        {ZengYi2021}
\bibfield{author}{\bibinfo{person}{Yiming Zeng}, \bibinfo{person}{Yaodong Huang}, {et~al\mbox{.}}} \bibinfo{year}{2021}\natexlab{}.
\newblock \showarticletitle{{Privacy-Preserving Decentralized Edge Caching in 5G Networks}}. In \bibinfo{booktitle}{\emph{IEEE 14th International Conference on Cloud Computing (CLOUD)}}. \bibinfo{publisher}{IEEE}, \bibinfo{pages}{189--199}.
\newblock


\bibitem[\protect\citeauthoryear{Zhang, Tan, Li, Han, Jiang, et~al\mbox{.}}{Zhang et~al\mbox{.}}{2022b}]%
        {Zhang2022d}
\bibfield{author}{\bibinfo{person}{Chi Zhang}, \bibinfo{person}{Haisheng Tan}, {et~al\mbox{.}}} \bibinfo{year}{2022}\natexlab{b}.
\newblock \showarticletitle{{Online File Caching in Latency-Sensitive Systems with Delayed Hits and Bypassing}}. In \bibinfo{booktitle}{\emph{IEEE Conference on Computer Communications (INFOCOM)}}. \bibinfo{publisher}{IEEE}, \bibinfo{pages}{1059--1068}.
\newblock
\showISBNx{9781665458221}
\showISSN{0743166X}


\bibitem[\protect\citeauthoryear{Zhang, Qu, Chen, Wang, Zhan, et~al\mbox{.}}{Zhang et~al\mbox{.}}{2021}]%
        {Zhang2021}
\bibfield{author}{\bibinfo{person}{Jie Zhang}, \bibinfo{person}{Zhihao Qu}, {et~al\mbox{.}}} \bibinfo{year}{2021}\natexlab{}.
\newblock \showarticletitle{Edge Learning: The Enabling Technology for Distributed Big Data Analytics in the Edge}.
\newblock \bibinfo{journal}{\emph{Comput. Surveys}} \bibinfo{volume}{54}, \bibinfo{number}{7}, Article \bibinfo{articleno}{151} (\bibinfo{date}{Jul} \bibinfo{year}{2021}), \bibinfo{numpages}{36}~pages.
\newblock
\showISSN{0360-0300}


\bibitem[\protect\citeauthoryear{Zhang, Hu, Liang, Li, and Gupta}{Zhang et~al\mbox{.}}{2023a}]%
        {Zhang2023}
\bibfield{author}{\bibinfo{person}{Shiwen Zhang}, \bibinfo{person}{Biao Hu}, {et~al\mbox{.}}} \bibinfo{year}{2023}\natexlab{a}.
\newblock \showarticletitle{{A Caching-based Dual K-Anonymous Location Privacy-Preserving Scheme for Edge Computing}}.
\newblock \bibinfo{journal}{\emph{IEEE Internet of Things Journal}} \bibinfo{volume}{10}, \bibinfo{number}{11} (\bibinfo{date}{Jan} \bibinfo{year}{2023}), \bibinfo{pages}{9768 --9781}.
\newblock


\bibitem[\protect\citeauthoryear{Zhang, Li, Tan, Peng, and Wang}{Zhang et~al\mbox{.}}{2019}]%
        {Zhang2019b}
\bibfield{author}{\bibinfo{person}{Shaobo Zhang}, \bibinfo{person}{Xiong Li}, {et~al\mbox{.}}} \bibinfo{year}{2019}\natexlab{}.
\newblock \showarticletitle{{A caching and spatial K-anonymity driven privacy enhancement scheme in continuous location-based services}}.
\newblock \bibinfo{journal}{\emph{Future Generation Computer Systems}}  \bibinfo{volume}{94} (\bibinfo{date}{May} \bibinfo{year}{2019}), \bibinfo{pages}{40--50}.
\newblock
\showISSN{0167739X}


\bibitem[\protect\citeauthoryear{Zhang, Wang, Li, Guo, Pei, et~al\mbox{.}}{Zhang et~al\mbox{.}}{2018}]%
        {Zhang2018}
\bibfield{author}{\bibinfo{person}{Xinyue Zhang}, \bibinfo{person}{Jingyi Wang}, {et~al\mbox{.}}} \bibinfo{year}{2018}\natexlab{}.
\newblock \showarticletitle{{Data-Driven Caching with Users' Local Differential Privacy in Information-Centric Networks}}. In \bibinfo{booktitle}{\emph{IEEE Global Communications Conference (GLOBECOM)}}. \bibinfo{publisher}{IEEE}, \bibinfo{pages}{1--6}.
\newblock
\showISBNx{9781538647271}


\bibitem[\protect\citeauthoryear{Zhang, Xiao, Zhou, Hu, Wu, et~al\mbox{.}}{Zhang et~al\mbox{.}}{2023b}]%
        {zhang2023a}
\bibfield{author}{\bibinfo{person}{Xianzhi Zhang}, \bibinfo{person}{Linchang Xiao}, {et~al\mbox{.}}} \bibinfo{year}{2023}\natexlab{b}.
\newblock \bibinfo{title}{cRVR: A Stackelberg Game Approach for Joint Privacy-Aware Video Requesting and Edge Caching}.
\newblock
\newblock
\showeprint[arxiv]{cs.NI/2310.12622}
\urldef\tempurl%
\url{https://arxiv.org/abs/2310.12622}
\showURL{%
\tempurl}


\bibitem[\protect\citeauthoryear{Zhang, Zhong, Fan, Bolodurina, and Cui}{Zhang et~al\mbox{.}}{2022c}]%
        {Zhang2022b}
\bibfield{author}{\bibinfo{person}{Xiaoyu Zhang}, \bibinfo{person}{Hong Zhong}, {et~al\mbox{.}}} \bibinfo{year}{2022}\natexlab{c}.
\newblock \showarticletitle{{CBACS: A Privacy-Preserving and Efficient Cache-Based Access Control Scheme for Software Defined Vehicular Networks}}.
\newblock \bibinfo{journal}{\emph{IEEE Transactions on Information Forensics and Security}}  \bibinfo{volume}{17} (\bibinfo{date}{May} \bibinfo{year}{2022}), \bibinfo{pages}{1930--1945}.
\newblock
\showISSN{15566021}


\bibitem[\protect\citeauthoryear{Zhang, Zhou, Wu, Hu, Zheng, et~al\mbox{.}}{Zhang et~al\mbox{.}}{2022d}]%
        {Zhang2022}
\bibfield{author}{\bibinfo{person}{Xianzhi Zhang}, \bibinfo{person}{Yipeng Zhou}, {et~al\mbox{.}}} \bibinfo{year}{2022}\natexlab{d}.
\newblock \showarticletitle{{Optimizing Video Caching at the Edge: A Hybrid Multi-Point Process Approach}}.
\newblock \bibinfo{journal}{\emph{IEEE Transactions on Parallel and Distributed Systems}} \bibinfo{volume}{33}, \bibinfo{number}{10} (\bibinfo{date}{Oct} \bibinfo{year}{2022}), \bibinfo{pages}{2597--2611}.
\newblock
\showISSN{15582183}


\bibitem[\protect\citeauthoryear{Zhang, Zhou, Wu, Sheng, Hu, et~al\mbox{.}}{Zhang et~al\mbox{.}}{2024}]%
        {zhang2024}
\bibfield{author}{\bibinfo{person}{Xianzhi Zhang}, \bibinfo{person}{Yipeng Zhou}, {et~al\mbox{.}}} \bibinfo{year}{2024}\natexlab{}.
\newblock \bibinfo{title}{PPVF: An Efficient Privacy-Preserving Online Video Fetching Framework with Correlated Differential Privacy}.
\newblock
\newblock
\showeprint[arxiv]{cs.MM/2408.14735}
\urldef\tempurl%
\url{https://arxiv.org/abs/2408.14735}
\showURL{%
\tempurl}


\bibitem[\protect\citeauthoryear{Zhang, Cao, Wang, Xiao, and Guan}{Zhang et~al\mbox{.}}{2022a}]%
        {Zhang2022a}
\bibfield{author}{\bibinfo{person}{Zizhen Zhang}, \bibinfo{person}{Tengfei Cao}, {et~al\mbox{.}}} \bibinfo{year}{2022}\natexlab{a}.
\newblock \showarticletitle{{VC-PPQ: Privacy-preserving Q-learning Based Video Caching Optimization in Mobile Edge Networks}}.
\newblock \bibinfo{journal}{\emph{IEEE Transactions on Network Science and Engineering}} \bibinfo{volume}{9}, \bibinfo{number}{6} (\bibinfo{date}{Aug} \bibinfo{year}{2022}), \bibinfo{pages}{4129--4144}.
\newblock
\showISSN{23274697}


\bibitem[\protect\citeauthoryear{Zhao, Mopuri, and Bilen}{Zhao et~al\mbox{.}}{2020}]%
        {Zhao2020}
\bibfield{author}{\bibinfo{person}{Bo Zhao}, \bibinfo{person}{Konda~Reddy Mopuri}, {et~al\mbox{.}}} \bibinfo{year}{2020}\natexlab{}.
\newblock \bibinfo{title}{iDLG: Improved Deep Leakage from Gradients}.
\newblock
\newblock
\showeprint[arxiv]{cs.LG/2001.02610}
\urldef\tempurl%
\url{https://arxiv.org/abs/2001.02610}
\showURL{%
\tempurl}


\bibitem[\protect\citeauthoryear{Zhao, Zhou, Chen, Zhou, Zhang, et~al\mbox{.}}{Zhao et~al\mbox{.}}{2023}]%
        {Zhao2023}
\bibfield{author}{\bibinfo{person}{Kongyange Zhao}, \bibinfo{person}{Zhi Zhou}, {et~al\mbox{.}}} \bibinfo{year}{2023}\natexlab{}.
\newblock \showarticletitle{EdgeAdaptor: Online Configuration Adaption, Model Selection and Resource Provisioning for Edge DNN Inference Serving at Scale}.
\newblock \bibinfo{journal}{\emph{IEEE Transactions on Mobile Computing}} \bibinfo{volume}{22}, \bibinfo{number}{10} (\bibinfo{year}{2023}), \bibinfo{pages}{5870--5886}.
\newblock


\bibitem[\protect\citeauthoryear{Zhao, Zhou, Jiao, Cai, Xu, et~al\mbox{.}}{Zhao et~al\mbox{.}}{2024}]%
        {Zhao2024}
\bibfield{author}{\bibinfo{person}{Kongyange Zhao}, \bibinfo{person}{Zhi Zhou}, {et~al\mbox{.}}} \bibinfo{year}{2024}\natexlab{}.
\newblock \showarticletitle{Taming Serverless Cold Start of Cloud Model Inference With Edge Computing}.
\newblock \bibinfo{journal}{\emph{IEEE Transactions on Mobile Computing}} \bibinfo{volume}{23}, \bibinfo{number}{8} (\bibinfo{year}{2024}), \bibinfo{pages}{8111--8128}.
\newblock


\bibitem[\protect\citeauthoryear{Zheng, Liu, Huang, Zhang, and Yang}{Zheng et~al\mbox{.}}{2022}]%
        {Zheng2022}
\bibfield{author}{\bibinfo{person}{Chong Zheng}, \bibinfo{person}{Shengheng Liu}, {et~al\mbox{.}}} \bibinfo{year}{2022}\natexlab{}.
\newblock \showarticletitle{{Unsupervised Recurrent Federated Learning for Edge Popularity Prediction in Privacy-Preserving Mobile-Edge Computing Networks}}.
\newblock \bibinfo{journal}{\emph{IEEE Internet of Things Journal}} \bibinfo{volume}{9}, \bibinfo{number}{23} (\bibinfo{year}{2022}), \bibinfo{pages}{24328--24345}.
\newblock
\showISSN{23274662}


\bibitem[\protect\citeauthoryear{Zhong, Li, and Liao}{Zhong et~al\mbox{.}}{2021}]%
        {Zhong2021}
\bibfield{author}{\bibinfo{person}{Yuqing Zhong}, \bibinfo{person}{Zhaohua Li}, {et~al\mbox{.}}} \bibinfo{year}{2021}\natexlab{}.
\newblock \showarticletitle{{A Privacy-Preserving Caching Scheme for Device-to-Device Communications}}.
\newblock \bibinfo{journal}{\emph{Security and Communication Networks}}  \bibinfo{volume}{2021} (\bibinfo{date}{Jan} \bibinfo{year}{2021}), \bibinfo{pages}{10958}.
\newblock
\showISSN{19390122}


\bibitem[\protect\citeauthoryear{Zhou, Wang, Xu, and Wu}{Zhou et~al\mbox{.}}{2019}]%
        {Zhou2019}
\bibfield{author}{\bibinfo{person}{Pan Zhou}, \bibinfo{person}{Kehao Wang}, {et~al\mbox{.}}} \bibinfo{year}{2019}\natexlab{}.
\newblock \showarticletitle{{Differentially-private and trustworthy online social multimedia big data retrieval in edge computing}}.
\newblock \bibinfo{journal}{\emph{IEEE Transactions on Multimedia}} \bibinfo{volume}{21}, \bibinfo{number}{3} (\bibinfo{date}{Mar} \bibinfo{year}{2019}), \bibinfo{pages}{539--554}.
\newblock
\showISSN{15209210}


\bibitem[\protect\citeauthoryear{Zhou, Liu, Fu, Wu, Wang, et~al\mbox{.}}{Zhou et~al\mbox{.}}{2023}]%
        {Zhou2023}
\bibfield{author}{\bibinfo{person}{Yipeng Zhou}, \bibinfo{person}{Xuezheng Liu}, {et~al\mbox{.}}} \bibinfo{year}{2023}\natexlab{}.
\newblock \showarticletitle{Optimizing the Numbers of Queries and Replies in Convex Federated Learning With Differential Privacy}.
\newblock \bibinfo{journal}{\emph{IEEE Transactions on Dependable and Secure Computing}} \bibinfo{volume}{20}, \bibinfo{number}{6} (\bibinfo{date}{Nov} \bibinfo{year}{2023}), \bibinfo{pages}{4823--4837}.
\newblock


\bibitem[\protect\citeauthoryear{Zhu, Xu, Li, Wang, and You}{Zhu et~al\mbox{.}}{2021}]%
        {Zhu2021}
\bibfield{author}{\bibinfo{person}{Pengcheng Zhu}, \bibinfo{person}{Jun Xu}, {et~al\mbox{.}}} \bibinfo{year}{2021}\natexlab{}.
\newblock \showarticletitle{{Learning-Empowered Privacy Preservation in beyond 5G Edge Intelligence Networks}}.
\newblock \bibinfo{journal}{\emph{IEEE Wireless Communications}} \bibinfo{volume}{28}, \bibinfo{number}{2} (\bibinfo{date}{Apr} \bibinfo{year}{2021}), \bibinfo{pages}{12--18}.
\newblock
\showISSN{15580687}


\end{thebibliography}

\newpage
\appendix

{\color{black}
\section*{Appendix}
As supplementary information, we present total five comprehensive tables to facilitate readability. All information is referenced in the main text. We first have compiled a summary of commonly used abbreviations for the solutions in Table~\ref{tab:abbreviations}.
\begin{table}[h]
    \caption{List of Common Abbreviations in this Paper.}
    \label{tab:abbreviations}
    \rowcolors{2}{white}{gray!25} 
    \begin{tabular}{p{2.5cm}<{\centering}p{4cm}||p{2.5cm}<{\centering}p{4.5cm}}
        \toprule
        \textbf{Abbreviation} &
        \textbf{Meaning}&
        \textbf{Abbreviation} &
        \textbf{Meaning}\\
        \midrule
        CDN & Content Delivery Network & CP(s) & Content Provider(s) \\
        EC(s) & Edge Cache(s) & ES(s) & Edge Server(s) \\
        EN(s) & Edge Node(s) & D2D & Device-to-Device \\
        (L)DP & (Local) Differential Privacy & 
        DL & Deep Learning \\
        FL &  Federated Learning&
        HE & Homomorphic Encryption \\
        ICN & Information-Centric Network&
        IoV & Internet of Vehicles \\
        ISP(s) & Internet Service Provider(s)&
        LBS & Location-Based Services \\
        ML & Machine Learning&
        PIR & Private Information Retrieval \\
        POI(s) & Point of Interest(s)  &
        PPEC & Privacy-Preserving Edge Caching \\
        RSU(s) & Roadside Unit(s)  &
        (D)RL & (Deep) Reinforcement Learning \\
        (S)BS(s) & (Small) Base Station(s)&
        SS & Secret Sharing \\
        TTP & Trusted Third Party&
        TDC & Trusted Distributed Computing \\
        \bottomrule
    \end{tabular}
\end{table}

For easy reference, we also present a classification matrix for the solutions introduced in this survey based on countermeasures and privacy data in the realm of edge caching in Table~\ref{Tab: countermeasures}.

\begin{table}[h]
\caption{Method classification based on countermeasures and privacy types.}
\renewcommand{\arraystretch}{1.1}
\resizebox{1\linewidth}{!}{
\begin{tabular}{|m{1.5cm}<{\centering}|m{2.1cm}<{\centering}||m{3cm}<{\centering}|m{1.9cm}<{\centering}|m{1.9cm}<{\centering}|m{1.7cm}<{\centering}|m{1.5cm}<{\centering}|m{1.6cm}<{\centering}|}
\toprule
\hline
\textbf{Classes}&\textbf{Methods}&\textbf{Request Record}&\textbf{Personal Information}&\textbf{Location}&\textbf{Extracted Knowledge}&\textbf{Private Content}&\textbf{Content Popularity}\\ \hline
\midrule
\multirowcell{3}{Noise-\\
based}
    &{DP}&\cite{Zhang2018, Wang2019, Zhou2019, Guo2022,Zhang2022a,Sivaraman2021,Xiong2022, zhang2024}&{\cite{Zhu2021,Zeng2020, ZengYi2021}}&{\cite{Zhang2022a}}&{\cite{Yu2021b, LuYun2020, Jiang2023, NAIR2023}}&{\cite{WangHu2022}}&{\cite{Yu2021b}}\\    \cline{2-8}
    &Obfuscation&\cite{Wu2016,Nikolaou2016,Qian2020,zhang2024,zhang2023a}&\cite{Wang2022}&\cite{Zhang2019b,Amini2011,GUYi-mingBAIGuang-weiSHENHang}&\cite{Wang2022a,Wang2022}&/&/\\   \cline{2-8}
	& Anonymity&\cite{Cui2020b}&\cite{Zhang2022b, Xue2019, Xue2018, Nguyen2023}&\cite{Nisha2022,Cui2020b,Sen2018,Hu2018,Yang2016, Zhang2023}&/&/&/\\
	 \hline
\multirowcell{3}{TDC-based}
        &{FL}&/&{\cite{Qiao2022,Wang2022}}&/&\cite{Cui2022,Qiao2022,Liu2022,Yu2021b,Zheng2022,Chen2022,Wang2022a,Wang2022,Saputra2022,Cheng2021,Li2020a,Wang2020,Yu2018,Wang2019a,Yu2020,Qi2020,Yu2020a, LuYun2020, Jiang2023, NAIR2023}&/&{\cite{Li2020a}}\\ \cline{2-8}
	&SS&\cite{Acs2019, Schlegel2022}&/&/&/&\cite{Pu2019}&\cite{Andreoletti2019}\\\cline{2-8}
	&\multirowcell{2}{Blockchain}&\multirowcell{2}{\cite{Qian2020}}&\multirowcell{2}{\cite{Lei2020,Dai2020,Vu2019, LiuJi2020}}&\multirowcell{2}{/}&\multirowcell{2}{\cite{Cui2022}}&\multirowcell{2}{/}&\multirowcell{2}{/}\\\cline{1-1}
    &&&&&&&\\\cline{2-8}
\multirowcell{2}{Cryptology\\-based}
		& Encryption  Communication&\cite{Leguay2017,Yuan2016a,Cui2020c,Jiang2020}&{\cite{Xue2019,Xue2018,Zhang2022b}}&/&{\cite{Chen2022}}&{\cite{Xu2019,Pu2019}}&{\cite{Cui2020,Araldo2018}}\\ \cline{2-8}
		& HE&\cite{Kong2019,Cui2020}&\cite{Kong2019}&/&\cite{Saputra2022}&/&/\\  \cline{2-8}
		& PIR&\cite{Tong2022,Yan2021,Kumar2019}&/&/&/&\cite{Tong2022}&/\\   \hline
\multirowcell{2}{Others}
		& Optimization&\cite{Sivaraman2021,Xiong2022,Hassanpour2023,Hassanpour2021}&/&/&/&\cite{Xu2019,Xu2020,Shi2018}&\cite{Andreoletti2019a}\\  
  \cline{2-8}
	    & {Access Control}&{\cite{Cui2020c,Jiang2020}}&{\cite{Lei2020, Zhang2022b, Cui2020c, Nguyen2023}}&/&/&{\cite{Xue2019,Xue2018,Cui2020c}}&/\\
         \hline
\bottomrule
\end{tabular}
}
\label{Tab: countermeasures}
\end{table}

In addition to the comprehensive introduction of major solutions for protecting user privacy in Section~\ref{sec: user privacy}, we also provide a supplementary introduction in Table~\ref{Tab: user privacy}. This table briefly outlines other solutions for safeguarding user privacy in edge caching systems that were not discussed in detail in Section~\ref{sec: user privacy}.

\begin{table*}[h]
\renewcommand\arraystretch{1.3}
\caption{A brief supplement of solutions to protect user privacy in the edge caching systems.}
\resizebox{1\linewidth}{!}{
\begin{tabular}{|m{60pt}<{\centering}| m{20pt}<{\centering}| m{50pt}<{\centering}| m{60pt}<{\centering}|m{225pt}<{\centering}|m{45pt}<{\centering}|}
\hline
\textbf{User Privacy}&\textbf{Refs.}&\textbf{Edge Cache Entities}&\textbf{Mitigation Methods}&\textbf{Key Ideas}&\textbf{Potential Attackers}\\ \hline
\midrule
\multirowcell{10}{Request Traces}
&\cite{Zhang2018}
&APs
&LDP
&Add LDP noise to the users' preference content information.
&CP
\\\cline{2-6}

&\cite{zhang2024}
&ESs
&LDP / Obfuscation
&Integrate the obfuscation and correlated differential privacy methods to generate content fetching and caching strategy.
&CP
\\\cline{2-6}

&\cite{ZengYi2021}
&SBSs
&LDP
& Add LDP noise to the caching policy when the spread of the caching policy is needed.
&Other SBSs
\\\cline{2-6}

&\cite{zhang2023a}
& ESs / Users Devices 
& Obfuscation / Opt.-based
& Develop a Stackelberg game to optimize the redundant request and caching strategy.
& CP / ESs / Other Users
\\\cline{2-6}

&\cite{Leguay2017}
&ESs
&Encrypt. Comm. / Pseudonyms
&Cache symmetrically encrypted content with pseudo-identifiers.
&ESs
\\\cline{2-6}

&\cite{Yan2021}
&Users Devices
&PIR
&Propose a collaborating caching scheme with encoding methods based on PIR.
&Other Users
\\\cline{1-6}

\multirowcell{3}{Request Traces /\\ Personal\\ Information}
&\cite{Jiang2020}
&ESs / Vehicles 
&Encrypt. Comm. / Access Control
&Design a double-layer encryption scheme to achieve access control and data integrity verification in the edge cache of IoV.
&Other Vehicles
\\\cline{2-6}

&\cite{Nguyen2023}
&ESs
&Opt.-based
&Introduce a novel distributed game-theoretic technique for collaborations among CP and ESs.
&CP
\\\cline{1-6}

\multirowcell{3}{Location}
&\cite{Yang2016}
&User Devices
&Anonymity 
&Disturb the real POI with $k$-anonymity method during interaction with LBS.
&LBS
\\\cline{2-6}

&\cite{Cui2020b}
&User Devices
&Anonymity 
&Combine peer-to-peer caching technique and $l$-diversity to reduce privacy exposure during interaction with LBS.
&LBS
\\\hline 
\bottomrule
\end{tabular}}
\label{Tab: user privacy}
\end{table*}

Additionally, Table~\ref{tab: solutions for knowledge privacy} provides a detailed classification of protection methods used to safeguard knowledge privacy in federated learning (FL) systems. The table highlights the effectiveness of various approaches in protecting different types of data, including the training dataset, model or gradient data, and other machine learning data. These methods encompass the original FL framework, combinations of FL with additional privacy protection techniques, and the use of noise-adding frameworks. By outlining these approaches, the table offers a comprehensive overview of the techniques applied to safeguard privacy in extracted knowledge, ensuring robust protection in FL environments.

\begin{table*}[h]
\renewcommand\arraystretch{1.5}
\caption{Protection methods of private information in extracted knowledge.}
\resizebox{1\linewidth}{!}{
\centering
\begin{tabular}{|m{5.1cm}||m{3.5cm}<{\centering}|m{1.8cm}<{\centering}|m{2cm}<{\centering}|m{2.5cm}<{\centering}|}
\toprule
\hline
\centering\textbf{Methods}&\centering\textbf{References}&\textbf{Training Dataset}&\textbf{Model or Gradient Data}&\textbf{Other Machine Learning Data}\\ \hline
\midrule
Origin FL framework 
&\cite{Yu2018,Wang2019a,Wang2020,Liu2022,Yu2020,Yu2020a,Li2020a}
&\Checkmark
&\XSolid  
&\XSolid 
\\\hline 
Combination of origin FL framework and other privacy protection methods
&\cite{Qi2020,Cheng2021,Zheng2022,Wang2022a,Saputra2022}
&\Checkmark
&\XSolid 
&\Checkmark
\\\hline
Noising FL framework 
&\cite{Yu2021b}
&\Checkmark
&\Checkmark
&\XSolid 
\\\hline 
Combination of noising FL framework and other privacy protection methods
&\cite{Wang2022,Cui2022,Chen2022}
&\Checkmark
&\Checkmark
&\Checkmark
\\\hline
\bottomrule
\end{tabular}
}
\label{tab: solutions for knowledge privacy}
\end{table*}

Lastly, we provide a supplementary overview of solutions for safeguarding knowledge privacy in edge caching systems in Table~\ref{Tab: knowledge privacy}. This table briefly summarizes additional solutions that were not discussed in detail in Section~\ref{sec: knowledge privacy}, offering a broader perspective on the various approaches used to protect knowledge privacy in edge caching environments.

\begin{table*}[h]
\renewcommand\arraystretch{1.5}
\caption{A brief supplement of solutions to protect knowledge privacy in the edge caching systems.}
\resizebox{1\linewidth}{!}{
\begin{tabular}{|m{38pt}<{\centering}| m{20pt}<{\centering}| m{50pt}<{\centering}| m{42pt}<{\centering}|m{245pt}<{\centering}|m{40pt}<{\centering}|}
\toprule
\hline
\textbf{Protected Entities}&\textbf{Refs.}&\textbf{Edge Cache Entities}&\textbf{Mitigation Methods}&\textbf{Key Ideas}&\textbf{Potential Attackers}\\ \hline
\midrule
Users
&\cite{Li2020a}
&BSs / User Devices
&Distributed ML
&Proposed a weighted distributed DRL model for edge caching replacement in D2D networks.
&BSs
\\\cline{1-6}
Vehicle Devices
&\cite{Yu2020a}
&RSUs / Vehicles
&FL
&Designed a peer-to-peer-based FL framework for proactive caching in vehicular edge networks.
&BS / RSUs 
\\\cline{1-6}
IoT Devices 
&\cite{Wang2020}
&BSs
&FL
&Proposed an FL-based cooperative edge caching framework with the DRL technique.
&BSs
\\\cline{1-6}
Users 
&\cite{ZengYi2021}
&User Devices
&FL / Blockchain
& Proposed a privacy-preserving D2D caching method with the combination of FL framework and a two-layer blockchain structure.
&User Devices
\\\cline{1-6}
Users
&\cite{Qi2020}
&BSs
&Noised-based FL
& Proposed FL-based method to predict content popularity with obfuscated feature information.
&BSs
\\\hline 
\bottomrule
\end{tabular}}
\label{Tab: knowledge privacy}
\end{table*}
}
\end{document}